\documentclass{aastex62} 

\newcommand{\calm}{\mbox{${\cal M}$}}
\newcommand{\caln}{\mbox{${\cal N}$}}
\newcommand{\calc}{\mbox{${\cal C}$}}
\newcommand{\bftheta}{\mbox{\boldmath $\theta$}}
\newcommand{\bfbeta}{\mbox{\boldmath $\beta$}}

\newcommand{\bfmu}{\mbox{\boldmath $\mu$}}
\newcommand{\bfpsi}{\mbox{\boldmath $\psi$}}

\newcommand{\BS}{\mbox{\boldmath $S$}}
\newcommand{\bx}{\mbox{\boldmath $x$}}
\newcommand{\st}{\sigma_{ii}+\lambda_{ji}}
\newcommand{\BY}{\mbox{\boldmath $Y$}}
\newcommand{\bpsi}{\mbox{\boldmath $\psi$}}
\newcommand{\Y}{\mbox{\boldmath $Y$}}
\newcommand{\Z}{\mbox{\boldmath $Z$}}
\newcommand{\BZ}{\mbox{\boldmath $Z$}}
\newcommand{\beps}{\mbox{\boldmath $\varepsilon$}}

\received{April 19, 2018}
\revised{May 31, 2018}
\accepted{June 13, 2018}
\submitjournal{ApJS}

\shorttitle{Asteroseismic modeling of stars with a convective core from gravity modes}
\shortauthors{C.~Aerts et al.}

\begin{document} 

\title{\Large Forward asteroseismic modeling of stars with a convective core 
from gravity-mode
  oscillations:  parameter estimation and stellar model selection}

\correspondingauthor{Conny Aerts}
\email{conny.aerts@kuleuven.be}

\author{C.\ Aerts}
\affiliation{Institute of Astronomy, KU\,Leuven, Celestijnenlaan 200D,
B-3001 Leuven, Belgium}
\affiliation{Department of Astrophysics, IMAPP, Radboud University
Nijmegen, P.O.\ Box 9010, 6500 GL Nijmegen, The Netherlands}
\affiliation{Kavli Institute for Theoretical Physics, University of California, Santa
Barbara, CA 93106, USA}

\author{G.\ Molenberghs}
\affiliation{I-BioStat, Universiteit Hasselt, Martelarenlaan 42, B-3500 Hasselt,
  Belgium}
\affiliation{I-BioStat, KU\,Leuven, Kapucijnenvoer 35, B-3000 Leuven, Belgium}
\affiliation{Kavli Institute for Theoretical Physics, University of California, Santa
Barbara, CA 93106, USA}

\author{M.\ Michielsen}
\affiliation{Institute of Astronomy, KU\,Leuven, Celestijnenlaan 200D,
B-3001 Leuven, Belgium}

\author{M.\ G.\ Pedersen}
\affiliation{Institute of Astronomy, KU\,Leuven, Celestijnenlaan 200D,
B-3001 Leuven, Belgium}

\author{R.\ Bj\"orklund}
\affiliation{Institute of Astronomy, KU\,Leuven, Celestijnenlaan 200D,
B-3001 Leuven, Belgium}

\author{C.\  Johnston}
\affiliation{Institute of Astronomy, KU\,Leuven, Celestijnenlaan 200D,
B-3001 Leuven, Belgium}

\author{J.\ S.\ G.\ Mombarg}
\affiliation{Institute of Astronomy, KU\,Leuven, Celestijnenlaan 200D,
B-3001 Leuven, Belgium}


\author{D.\ M.\ Bowman}
\affiliation{Institute of Astronomy, KU\,Leuven, Celestijnenlaan 200D,
B-3001 Leuven, Belgium}

\author{B.\ Buysschaert}
\affiliation{Institute of Astronomy, KU\,Leuven, Celestijnenlaan 200D,
B-3001 Leuven, Belgium}
\affiliation{LESIA, Observatoire de Paris, PSL Research University, CNRS,
  Sorbonne Universit\'es, UPMC Univ.\ Paris 06, Univ. Paris Diderot, Sorbonne
  Paris Cit\'e, 5 place Jules Janssen, 92195, Meudon, France}

\author{P.\ I.\ P\'apics}
\affiliation{Institute of Astronomy, KU\,Leuven, Celestijnenlaan 200D,
B-3001 Leuven, Belgium}

\author{S.\ Sekaran}
\affiliation{Institute of Astronomy, KU\,Leuven, Celestijnenlaan 200D,
B-3001 Leuven, Belgium}

\author{J.\ O.\ Sundqvist}
\affiliation{Institute of Astronomy, KU\,Leuven, Celestijnenlaan 200D,
B-3001 Leuven, Belgium}

\author{A.\ Tkachenko}
\affiliation{Institute of Astronomy, KU\,Leuven, Celestijnenlaan 200D,
B-3001 Leuven, Belgium}

\author{K.\ Truyaert}
\affiliation{Institute of Astronomy, KU\,Leuven, Celestijnenlaan 200D,
B-3001 Leuven, Belgium}

\author{T.\ Van Reeth}
\affiliation{Institute of Astronomy, KU\,Leuven, Celestijnenlaan 200D,
B-3001 Leuven, Belgium}

\author{E.\ Vermeyen}
\affiliation{Institute of Astronomy, KU\,Leuven, Celestijnenlaan 200D,
B-3001 Leuven, Belgium}

\begin{abstract}
  We propose a methodological framework to perform forward asteroseismic
  modeling of stars with a convective core, based on gravity-mode oscillations.
  These probe the near-core region in the deep stellar interior.  The modeling
  relies on a set of observed high-precision oscillation frequencies of
  low-degree coherent gravity modes with long lifetimes and their observational
  uncertainties.  Identification of the mode degree and azimuthal order is
  assumed to be achieved from rotational splitting and/or from period spacing
  patterns. This paper has two major outcomes. The first is a comprehensive list
  and discussion of the major uncertainties of theoretically predicted
  gravity-mode oscillation frequencies based on linear pulsation theory, caused
  by fixing choices of the input physics for evolutionary models.  Guided by a
  hierarchy among these uncertainties of theoretical frequencies, we
  subsequently provide a global methodological scheme to achieve forward
  asteroseismic modeling.  We properly take into account correlations amongst
  the free parameters included in stellar models.  Aside from the stellar mass,
  metalicity and age, the major parameters to be estimated are the near-core
  rotation rate, the amount of convective core overshooting, and the level of
  chemical mixing in the radiative zones.  This modeling scheme allows for
  maximum likelihood estimation of the stellar parameters for fixed input
  physics of the equilibrium models, followed by stellar model selection
  considering various choices of the input physics. Our approach uses the
  Mahalanobis distance instead of the often used $\chi^2$ statistic and includes
  heteroscedasticity.  It provides estimation of the unknown variance of the
  theoretically predicted oscillation frequencies.
\end{abstract}


\section{Introduction}

Asteroseismology of low-mass stars with stochastically-excited pressure-mode
oscillations often relies on simple scaling of solar oscillation
frequencies.  The so-called
scaling relations are easy to apply and provide estimates of the stellar mass,
radius, and age, assuming that the stars under consideration behave similarly
to the Sun \citep[e.g.,][for reviews]{ChaplinMiglio2013,HekkerJCD2017}. If the
aim is to improve the input physics of stellar models, however, one must go
beyond application of scaling relations. This requires testing various
assumptions and choices for the input physics that enters the computation of
the stellar equilibrium models upon which one relies for the estimation of
theoretical oscillation frequencies. In practice, estimation of various
parameters that are left free in the computation of equilibrium models is
required, along with proper error assessment.  Subsequently, statistical model
selection must be applied after the parameter estimation has been achieved, in
order to evaluate the capacity of the input physics in explaining the
asteroseismic data.

Approaches to achieve parameter estimation followed by evaluation of the input
physics from stellar model selection in asteroseismology have been developed 
and improved,
particularly since space photometry became available.  An enlightening
astrostatistics tutorial of various statistical methods for asteroseismology is
available in \citet{Appourchaux2014}.  
Usually, parameter estimation is done from forward modeling by adopting a 
grid-based approach and considering millions of stellar structure  
models of different evolutionary stage. Early applications to
  low-mass stars with solar-like oscillations were made by, e.g., 
\citet[][]{Miglio2005,Quirion2010}
and to subdwarfs by \citet{Brassard2001}.
In those studies, a $\chi^2$ 
  comparison between observed and theoretically predicted frequencies 
was adopted  to achieve
  the parameter estimation, while selection of the best model input physics
  among various choices was
  limited so far.  Our focus is upon this latter aspect. 
Applications in the space
asteroseismology era have so far also 
been focused on low-mass stars with solar-like
oscillations \citep[e.g.,][to list a
few]{Gruberbauer2013,Appourchaux-etal2014,Deheuvels2016,Bellinger2016,
SilvaAguirre2017} and on
white dwarfs \citep[e.g.][]{Giammichele2017}. 
\citet{Gruberbauer2012} developed a Bayesian framework that
includes systematic uncertainties due to surface-effects that occur for
solar-like oscillations \citep[e.g.,][]{Ball2017,Trampedach2017}. The cause of
this effect is unknown, yet it is dominant over other choices of the input
physics when fitting oscillation frequencies and must therefore be treated
properly.  These unknown surface effects can be avoided by fitting frequency
separations and ratios thereof, rather than the frequency values
themselves. This has been done in most applications of solar-like oscillations
so far.

The potential of the detected oscillations in terms of probing power is totally
different for gravity modes than for pressure modes. The latter mainly probe
envelope physics, while gravity modes dominantly probe near-core physics, and
mixed modes in evolved stars have combined capacity.  The exploitation of
gravity-mode oscillations in stars born with a convective core and a radiative
envelope (i.e., spectral type from early F to O) faces different challenges but offers
new opportunities compared to solar-like oscillations.  Gravity modes are
strongly affected, and thus optimally suited to probe three phenomena in the
deep stellar interior: convective core overshooting, near-core rotation and
chemical mixing.  The aim of the current paper is twofold: (i)\ to assess the
major uncertainties in predicted theoretical frequencies for the case of gravity
modes with long lifetime, leading to the requirement of
heteroscedasticity in the statistical treatment; (ii)\ to provide a methodological
framework for forward modeling, taking into account that gravity-mode
oscillations depend non-linearly on the free parameters of stellar
equilibrium models 
\citep[e.g.,][]{Townsend2013}
and that these parameters are correlated 
\citep[e.g., Fig.\,11 in][]{Papics2014}.

While thousands of low-mass stars were observed with $\mu$mag level precision by
{\it Kepler\/} \citep[see, e.g.,][for reviews]{ChaplinMiglio2013,HekkerJCD2017},
the satellite observed far fewer intermediate- and high-mass stars with
gravity-mode oscillations, because the exoplanet hunting from planetary transits
required observing a
large sample of cool low-mass stars as potential hosts.
For this reason, few stars with mass above 2\,M$_\odot$ in the
Field-of-View of the nominal mission were monitored.  Moreover, despite notable
new detections of non-radial oscillations in tens of such stars made with the
MOST \citep[e.g.,][]{Walker2005,Saio2006,Aerts2006,Cameron2008,Zwintz2013} and
CoRoT \citep[e.g.,][]{Poretti2009,Neiner2009,Degroote2010,Briquet2011,
  Papics2012,Zwintz2014} satellites, the time-series photometry assembled by
these two missions was limited from weeks to a few months.  While this is
sufficient to detect gravity modes, it is often insufficient to identify them in
terms of their spherical harmonic mode wavenumbers $(l,m)$.  Such mode
identification is a necessary pre-requisite to model the interior physical
properties of stars and to improve the input physics of theoretical models.

The bias towards asteroseismology of low-mass stars is also related to the time
scales of the oscillations.  Solar-like oscillations of low-mass stars are
excited stochastically in their outer convective envelope and have periodicities
ranging from minutes for the core-hydrogen burning phase to hours for red
giants.  This stands in sharp contrast to the coherent gravity modes with
periods up to several days, leading to complex beating patterns among the modes
that may cover several years
\citep[e.g.,][]{Kurtz2014,Saio2015,VanReeth2015,Bowman2016,Papics2017}.  
Such modes are
self-driven by a heat- or flux-blocking mechanism \citep[e.g.][Chapters 2 and
3]{Aerts2010} and have lifetimes of tens to millions of years, i.e., 
long compared
with the duration of the data sets
(cf.\ Fig.\,\ref{AD-nonAD} discussed below).
Such gravity modes occur in stars with
masses from roughly 1.4 to 40\,$M_\odot$. Our long-term aim to improve the
evolutionary models of such intermediate- and high-mass stars using
asteroseismology is motivated by the fact that they are the dominant suppliers
of the chemicals in galaxies, while their evolution models are far more
uncertain than those of low-mass stars.

The evolution of stars born with a well-developed convective core is appreciably
affected by the size and properties of their convective core region, the
interior rotation, and the transport of chemicals and angular momentum in the
radiative envelope.  Asteroseismic (hereafter referred to as seismic) modeling
of such stars had to await proper and unique mode identification for a
sufficient number of detected non-radial oscillations.  Early achievements were
focused on the identification of a few low-order pressure modes detected in
extensive ground-based data of a few bright A and B-type stars \citep[e.g.,][to
list a few
examples]{Breger1999,Aerts2003,DeRidder2004,Handler2006,Handler2009a}.  At best,
these studies provided a rough estimate of the core overshooting and
core-to-envelope rotation because the pressure modes have limited probing power
near the stellar core
\citep[e.g.,][]{Dupret2004,Dziembowski2008,Handler2009b,Briquet2012}.  Since the
4-year nominal {\it Kepler\/} mission, seismic modeling based on numerous
identified gravity-mode oscillations can be achieved with sufficiently high
precision to improve the input physics of stellar evolution models of stars with
$M\gtrsim 1.4\,$M$_\odot$. 
Meanwhile, several stars have been modeled
\citep{Kurtz2014,Saio2015,Moravveji2015,Moravveji2016,SchmidAerts2016,VanReeth2016,Sowicka2017,Kallinger2017,Szewczuk2018},
but at a very different level of depth, relying on a variety of fixed choices for the
input physics, and making diverse assumptions on
the importance of the free parameters of the stellar models.
Now that suitable data for such inference studies are available, we need an
appropriate statistical methodology to perform seismic modeling, which goes
beyond the methods currently used.  This is the topic of the current paper. We
focus on core-hydrogen burning stars for which the {\it Kepler\/} mission
provided suitable data, but our methods are readily applicable to stars with a
convective core in more advanced stages of stellar evolution.

\section{\label{ingredients}The physical ingredients 
for stellar oscillation computations}

\subsection{Preliminaries}

The seismic modeling presented here involves comparing the values of observed
gravity-mode oscillation frequencies with those predicted from theoretical
stellar structure models based on chosen input physics for a given set of
stellar parameters. We opt to focus the application on 
gravity modes because they have the highest mode energy in the
near-core regions that we wish to probe \citep[e.g., Fig.\,2 in][for typical
examples]{Triana2015,VanReeth2016}.  These modes are far less sensitive to poor
descriptions of the physical conditions in the outer envelope of a star, so we
are not in need of corrections for surface effects as in the case of high-order
pressure modes \citep[e.g.,][]{Ball2017,Trampedach2017}.

We take an {\it observationally-driven\/} approach in the sense that we use all
detected oscillation frequencies, irrespective of the excitation mechanism, as
long as the mode frequency has unambiguous mode identification in terms of its
spherical harmonic wavenumbers $l$ and $m$.  Together with the mode frequency
and the radial and horizontal amplitude, wavenumbers define the displacement
vector due to the mode (see Eq.\,(3.132) in \citet{Aerts2010}, and Section 3.3
in that monograph for a general description of linear stellar
oscillations). Although we have a measurement of the surface-integrated
amplitude of the modes in the line-of-sight, theories to predict the
  intrinsic amplitudes for coherent modes remain challenging.  While amplitude
  predictions for radial pulsators based on first principles are available to
  some extent \citep[e.g.][]{Smolec2007, Geroux2013}, amplitude-saturation
  mechanisms based on non-linear mode coupling of non-radial modes have only
  been developed for pressure modes in main-sequence A-type stars
  \citep[e.g.,][]{Dziembowski1985,Buchler1997,Nowakowski2005} and not for
  gravity modes in rotating stars.  The predictive power of these available
  non-linear mode-coupling computations is insufficient to rely on them in
  forward modeling.   Moreover, although generally successful, details of the
excitation mechanisms are insufficiently understood to rely on them for seismic
modeling because modes predicted to be excited are often not observed or vice
versa. Hence, a secure way to proceed in evaluating the input physics of stellar
structure models from detected oscillations is to rely solely on their measured
frequency values and visit the mode excitation problem after the forward
  modeling has delivered the most likely models. The current paper focuses only
  on the modeling of detected frequencies and parameter estimation; 
for extensive excitation computations of
  intermediate-mass and high-mass pulsators, we refer to the recent paper by
  \citet{Szewczuk2017}.

As further discussed below,
coherent gravity modes have long lifetimes compared
to the duration of the data sets. It is hence fully justified to assume they
cause a delta-function convolved with the spectral window function
in a Fourier transform and the observational frequency uncertainties
are inversely proportional to the total time base of the data.  Here, we adopt the
notation that $f_i^\ast$ is the observed cyclic oscillation frequency of
mode $i$, which has known error $\varepsilon_i^\ast$, for $i=1, \ldots, n$.

The typical uncertainty of measured oscillation frequencies derived from data
with high duty cycle that does not suffer from aliasing, as is the case for
contemporary high-cadence uninterrupted space photometry, is well below the
resolving power of the data set. This resolving power is about 2.5 times the
Rayleigh limit of the data set \citep{Loumos1978}, the latter being equal to the
inverse of the total time base of the data. The Rayleigh limit is
0.00068\,d$^{-1}$ (0.0079$\mu$Hz) for the 4-year nominal {\it Kepler\/} data
\citep[e.g.,][]{Bowman2017} and 0.0066\,d$^{-1}$ (0.077$\mu$Hz) for a CoRoT
150-d long run \citep[e.g.,][]{Degroote2009}.  Recent BRITE photometry has time
spans between roughly 100 to 180\,d and leads to frequency resolutions similar
to CoRoT long runs for the brightest stars in the sky \citep{Pablo2016}.  

Each oscillation frequency has its own measurement error $\varepsilon_i^\ast$,
because it is dependent on the measured mode amplitude and on the noise
properties of the data in that particular frequency regime, aside from the total
time base \citep[see, e.g., Eq.\,(5.52) in][]{Aerts2010}. For all significant
frequencies detected from well sampled uncorrelated data that do not suffer from
systematic uncertainty caused by aliasing, the frequency error is at least a
factor of ten below the Rayleigh limit.  An even larger factor applies to
oversampled space photometry, but one has to correct for the correlated nature
of the data \citep{Degroote2009}.  Methods of frequency analysis and
  prewhitening applied to heat-driven pulsators are often based on the
  assumptions of uncorrelated, homoscedastic white Gaussian noise.
  \citet{Degroote2009} showed that these assumptions are not strictly true for
  the CoRoT asteroseismology data (32\,s sampling), but that the deviations
  thereof are often suffciently small to still apply the methods, provided that
  a correction for correlated data is used. However, the prewhitening process
  inherently implies introduction of uncertainties connected with the limited
  resolving power of the data set, even for the 4-year long {\it Kepler\/} data,
  and may result in dependencies among the frequencies and their errors. For
  these reasons, a good approach is to allow for heteroscedasticity in the
  measurement errors whenever frequencies deduced from a prewhitening procedure
  are used in forward modeling.

For CoRoT/BRITE and {\it Kepler\/} data of
core-hydrogen burning gravity-mode pulsators, the uncertainties of the
measured frequencies that satifsy the commonly used significance criterion of
having an amplitude above four times the noise level \citep{Breger1993} are
below 0.001 and 0.0001\,d$^{-1}$, respectively.  Any comparison with
theoretically predicted frequencies based on stellar models hence requires the
frequencies to be computed at such levels.
For all the above reasons, we consider 0.001\,d$^{-1}$ to be a typical 
frequency error derived from a quasi-uninterrupted half- to one-year
   light curve and we allow for heteroscedasticity in the statistical methodology.

\subsection{Stellar structure models}

In order to compute theoretical cyclic frequencies of oscillations, denoted here
as $f_i^{\rm th}$, we rely on stellar structure models in hydrostatic
equilibrium and perturb them in a linear approach.  In this work, we use
spherically symmetric 1D stellar models for single stars, computed with the
publicly available MESA code; we refer to the extensive papers by
\citet{Paxton2011,Paxton2013,Paxton2015,Paxton2018} for a full description of
the suite of routines, the available choices of the input physics and the list
of the numerous free parameters that one can use, as well as the code-of-conduct
adopted by the MESA development team.  Of course, our methodology can also be
applied to any other stellar evolution code than the one we use here to
illustrate it. As a word of caution, we stress that any user must check
carefully that the stellar structure models computed numerically comply to a
high level of precision with the equations of stellar structure upon which they
are based in the first place.  Performing checks (e.g., on choosing mesh points
and tolerances for convergence) to make sure that the models are astrophysically
appropriate and that the oscillations are computed with sufficient
  numerical accuracy is the responsibility of the user (not of the code
developers!). This is particularly relevant for applications to gravity modes,
whose mode cavities are highly sensitive to discontinuities in the interior
profiles of the physical quantities.  Typically, gravity mode predictions
  require some 5,000 to 10,000 meshpoints within the star to meet the
  observational frequency errors, not necessarily equidistantly distributed.
Hence, gravity-mode computations need to be done
  with a refined mesh compared to computations that only require smooth
  evolutionary tracks in the Hertzsprung-Russell diagram or predictions of
  surface abundances (for which $\sim$1,000 mesh points is usually sufficient). 
An appropriate mesh is particularly important in
  the transition regions between convective and radiative zones, to ensure
  that the Bruntt-V\"ais\"al\"a (BV) frequency and $\mu-$gradients are well
  computed as these determine the propagation cavity of the gravity
  modes. A typical MESA inlist based on 5,000 meshpoints
is provided through the links in  Appendix\,\ref{Inlists}.

Less than 10\% of all early-F to O-type stars have a surface magnetic field at
the current observational threshold of spectropolarimetry \citep[typically a few
Gauss,\ ][]{Wade2016}. Moreover, it is currently unknown whether magnetic fields
occur at the interface of the convective core and radiative envelope, let alone
what their shape and strength is. Theoretical predictions, e.g., as in
\citet{Braithwaite2009}, are too uncertain to be used as input physics. Rather
they can be evaluated with asteroseismology.  In this way, it was recently found
that the effect of the Lorentz force on oscillation frequencies is far below that
of the Coriolis force in a magnetic gravity-mode pulsator
\citep{Buysschaert2018}.

Ignoring rotational effects in equilibrium models
is less justified than magnetism, because a large fraction of early-F to O stars
are fast rotators \citep{ZorecRoyer2012}.
Interior rotation or magnetism of stars introduce a multitude of
  instabilities and accompanying transport of elements and of angular momentum
  \citep[][for a comprehensive description]{Heger2000}.  Each of these phenomena
  has been implemented in MESA, with its own diffusive mixing and angular
  momentum transport coefficient as a free parameter.  However, these
  ingredients are subject to considerable uncertainties because they 
remain essentially
  uncalibrated. The theoretical descriptions of these instabilities give rise to
  steep and discontinuous local fluctuations in the chemical mixing profile
  throughout the radiative envelope.  While the consequences of these
  fluctuations remain invisible in evolutionary tracks, they affect the gravity
  modes appreciably.  Because their nature may be numerical rather than physical
  \citep[e.g., Figs 28 and 29 in][]{Paxton2013}, we performed numerous tests on
  the effect of each of these instabilities separately, as well as jointly, on
  gravity-mode predictions \citep{Truyaert2016}.  From these tests, we deduce
  that the current prescriptions of the instabilities induced by rotation
  or magnetism are not of practical use for forward seismic modeling of gravity
  modes. Either the discontinuities induced by the instabilities are physical
  and we conclude from the gravity-mode computations that real stars with such
  detected oscillations do not behave according to the theoretical predictions, 
or they are numerical in nature  and thus the MESA 
models that include them are not applicable.  It is therefore better
  not to rely on the theoretical uncalibrated
descriptions for these instabilities, but rather to
  estimate the global level of macroscopic chemical 
mixing in the radiative envelope, as
  of now denoted as $D_{\rm ext}(r)$, from forward seismic modeling. We therefore 
assume that the magnetic and centrifugal forces are negligible for the
computation of the equilibrium models along the evolutionary track. 

Deviations
from spherical symmetry due to magnetism or rotation are thus only treated at
the level of the non-radial oscillations.  This common approach allows us to
evaluate the quality of the physical ingredients of the stellar models by
exploiting the differences between $f_i^{\rm th}$ and $f_i^\ast$, keeping
in mind $\varepsilon_i^\ast$. This approach is suitable as long as the
departure from spherical symmetry of the equilibrium models is modest and linear
pulsation theory is applicable
\citep[e.g.][]{Ledoux1951,Saio1981,Dziembowski1996,Townsend2003a}.
Current pulsation computations taking full account of the
deformation due to rotation are at best 2D and limited in the
model assumptions \citep[see][for pioneering work]{Ballot2012,Rieutord2016}. One
of our aims is to provide guidance to future 2D models from application of our
asteroseismic methodology. At present, such applications should be restricted to
stars rotating at less than about half of their critical velocity.

Equilibrium models of stellar structure for the core-hydrogen burning phase
require as basic input parameters the birth mass $M$ and the 
mass fractions representing the initial chemical composition:
$(X,Z)$ or $(X,Y)$, where $X$ represents hydrogen, $Y$ helium, and $Z$ the
  ensemble of all
  other chemical elements. Here, we work with $(M,X,Z)$.
For these three
parameters, evolutionary models are computed
as of the zero-age main sequence (ZAMS), for
which $X_c=X$ with $X_c$ the hydrogen mass fraction in the convective core. 
The latter is a
proxy for the stellar age during the core-hydrogen burning stage, which is the
application we focus upon here.

\section{\label{pulsation-error} Theoretical frequency
  uncertainties stemming from pulsation theory}

We assume that a ``sparse'' grid of equilibrium models for $(M,X,Z)$ has
  been computed, with typical grid steps for intermediate-mass stars as in
  Table\,1 of \citet{VanReeth2016}.  We stress that the application of forward
  modeling requires pulsation computations to be done along the full
  evolutionary track from ZAMS to TAMS, with a relatively small step in $X_c$
  (typically 0.001). Moreover, it is essential to keep in mind that there is a
  large difference in mode density as a function of $X_c$ when performing
  modeling.  Indeed, as discussed in detail by \citet[][see Figs\,2 and 3 and
  the discussion in Sect.\,3.2]{Buysschaert2018}, unconstrained forward modeling
  based on frequency fitting without knowing the identification of the modes
  artificially gives preference to near-TAMS models because there are many more
  modes for a fixed frequency range at evolved stages ($X_c\leq 0.1$) than at
  earlier stages. For evolutionary stages beyond hydrogen burning, it is even
  more critical to have sufficiently small time steps given the much shorter
  timescales. Here, we only treat the case from ZAMS to TAMS.

For chosen time steps along the evolutionary track, or
equivalently $X_c$ values, the equilibrium models are perturbed in a linear
framework to compute the 
3D eigenvectors of the oscillation modes and their accompanying  
eigenfrequencies. In this way, one obtains  
the spectrum of linear non-radial oscillations at a particular $X_c$ value. 
Aside
from these four basic parameters $(M,X,Z,X_c)$, a multitude of choices for the
input physics occurs. In Sect.\,\ref{models-error}, we offer a hierarchy among
choices in the specific case of stars with a convective core on the main
sequence, from the viewpoint of the uncertainties for the theoretical
eigenfrequencies of gravity modes.

\subsection{The traditional approximation}

In practice, once we have an equilibrium 1D stellar structure model, its
non-radial oscillation mode frequencies are computed with the publicly available
code GYRE \citep{Townsend2013,Townsend2018}.
  We adopt an adiabatic framework, the Cowling approximation
  \citep{Cowling1941}, a static atmosphere model and the standard inner and
  outer boundary conditions as outlined in Appendix\,A of
  \citet{Townsend2013}. A typical GYRE input list is given in
  Appendix\,\ref{Inlists}.  Furthermore, we use the traditional approximation
  \citep[e.g.,][hereafter denoted as
  TA]{Eckart1960,Chapman1970,Unno1989,Bildsten1996,Townsend2003a} for the
  computation of the gravity modes.
  Mathematically, the TA is only valid for high-order gravity modes, because it
  requires the mode frequency to remain far below the BV frequency and the
  tangential eigenvector component to be much larger than the radial
  component. Only in that case can the radial component of the Coriolis force
  be neglected with respect to the buoyancy force in the linearized momentum
  equation. However, the TA also provides a good approximation
  for low-order gravity modes \citep[e.g., Sect.\,4.2 in][]{Lai1997}.

The TA allows oscillation
modes to be computed from the Laplace tidal equations using an analytical
approach \citep[see, e.g.,][for comprehensive
descriptions]{Townsend2003b,Mathis2013LNP} and offers an excellent framework for
the computation of gravity-mode oscillations of slowly to moderately rotating
stars that are not too deformed from spherical symmetry 
by rotation.  From polytropic
models, \citet{Ballot2010,Ballot2012} compared 1D and 2D treatments to predict
gravity-mode frequencies and showed the TA to be acceptable in the
super-inertial regime: $f_i^\ast\lesssim 2 f_{\rm rot}$, with $f_{\rm rot}$
the cyclic rotation frequency of the star ($\Omega_{\rm rot}/2\pi$, often called
the rotation rate). This remains valid even when the rotation rate is a significant
fraction of the critical rotation rate (recall that the centrifugal distortion is
proportional to $\Omega_{\rm rot}^2$). 

The limits of the TA were recently re-evaluated for {\it Kepler\/} gravity-mode
pulsators of spectral type F by \citet{Ouazzani2017}. Formally, these
  authors found the TA to be appropriate for retrograde gravity modes with
  frequencies up to at least a quarter of the critical rotation rate; for
  prograde and zonal modes the TA can be used up to about half the critical
  rate.  However, the TA was even applied to the Be star HD\,163868 rotating
  near-critically \citep{Dziembowski2007} and compared with a second-order
  perturbative treatment developed by \citet{LeeBaraffe1995} for several Be
  stars based on MOST photometry \citep{Cameron2008} and for the F-type star
  KIC\,5608334 from {\it Kepler\/} photometry \citep{Saio2018b}. Despite a
  different behavior in the mode trapping \citep[Fig.\,6 in][]{Saio2018b}, both
  treatments were found to be similarly appropriate when compared with the
  observational uncertainties of the detected frequencies in the 4-year {\it
    Kepler\/} photometry of the $\gamma\,$Dor star KIC\,5608334.  Hence, it is
  very sensible to perform forward seismic modelling in the TA, given its
  computational convenience.

The TA is not needed for  stars that rotate well below their critical
rate \citep[][and Fig.\,\ref{TA-nonTA} below]{Reese2006}. For
those, first-order rotational splitting in the Ledoux approximation
\citep{Ledoux1951} is sufficient to predict the theoretical frequencies. Such
low rotation rates occur for a minority of intermediate-mass and high-mass
stars. Nevertheless, pulsators with ultra-slow rotation and high
amplitudes have been found \citep[e.g.,][]{Aerts2003,Kurtz2014,Triana2015}.
\citet{SchmidAerts2016} made a comparison between the theoretical frequency
values for a {\it Kepler\/} binary with two F-type hybrid pulsators and found the
Ledoux splitting to be inadequate for pulsation frequency matching of gravity
modes when $f_{\rm rot}\gtrsim 0.1\,$d$^{-1}$. Here, we conduct a more systematic
test by providing an overview of the differences between oscillation frequencies
computed with and without the Coriolis force, for a range of stellar
masses.  

We computed theoretical frequencies for zonal dipole modes for a set of 12
baseline models with parameters listed in Table\,\ref{baseline}.  While
  any forward seismic modeling application to a real star needs to consider the
  full evolutionary stage from the ZAMS to the TAMS,
  we illustrate the differences between the mode frequencies/periods 
of these benchmark  models versus those of 
models of the same $X_c$ but with different input physics or
  different pulsational computations for only three snapshots: close to ZAMS
  ($X_c=0.7$), close to TAMS ($X_c=0.1$) and about halfway through the
  core-hydrogen burning ($X_c=0.4$). We provide
a few example period spacing
  patterns in Fig.\,\ref{DeltaP-Rotation} below,
and histograms of the frequency differences
  for all comparisons we made. The masses of the benchmark models 
were chosen so as to encapsulate roughly the mass range for which high-order
  gravity modes have been detected so far.
The lowest mass models have 1.4$\,M_\odot$ with a growing convective core
and a thin convective outer envelope throughout the main sequence, while the three
other masses of 2.8$\,M_\odot$, 14$\,M_\odot$, and 28$\,M_\odot$ have a
  shrinking core and only develop a
convective envelope past the TAMS. The two highest-mass models are subject to a
considerable radiation-driven wind throughout their evolution.
\begin{table}
\caption{\label{baseline}Parameters of 12 benchmark models used 
to estimate theoretical
  frequency uncertainties of their 660 zonal dipole modes with radial order
  $n_{pg}\in [-50,+5]$, where $n_{pg}<0$ for gravity modes and $n_{pg}>0$ for
  pressure modes.} 
\begin{tabular}{ccrr}
\hline
Mass  (M$_\odot$) & $X_c$  & $\Omega_{\rm crit}$ (rad\,d$^{-1}$) & 
$\dot{M} (10^{-9}\,$M$_\odot$/yr)\\
\hline
1.4$\,M_\odot$ & 0.7 & 22.8 & 0.0 \\
1.4$\,M_\odot$ & 0.4 & 16.6 & 0.0 \\
1.4$\,M_\odot$ & 0.1 & 12.1 & 0.0 \\
\hline
2.8$\,M_\odot$ & 0.7 & 19.5 & 0.00099 \\
2.8$\,M_\odot$ & 0.4 & 11.1 & 0.00135 \\
2.8$\,M_\odot$ & 0.1 & 5.4 & 0.00066 \\
\hline
14$\,M_\odot$ & 0.7 & 11.0 & 7.0 \\
14$\,M_\odot$ & 0.4 & 6.3 & 13.4 \\
14$\,M_\odot$ & 0.1 & 2.8 & 250.4 \\
\hline
28$\,M_\odot$ &  0.7 & 8.7 & 303.0\\
28$\,M_\odot$ &  0.4 & 4.8 & 539.0\\
28$\,M_\odot$ &  0.1 & 1.6 & 417.0\\
\hline
\end{tabular}
\tablecomments{Fixed choices for the input parameters and 
  input physics of these benchmark models are: $\alpha_{\rm MLT}=2.0$, 
  exponential core convective overshooting with $f_{\rm ov}=0.02$ and the
  radiative temperature gradient in the overshoot zone
  (undershooting below the convective outer envelope is not included), 
  $Z=0.014$, chemical mixture in
  \protect\citet{Przybilla2013}, OPAL opacities, the MESA EOS, 
  the NACRE rates, the basic rate
  network based on the eight isotopes: $^1$H, $^3$He, $^4$He, $^{12}$C, $^{14}$N,
  $^{16}$O, $^{20}$Ne, and $^{24}$Mg \citep{Paxton2011}, 
  no rotation, no atomic
  diffusion,  and an Eddington gray atmosphere. 
  The oscillation
  frequencies are computed in the adiabatic approximation while ignoring the
  Coriolis force; $\Omega_{\rm crit}$  stands for the angular critical rotation rate. We use
  the so-called predictive mixing scheme of MESA (available as of version 10000) 
  to determine the position of the
  convective core boundary \citep{Paxton2018}.}
\end{table}

\begin{figure}
\begin{center}
\includegraphics[width=3.5in]{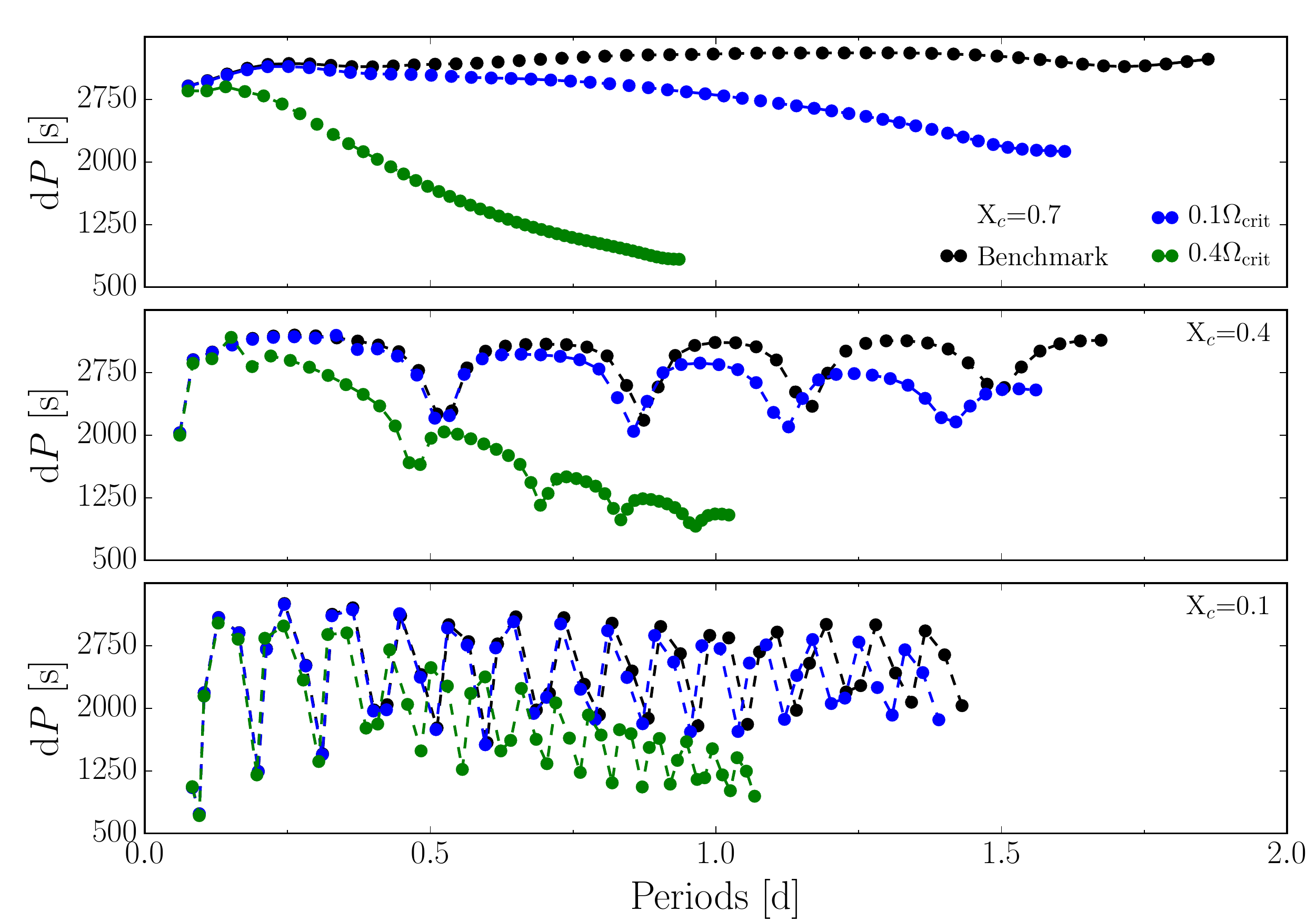}
\includegraphics[width=3.5in]{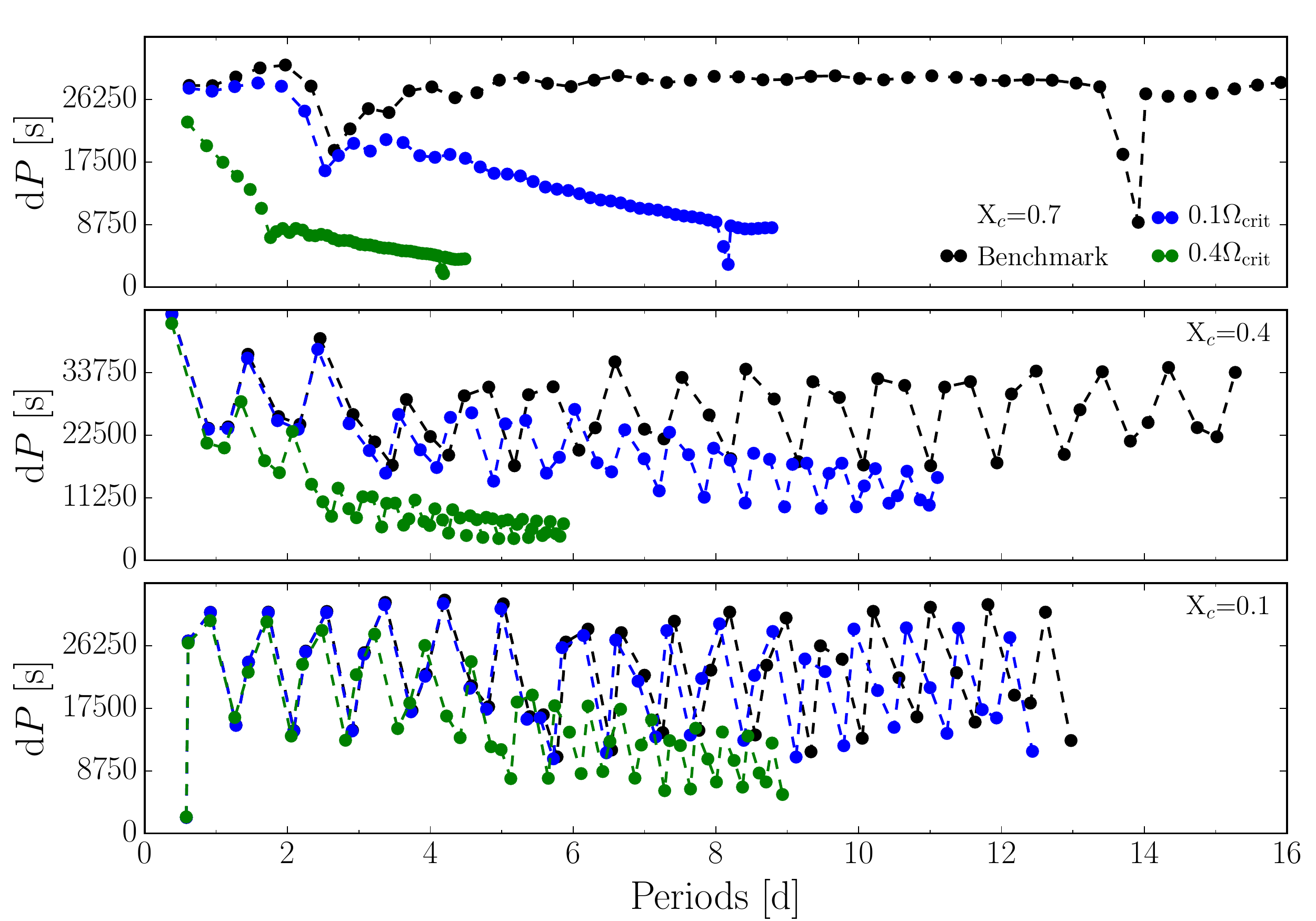}
 \caption{\label{DeltaP-Rotation} Gravity-mode period spacing patterns for a
   1.4\,M$_\odot$ (left) and a 28\,M$_\odot$ (right) model at $X_c= 0.7, 0.4, 0.1$,
 for two indicated rotation rates taken into account
   in the pulsation computations versus those for a non-rotating (benchmark) 
model. Mode trapping occurs for all cases, except the 1.4\,M$_\odot$ model at
$X_c=0.7$.}
\end{center}
\end{figure}

\begin{figure}
\begin{center}
 \includegraphics[width=3.5in]{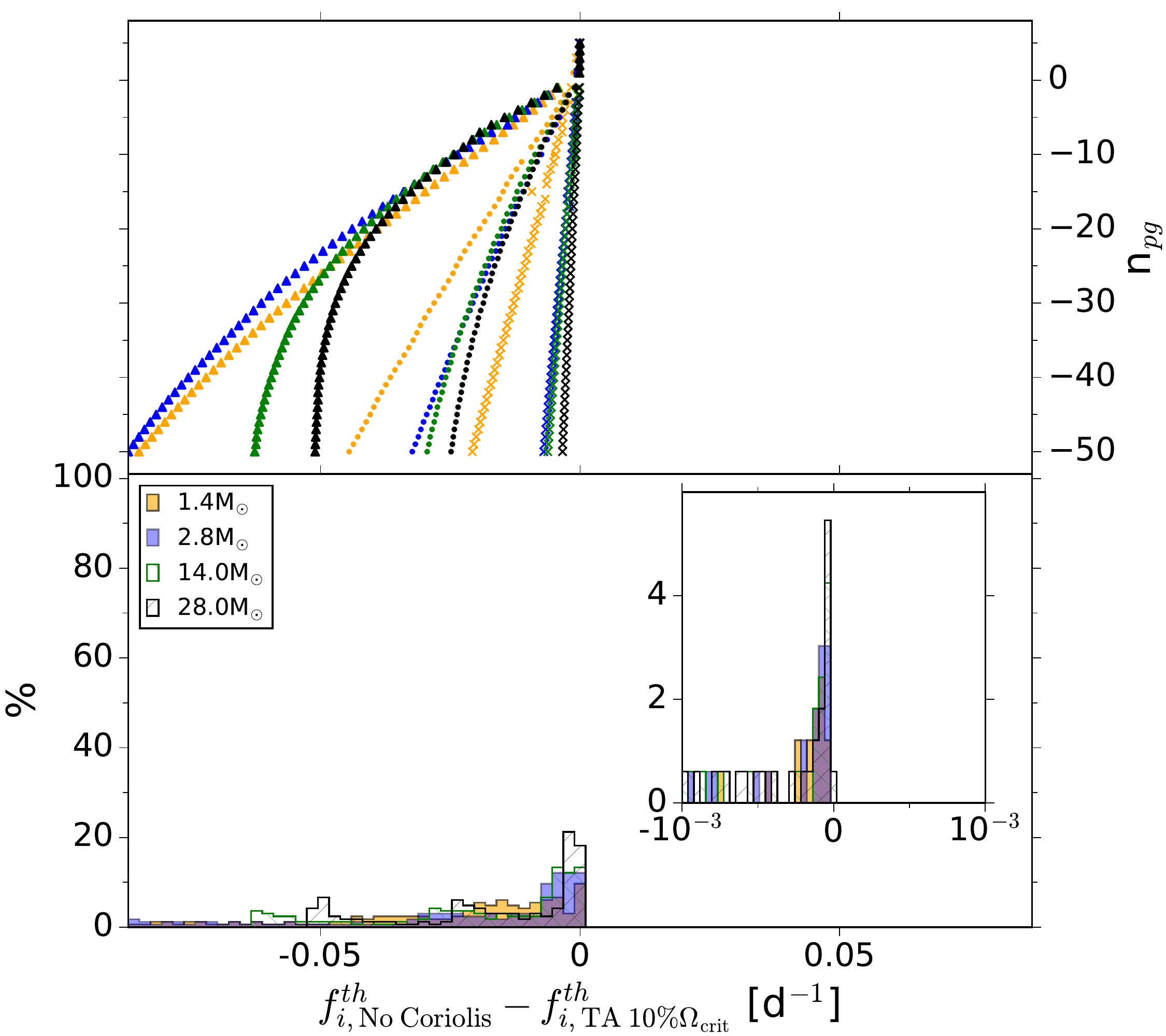}
 \includegraphics[width=3.5in]{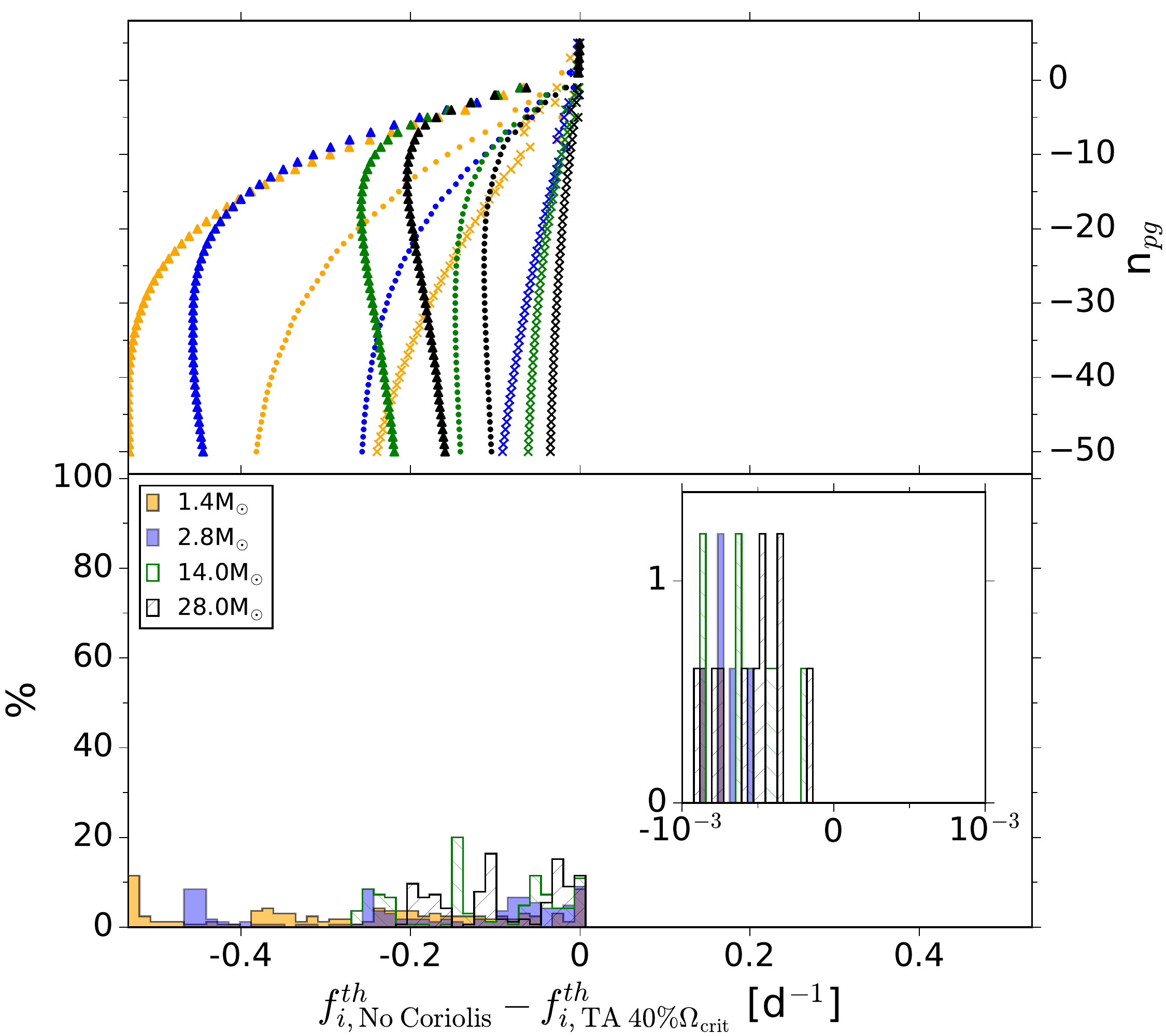}
 \caption{\label{TA-nonTA} Frequency differences for dipole zonal modes of the
   same radial order $n_{pg}$, for the 12 benchmark models with parameters
   listed in Table\,\protect\ref{baseline}. The frequency differences
   $\Delta f_i^{\rm th}$ are equal to the frequencies computed without taking
   the Coriolis force into account minus those where the Coriolis force is
   treated in the TA, for the case of a rotation rate equal to 10\% (left) and
   40\% (right) of the critical rate. The bottom panels give the global
   distribution of  all $\Delta f_i^{\rm th}$-values adopting 81 frequency
   bins, while the inset is a zoom for 51 bins focusing on an interval
   corresponding with all $\Delta f_i^{\rm th}$ that fall within a
     typical frequency error derived from a quasi-uninterrupted half- to
     one-year light curve. The upper panel shows the values of
   $\Delta f_i^{\rm th}$ as a function of the radial order $n_{pg}$, 
     $X_c=0.7$ as triangles, $X_c=0.4$ as circles, and $X_c=0.1$ 
as crosses.  See also
     Table\,\ref{percentages} for the total percentage 
of the frequencies shown in the
      inset per mass and per evolutionary stage.}
\end{center}
\end{figure}

Figure\,\ref{DeltaP-Rotation} shows 
the period
spacing patterns for the dipole zonal modes of the 
models with 1.4\,M$_\odot$ and 28\,M$_\odot$, for the three considered 
evolutionary stages and for the indicated rotation rate.
It can be seen that rotation causes a downward tilted
period spacing pattern, a property well known from theory
\citep[e.g.,][]{Bouabid2013} and also observed in {\it Kepler\/} data of 
pulsating F and
B stars
\citep[e.g.,][]{VanReeth2015,Papics2017,Ouazzani2017}.
In Fig.\,\ref{TA-nonTA} we compare the dipole zonal frequencies of modes with
the same radial order $n_{pg}$, computed without the Coriolis force versus in
the TA.  It can be seen in Fig.\,\ref{TA-nonTA} that the inclusion of the
Coriolis force increases the gravity-mode frequencies for the entire mass
range. The frequency differences increase for increasing radial order.
Almost all frequency differences are larger than 
the typical frequency error expected for a light curve spanning 
about a half to one year
(see
Table\,\ref{percentages}).  We conclude that the Coriolis force cannot be
neglected in forward seismic 
modeling of gravity modes and that rotational effects
must be included to
compute theoretical gravity-mode frequencies, even for stars rotating as slowly
as 10\% of their critical rotation rate.

\citet{SchmidAerts2016} treated two F-type super-inertial gravity-mode pulsators
in a non-synchronized close binary, with near-core rotation rates of 0.09 and
0.14\,d$^{-1}$ and spin parameters in the near-core region
($2\Omega_{\rm rot}/f_{\rm corotating}$) ranging from 0.14 to 0.26.  They
assessed that the impact of ignoring the Coriolis force on the seismically
modeled isochrone age is about 5\%; the radii are affected by about 3\%, while the
masses remain unaffected for these rotation rates. These two pulsators can hence
be considered as limiting cases for the validity of ignoring the Coriolis force.

Computations of frequencies in the inertial frame
of reference based on the TA require estimation of the rotation frequency,
$f_{\rm rot}$, in addition to identification of the mode degree $l$ and
azimuthal order $m$ for each frequency \citep{VanReeth2016,Ouazzani2017}. This
implies that forward seismic modeling of gravity-mode frequencies is at least a 5D
optimization problem, since $f_{\rm rot}$ must be estimated, along with
$(M,X,Z,X_c)$.  It was shown by \citet{VanReeth2016} and by 
\citet{Ouazzani2017} that the slope
of period spacing series of gravity modes of consecutive radial order $n_{pg}$ in
F-type pulsators allows for the simultaneous identification of the mode
wavenumbers $(l,m)$ and the derivation of $f_{\rm rot}$. This method also works
for B-type pulsators \citep{Papics2017} and is adopted here as an essential step in
our seismic modeling scheme (cf.\ Sect.\,\ref{scheme}).
\begin{table}
\caption{\label{percentages}
Percentage of the 660 zonal dipole modes of the models in Table\,\ref{baseline}
that fall within a frequency error typical of a half to a one-year data set 
when changing one
aspect of the physics. The first three rows concern changes in the
  computation of the pulsation modes, 
while all other rows are results for changes in the input physics of the equilibrium
models. }
\tabcolsep=2pt
\begin{tabular}{r|llll}
\hline
Comparison & 1.4\,M$_\odot$ & 2.8\,M$_\odot$ & 14\,M$_\odot$ & 28\,M$_\odot$\\
\hline
No Coriolis $\leftrightarrow$ TA@10\%$\Omega_{\rm crit}$ & 10\% (11,9,9) & 12\%
                     (16,11,9) & 13\% (22,9,9) & 17\% (33,9,9) \\
No Coriolis $\leftrightarrow$ TA@40\%$\Omega_{\rm crit}$ & 1\% (2,2,0) & 3\% (2,7,0)
                                             & 5\% (4,7,4) & 6\% (7,7,4) \\
Adiabatic $\leftrightarrow$ Non-adiabatic & 100\% (100,100,100) & 
96\%  (93,96,100) & 83\% (100,77,63) & 96\% (100,91,89)\\
\hline
Ledoux $\leftrightarrow$ Schwarzschild & 86\% (67,91,100) & 80\% (80,71,89) & 
100\% (100,100,100) & 99\% (96,100,100) \\
$\alpha_{\rm MLT}=2.0$ $\leftrightarrow$ $\alpha_{\rm MLT}=1.0$ & 10\% (8,6,15) & 
7\% (2,6,13) & 73\% (55,73,91) & 81\% (59,82,100) \\
Gray $\leftrightarrow$ Paczynski & 59\% (44,42,91) & 90\% (89,82,100) & 
93\% (89,91,100) & 97\% (91,100,100) \\
Gray $\leftrightarrow$ FASTWIND & -- & -- &  94\% (93,94,94) & 95\% (95,94,96) \\
Precomputed ZAMS $\leftrightarrow$ From Hayashi & 7\% (0,0,22) & 27\% (0,2,80) 
& 33\% (6,8,82) & 45\% (20,27,82) \\
Standard network $\leftrightarrow$ Fe, Ni added & 98\%
     (98,96,100) & 90\% (82,87,100) & 93\% (89,91,100) & 100\% (100,100,100) \\
OPAL $\leftrightarrow$ OP &  41\% (6,51,67) & 0\% (0,0,0) & 37\% (20,41,51) & 
53\% (26,56,74) \\
Przybilla $\leftrightarrow$ Asplund & 2\% (7,0,0) & 0\% (0,0,0) & 33\% (42,51,7)
                                                             & 49\% (56,69,22) \\
No Atomic Diffusion $\leftrightarrow$ Settling \& Levitation & --
& 18\%  & 19\%  & 43\% \\
\hline
\end{tabular}
\tablecomments{The frequency differences are computed for models that differ
  less than 0.001 in $X_c$.  These percentages are visually represented in histograms in 
  the upper right  panels of Figs\,\ref{TA-nonTA} -- \ref{diffusion}. 
The higher the percentage, the
  more justified it is to fix this phenomenon in the input physics of the stellar
  models for forward seismic modeling. The three values in brackets denote the
  sub-percentages for $X_c=(0.1, 0.4, 0.7)$, to reveal the dependence
  on evolutionary stage. In the case of the models with 
  atomic diffusion, the reaction network  was 
  extended with $^{56}$Fe and $^{58}$Ni. 
For the bottom row, the combined effect of 
gravitational settling, concentration diffusion,  
thermal diffusion, and radiative levitation was considered; these
comparisons were limited to $X_c=0.7$ for CPU  reasons.}  
\end{table}

\subsection{The adiabatic approximation}

\begin{figure}
\begin{center}
 \includegraphics[width=3.5in]{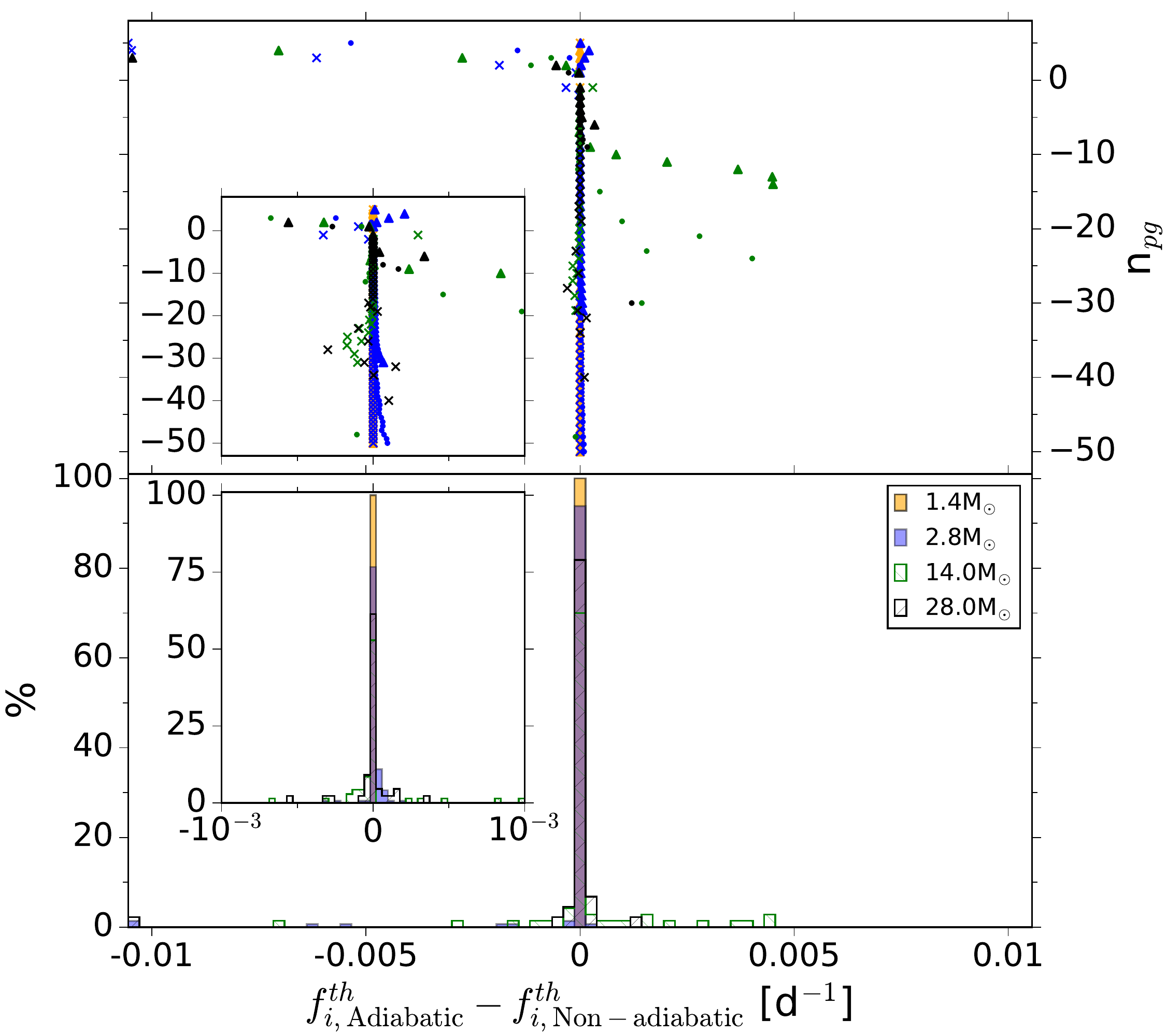}
 \caption{\label{AD-nonAD} Same as Fig.\,\ref{TA-nonTA}, but for adiabatic
   versus non-adiabatic frequencies for modes of the same radial order $n_{pg}$,
   showing that almost all frequency differences fall within the frequency
   precision typical for an uninterrupted 1-year light curve.}
\end{center}
\end{figure}
\begin{figure}
\begin{center}
\includegraphics[width=3.5in]{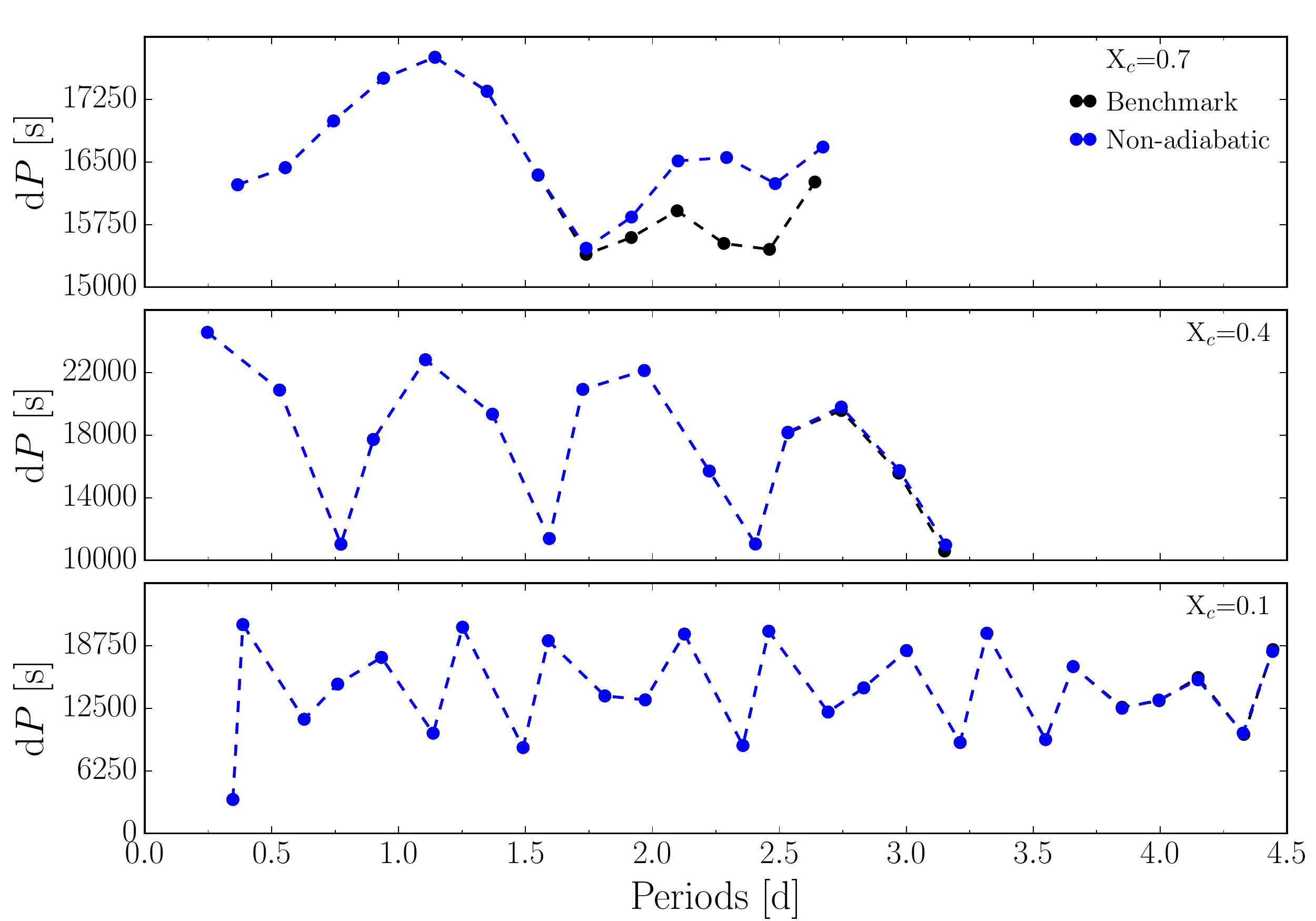}
\caption{\label{DeltaP-NAD} Adiabatic and non-adiabatic gravity-mode period
  spacing patterns for a 14\,M$_\odot$ model at $X_c= 0.7, 0.4, 0.1$. When
  invisible, there is no difference between the black and the blue dots. }
\end{center}
\end{figure}
\citet{Moravveji2015} performed the first forward seismic modeling of a slowly
pulsating B star. It was based on 19 detected dipole zonal modes with radial
order $n_{pg}$ ranging from -14 to -32 for the B8.3V star KIC\,10526294
\citep[see also][]{Triana2015}. \citet{Moravveji2015} found that the adiabatic
and non-adiabatic approximations to compute the mode frequencies are
equivalent for the frequency precision yielded by a nominal {\it Kepler\/}
light curve for these gravity modes in that mass regime.

Figure\,\ref{AD-nonAD} shows the frequency differences for the zonal dipole
modes of the 12 benchmark models described in Table\,\ref{baseline}.  
Comparison of the range of the abscissa axes
of Figs\,\ref{TA-nonTA} and
  \ref{AD-nonAD} reveals that 
the difference between adiabatic and non-adiabatic frequencies is almost always
smaller than a typical frequency error for a half- to one-year  
uninterrupted light curve and is 
much smaller than the effect of rotation. Moreover, various  
theoretical uncertainties caused by the choice of the input
physics, as discussed in the next section, give rise to much larger frequency
differences.   

The behaviour of some modes in Fig.\,\ref{AD-nonAD} attracts attention.  It
  mainly concerns modes of the 14\,M$_\odot$ models and some of the
  2.8\,M$_\odot$ models.  We show the period spacing patterns of the converged
  gravity modes of the model with 14\,M$_\odot$ in Fig.\,\ref{DeltaP-NAD}.  It
  can be seen that the modes at $X_c=0.7$ with periods above some 1.7\,d have
  deviating non-adiabatic behaviour. These deviations between the adiabatic and
  non-adiabatic approximation disappear at later evolutionary stages. 
Similar results occur,
  although more sporadically, for the pressure modes of the 2.8\,M$_\odot$
  model, but the deviations in frequency always remain below 0.012\,d$^{-1}$.
  Our results are in agreement with those obtained earlier by J.\
  Daszy\'nska-Daszkiewicz in the framework of the HELAS project.\footnote{\tt
    http://helas.astro.uni.wroc.pl/deliverables.php?active=opalmodel\&lang=en}
  Her computations revealed that the difference between adiabatic and
  non-adiabatic frequencies of pressure modes increases as the radial order
  increases and can be as large as 0.05\,d$^{-1}$ for $n_{pg}=6$ or 7.  Indeed,
  high-order pressure modes are much more sensitive to the very outer envelope
  of the stars, where non-adiabatic effects occur, while the gravity modes are
  dominantly determined by the physics of the deep stellar interior, where the
  adiabatic approximation is appropriate.

Our non-adiabatic computations for the excited gravity modes reveal mode
  lifetimes ranging from about a million years for the 1.4\,M$_\odot$ models
  near the ZAMS to a minimum of only
3,000 years near the TAMS.  We do point out, however,
  that MESA does not offer a time-dependent convection treatment, so the mode
  excitation predictions are not optimal for this mass. We refer to
  \citet{Dupret2005} for more appropriate computations.  The 2.8\,M$_\odot$
  models reveal only a few excited modes close to the ZAMS (the $X_c=0.7$
  case), with a relatively short liftetime of some 2,500 years. On the other
  hand, the 14 and 28\,M$_\odot$ models have only a few gravity modes excited
  near the TAMS ($X_c=0.1$), with lifetimes of only some 10 years and 45\,000
  years, respectively.  These mode excitation results are in excellent agreement
  with those by \citet{Szewczuk2017}, who considered models with masses 
between
  2.5 and 19\,M$_\odot$ (these authors did not list the mode lifetimes).

The required CPU time for the computation of non-adiabatic frequencies is much
larger than for adiabatic computations. Moreover, uncertain physics is involved
in the non-adiabatic outer envelope of the star, while the frequencies of
detected gravity modes are dominantly determined by the physical properties of
the deep interior. For these reasons, we adopt the TA in the adiabatic approach
in the forward modeling, because such modeling has the purpose to deduce
  the global physical properties of the star. However, once the forward modeling
  results are known, instability computations should be performed for the best
  models, in order to test the mode excitation and the non-adiabatic frequencies
  of the detected modes, as a way to improve our knowledge of opacities and
  assess the properties of possible missing input physics.

\section{\label{models-error} Theoretical frequency
  uncertainties stemming from the equilibrium models}

As discussed in Sect.\,\ref{pulsation-error} and illustrated in
\citet{Moravveji2015,Moravveji2016}, the four basic parameters $(M,X,Z,X_c)$ to
be estimated from forward seismic modeling of stars born with a convective core are
insufficient.  Equilibrium models contain numerous additional parameters, some
of which have a significant impact on the oscillation frequencies while others do
not. This allows to estimate the parameters that have the largest 
effect on gravity-mode
frequencies, after careful choice of the input physics that leaves the gravity-mode
frequencies unchanged.

In order to make appropriate choices for which parameters to estimate and which
to keep fixed when describing the macro- and microphysics, we need to know which
processes affect the frequency values less than the resolving power of the data
set. Here, we provide a hierarchy in the multitude of choices to be made among,
e.g., the EOS, opacities, nuclear reaction rates, chemical mixtures, the
atmosphere model as a boundary condition, the starting model at the ZAMS, etc.
Whenever a different yet ``fixed choice'' of input physics is made without
parameters to be estimated, we adopt the terminology that we are dealing with a
{\it different stellar model\/} $\calm (\bftheta, \bfpsi)$ 
with free parameters $\bftheta$ and fixed physics $\bfpsi$.  For each
stellar model, estimation of the optimal value of its free parameters $\bftheta$
and the accompanying uncertainties of those parameters is achieved
by
minimizing the discrepancy between $f_i^{\rm th}$ predicted from
$\calm(\bftheta,\bfpsi)$ and the observations $f_i^\ast$. Subsequently,
{\it stellar model selection\/} is done by comparing the best
correspondence between observed and predicted frequencies according to a
well-defined criterion, for various choices of the input physics $\bfpsi$, using
the optimally estimated parameters $\bftheta$ within each stellar model. This
stellar model selection takes into account the number of free
parameters in each stellar model and applies a penalty accordingly.

In practical applications, the difference between parameter estimation for a
fixed type of stellar model $\calm(\bftheta,\bfpsi)$, versus selection of the
most appropriate input physics represented by 
$\bfpsi$, depends on the type of star.  It is
therefore essential that modelers have knowledge of the hierarchy among the
parameters to be estimated versus the physical properties that can be fixed.
For example, forward seismic modeling of pulsating subdwarfs 
or white dwarfs requires the
mass of the thin outer hydrogen layer as a free parameter to be estimated along
with the overall stellar mass, because it is vital to evaluate the mode trapping
\citep[e.g.,][]{Charpinet2011,Corsico2012}, while rotation can be treated at the
level of Ledoux splitting for such stars and does not require use of the TA
\citep[e.g.][]{Hermes2017,Giammichele2017}.  
Similarly, mass loss is considered not to be of importance for intermediate-mass
stars during core hydrogen burning.  For high-mass stars burning hydrogen in the
core, modelers should make sure that ignoring mass loss or fixing it to a
specified value is 
appropriate by considering a second stellar model, without adding the complexity
of having $\dot{M}$ as an extra parameter to estimate (cf.\
Fig.\,\ref{mass-loss} below).  Hence, two models $\calm (\bftheta, \bfpsi)$, 
one with and one
without mass loss, could be considered to check its effect on the seismic
parameter estimation, relative to the measurement error of the frequencies of
the detected modes. For stars beyond core-hydrogen burning, however, $\dot{M}$
could be included as parameter to estimate in $\bftheta$.

Many additional cases to those above can be discussed in the context of
considering different stellar models, the choices depending on the type of star
to be modeled.  Below we focus on the most important ingredients to consider as
different models for the specific case of core-hydrogen burning stars with a
convective core, whose most important parameters for the macrophysics to be
estimated will be discussed first.

\subsection{\label{parameter-estimation}Frequency errors 
due to choices of macrophysics}

Aside from the four basic parameters needed to compute the equilibrium models
during the main-sequence life of a star, $(M,X,Z,X_c)$, and an estimate of the
near-core rotation frequency $f_{\rm rot}$ as input for the TA oscillation
frequency computations, a multitude of parameters due to macrophysical phenomena
are included in state-of-the-art evolution codes. We evaluate the most important
ones, from the viewpoint of the uncertainty they cause in the computation of
theoretical gravity-mode oscillation frequencies.

\subsubsection{Convection and mixing parameters}

\begin{figure}
\begin{center}
\includegraphics[width=3.5in]{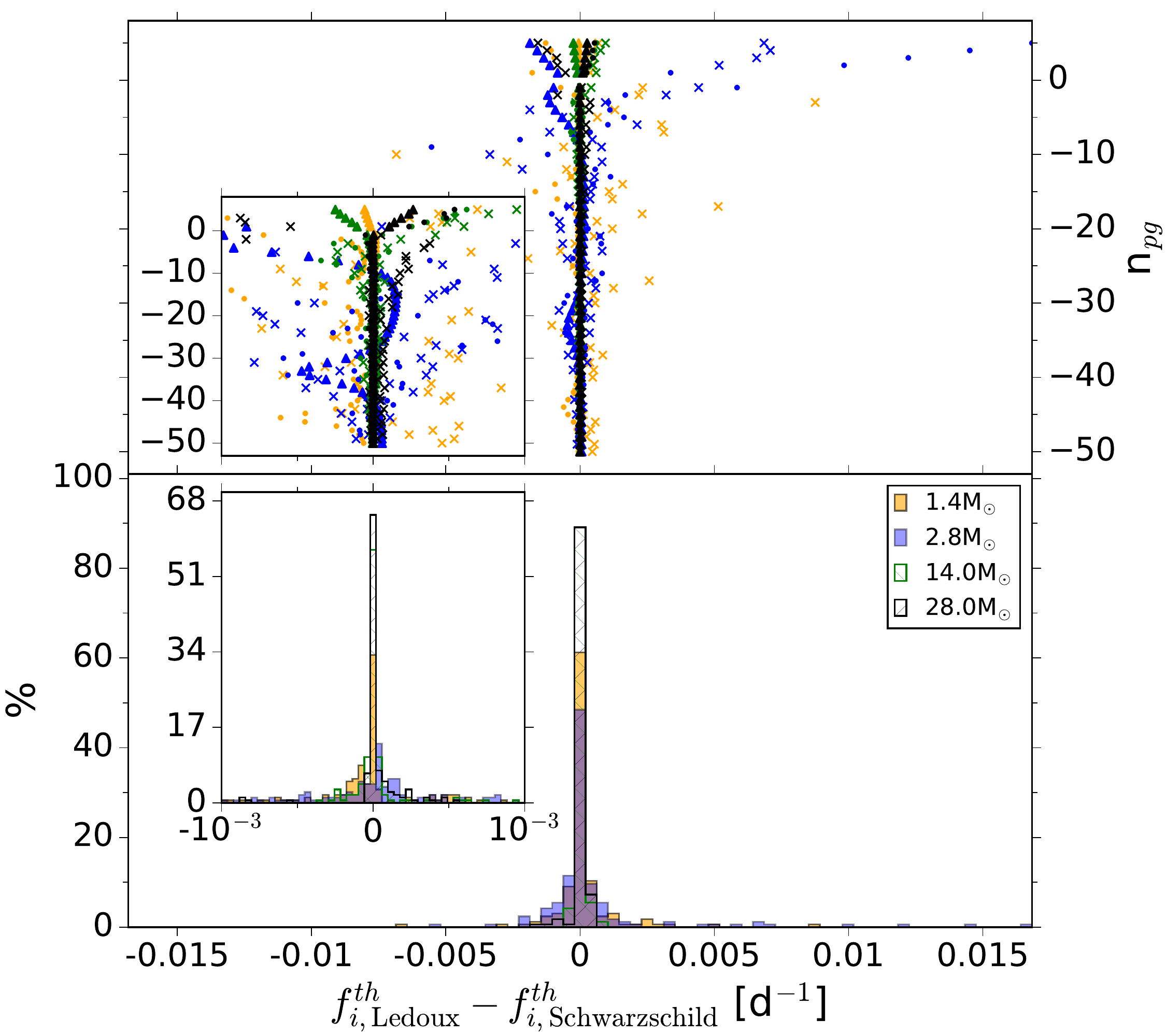}
\includegraphics[width=3.5in]{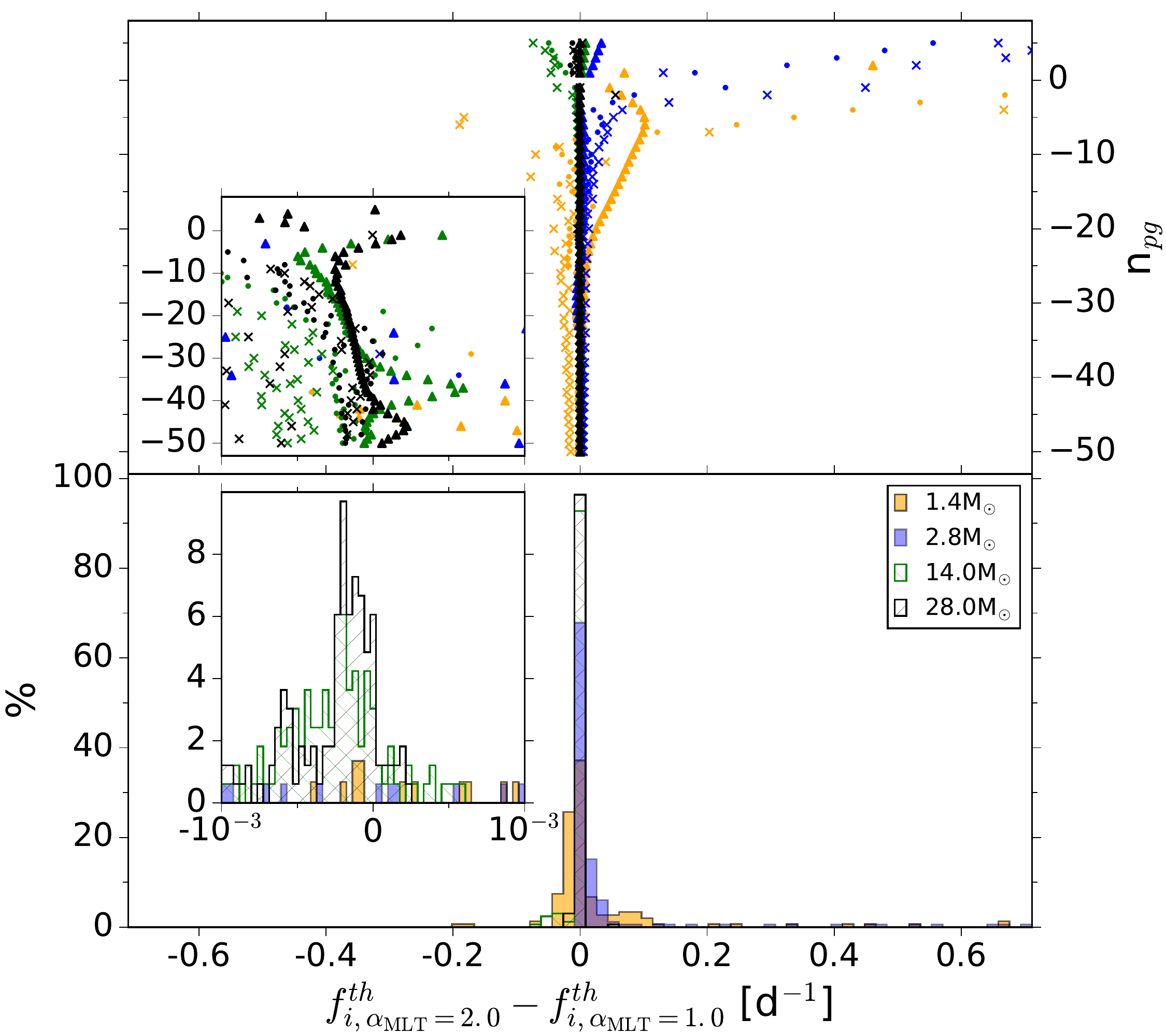}
\caption{\label{mlt} Same as Fig.\,\protect\ref{TA-nonTA} but for models with
  the Schwarzschild instead of Ledoux convection criterion (left) and for a
  mixing-length parameter $\alpha_{\rm MLT}=1.0$ instead of 2.0 (right).}
\end{center}
\end{figure}

In this work, we consider stars with $M\gtrsim 1.4\,$M$_\odot$ at birth. The
central nuclear burning for such stars is dominated by the CNO cycle, which is
highly temperature sensitive. Such stars have a convective core and its mass is
of major importance for the evolution of the star. For this reason, the core
mass is also a dominant parameter to estimate along with
$(M,X,Z,X_c,f_{\rm rot})$. The time scale of the convection in the core is much
shorter than the nuclear time scale, so we can assume full and instantaneous
mixing there.  Hence, the core has uniform chemical composition as long as
hydrogen burning is active.  We adopt the Ledoux criterion and use MESA's
predictive mixing scheme to locate the boundary of the convective core for the
equilibrium models in Table\,\ref{baseline}. We show in the left panel of
Fig.\,\ref{mlt} the difference in the oscillation frequencies if we instead
adopt the Schwarzschild criterion.  
It can be seen that igoring the $\mu$-gradient in the stability criterion
somewhat impacts
the gravity modes for the 1.4 and 2.8\,M$_\odot$ models, and more so 
at evolved rather
than at early evolutionary stages. The frequency differences 
always remain below 0.017\,d$^{-1}$. \citet{Paxton2018}
provide a thorough discussion on the subtle differences that may occur in the
inhomogeneous region beyond the core boundary due to the choice of the
convection criterion, depending on the stellar mass and evolutionary stage.
However, as becomes clear from Table\,\ref{percentages}, 
the implications for the modes due to these small differences
are inferior to those of several other effects connected with choices for
the input physics.

The observations of gravity-mode period spacings in a sample of F stars
\citep[e.g.,][]{VanReeth2015} and B stars \citep[e.g.,][]{Papics2017} reveal
detailed structure. Indeed, the period-spacing morphology appreciably deviates
from a constant value, even for stars that hardly rotate
\citep{Degroote2010,Papics2014}. 
Estimation of the core mass from
forward seismic modeling of core-hydrogen burning stars based on gravity
modes was so far only done for a few stars and all of them demanded the inclusion
of convective core overshooting and chemical mixing in the radiative envelope
\citep{Moravveji2015,Moravveji2016,SchmidAerts2016}.  Each of these two necessary
ingredients
are described by at least one parameter.  The overall mass in the core is
dependent on the treatment of the overshooting, while the chemical profiles
throughout the star are determined by the element transport in the radiative
envelope.  We describe the physical meaning of these two required ingredients
here.

As in many stellar evolution codes, convective energy transport in MESA relies
on the time-independent mixing-length theory, while time-dependent convection
(turbulent pressure) is ignored. Mixing-length theory is characterized by one
free parameter, $\alpha_{\rm MLT}$, representing the characteristic length scale
over which convective fluid elements travel before they dissipate; it is
expressed in local pressure scale heights and takes typical values between 1 and
2. For low-mass stars, $\alpha_{\rm MLT}$ determines the convective motions in
the outer convective envelope and one often fixes it to the solar value.
However, the intermediate-mass stars we consider here have very thin and shallow
convective envelopes. Moreover, along with the high-mass stars, they have a
fully mixed convective core, where physical circumstances are very different
than in Sun-like stars.  For this reason, $\alpha_{\rm MLT}$ is a free
parameter. We show in the right panel of Fig.\,\ref{mlt} the effect of changing
$\alpha_{\rm MLT}$ from 2.0 to 1.0. It can be seen from Fig.\,\ref{mlt} and
deduced from Table\,\ref{percentages} that this change has major impact for the
two lowest masses considered here, particularly at evolved stages. This is
in line with observational results for $\delta\,$Sct stars
\citep{BowmanKurtz2018}. The choice for $\alpha_{\rm MLT}$ is less important for
the two high-mass stars, particularly close to the ZAMS.  The results in
  Fig.\,\ref{mlt} are as expected since the two lower-mass models have shallow
  outer convection zones whose extent is determined by $\alpha_{\rm MLT}$, while
  the high-mass stars have radiative envelopes in their early life but develop
  convective envelopes near the TAMS. Given the results in
  Table\,\ref{percentages}, one should either estimate $\alpha_{\rm MLT}$,
  particularly for intermediate-mass stars, or consider at least two different
  fixed values to assess the impact on the forward modeling.

Complications to the above picture arise in the determination of the core
boundary because convective fluid elements ``overshoot'' the boundary between the
convective core and radiative envelope due to their inertia. The extent and
properties of this overshoot region, including the level of mixing of the
chemical elements inside it and its thermal structure, are of major importance
for the future evolution of the star \citep[][Section 30.4, for a thorough
discussion]{Kippenhahn2012}.  Gravity modes allow us to seismically probe the detailed
properties of this near-core overshoot region.  The parameters
describing the aspects of convective core overshooting are among the most
important ones for modeling of the stars in the mass range we
consider \citep[see Sect.\,3 in][for a thorough discussion]{Paxton2018}.

In MESA, convection is implemented as a diffusive process with an associated
mixing coefficient. In such a framework, the convective core has a large
diffusive mixing coefficient $D_{\rm mix}$, typically between $10^{11}$ and
$10^{17}$cm$^2$\,s$^{-1}$ during core-hydrogen burning for the mass regime we
consider here. As such, it is natural to describe the mixing due to core
overshooting as a diffusive process as well, with coefficient $D_{\rm ov}$
containing at least one dimensionless parameter \citep[Eq.\,(9)
in][]{Paxton2013}.  Several options are available to define the level and shape
of overshoot mixing, each characterized by a set of free parameters --
\citet[][Appendix B.7]{Paxton2013} and \citet{Pedersen2018}.  Among them is the
option to consider an exponentially decaying prescription described by the
parameter $f_{\rm ov}$, and the simpler step function overshoot for which the
core is extended into a fully mixed overshoot region from its boundary by an
amount $\alpha_{\rm ov}$.  

The temperature gradient in the overshoot region is unknown \citep[see][for a
thorough discussion on chemical mixing and temperature gradients]{Salaris2017}.
In the standard version of MESA, it is the radiative gradient but the code is
sufficiently flexible to change this into the adiabatic one (so-called
convective penetration) or any transition between these two.  These choices each
constitute a different stellar model $\calm (\bftheta, \bfpsi)$.  In regions
where the temperature gradient takes a value between the adiabatic one and the
adiabatic one increased with a positive $\mu$ gradient, semiconvective mixing
occurs. This phenomenon describes the growing oscillatory motion of a convective
fluid element in that region, which implies partial chemical mixing in that
zone.  The effect of semiconvection on the temperature gradient is unknown. In
MESA, semiconvective mixing in the vibrationally unstable region is again
implemented as a diffusive process, characterized by the dimensionless
coefficient $\alpha_{\rm sc}\in [0,1]$. The larger the coefficient, the faster
the matter mixes in the semiconvective zone.\footnote{In this work, we ignore
  thermohaline mixing (``fingering convection'') due to a negative
  $\mu$-gradient value in the models. This phenomenon of slow mixing typically
  occurs in evolved low-mass stars with shell burning. It can also occur during
  the hydrogen core burning but the mixing profile it would cause is not well
  known \citep{Deal2017}. Should a profile become available, this phenomenon
  could be included separately with yet another diffusive coefficient
  $\alpha_{\rm th}$, and could easily be included in our statistical formalism.}

As already discussed above, 
``extra'' macroscopic mixing $D_{\rm ext}$ 
occurs in the zone above the core
overshoot zone in the radiative envelope of intermediate-mass and high-mass 
stars.
Indeed, such extra mixing was required
to bring theoretical frequencies in agreement with observational data
\citep{Moravveji2015,Moravveji2016}. 
Coherent oscillation modes \citep{Aerts2014} and damped modes
\citep{RogersMcElwaine2017} may also induce mixing but these options have so far
hardly been taken into account in stellar evolution theory.  In order not to
introduce bias on the importance of the various phenomena giving rise to
element transport, we take a pragmatic approach and make use of one diffusive
mixing coefficient $D_{\rm ext}$ as free parameter in the forward seismic
modeling. This parameter represents the average overall effect of (rotationally
and/or magnetically and/or pulsationally induced) mixing in the radiative zones
of a star as probed by the gravity modes.  Only if this free parameter can be
estimated with good precision from seismic data for a considerable sample of
stars can we start to unravel the role of the various individual causes of
mixing at play in the stellar envelope.  As said, the first steps in this direction
were taken for a few stars \citep{Moravveji2015,Moravveji2016,SchmidAerts2016},
which showed the need to include $D_{\rm ext}\neq 0$ just adjacent to the core
overshoot region in the radiative envelope to explain the structure in the
observed period spacings.  Meanwhile, \citet{Pedersen2018} assessed if gravity
modes can distinguish between a constant $D_{\rm ext}$ versus an actual profile
$D_{\rm ext}(r)$ stemming from 2D simulations of internal gravity waves.
These authors found 
that a distinction can only be made on the
basis of combined observational restrictions from gravity modes and 
surface abundances. In first instance, it is therefore justified to
consider $D_{\rm ext}$ to be a single parameter, describing the mixing adjacent
to the core overshoot zone to be estimated.

As a side remark, we point out that extra turbulent mixing in the surface layers of
intermediate-mass stars is sometimes also considered in the context of
asteroseismology or galactic archeology. Such extra turbulent mixing in the
near-surface layers is usually assumed to have a density dependence. Aside
from this, envelope undershooting may also occur. While both these phenomena are
important for chemical tagging at advanced evolutionary stages of low-mass stars
\citep[e.g.,][]{Dotter2017},
they hardly affect the gravity mode period spacings \citep{Pedersen2018}. 
Indeed, mixing in the zone
adjacent to the core-overshoot zone is much more important for the fitting of
detected gravity-mode frequencies, because these modes mainly probe the
near-core regions.
At first instance, such surface mixing can therefore be
ignored in forward seismic modeling of gravity modes
but can be evaluated from high-precision surface abundances if available.

The transport of angular momentum during a star's evolution is also not calibrated
by observations, just as the element transport. Moreover, the level of its
inhibition by $\mu$-gradients is not known. Even for the simple case of low-mass
stars born with a radiative core, where the macroscopic mixing is thought to be
absent, current theory of angular momentum transport fails
\citep[e.g.,][]{Eggenberger2017}.  The importance of a convective core and
near-core properties during core hydrogen burning for appropriate understanding
and interpretation of angular momentum of red giants was recently highlighted
\citep{Tayar2013,Eggenberger2017}. Seismic modeling of intermediate-mass stars
from gravity modes during the core hydrogen burning is highly relevant to
understand the angular momentum of secondary clump red giants as well as more
massive blue supergiants.

Given the major uncertainties in the transport of chemicals and angular momentum
in current stellar evolution theory, an appropriate approach in terms of forward
seismic modeling is to estimate the chemical mixing by seismic inference from
gravity modes, assuming rigid rotation near the core as first step
\citep[cf.\ ][]{VanReeth2016,Aerts2017a,Ouazzani2017}.  
Subsequently, the rotation profile and
the accompanying angular momentum transport can be evaluated by considering
$f_{\rm rot} (r)$ instead of constant $f_{\rm rot}$ for the best seismic models,
once their mixing properties have been derived seismically. This approach has
been applied to the above mentioned 3.25\,M$_\odot$ star KIC\,10526294 by
\citet{Moravveji2015} and \citet{Triana2015}, pointing to the need of angular
momentum transport.  The required transport could meanwhile be explained by
internal gravity waves \citep{Rogers2015}. Such type of analyses needs to be
done for stars with various levels of near-core rotation to deduce a broad range
of $f_{\rm rot}(r)$ covering the entire hydrogen-burning phase, in order to be
able to predict angular momentum transport for the more evolved phases
\citep[][]{Aerts2017a,Ouazzani2018}.

In summary, near-core rotation, convective core overshooting and mixing in the
radiative envelope adjacent to the core overshoot zone must all be
included in the forward seismic modeling of gravity-mode pulsators with
$M\gtrsim 1.4\,$M$_\odot$, implying that the optimization problem is at least
7-dimensional (+7D), with $\bftheta = (M,X,Z,X_c,f_{\rm rot},D_{\rm ov},D_{\rm
  ext})$.
For intermediate-mass stars, one should additionally test the impact of a
  different mixing length value.

\subsubsection{Choice of a static atmosphere as boundary condition}

All stars experience mass loss during their life.  The occurrence and duration
of the episodes of mass loss are vastly different for stars with different
masses. Low- and intermediate-mass stars born with $M\lesssim 9\,$M$_\odot$
experience substantial mass loss due to a dust-driven wind on the asymptotic
giant branch, while more massive stars experience a radiation-driven wind during
and/or beyond core hydrogen burning until they explode as supernovae.  Two 
important parameters characterizing the wind are the mass-loss rate
$\dot{\rm M}$ and the wind velocity profile \citep{Kudrtizki2000}.
\begin{figure}
\begin{center}
 \includegraphics[width=3.5in]{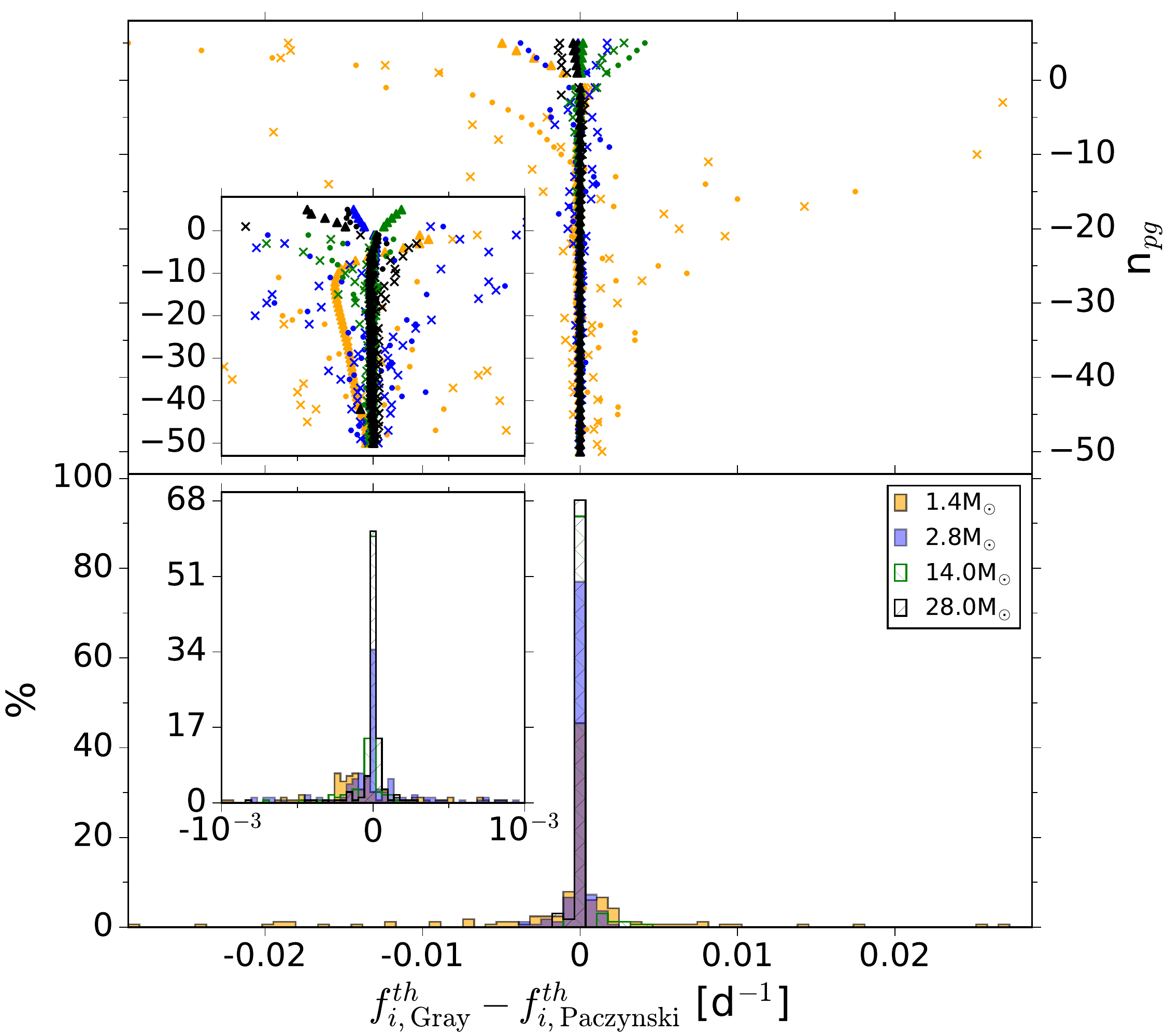}
 \includegraphics[width=3.5in]{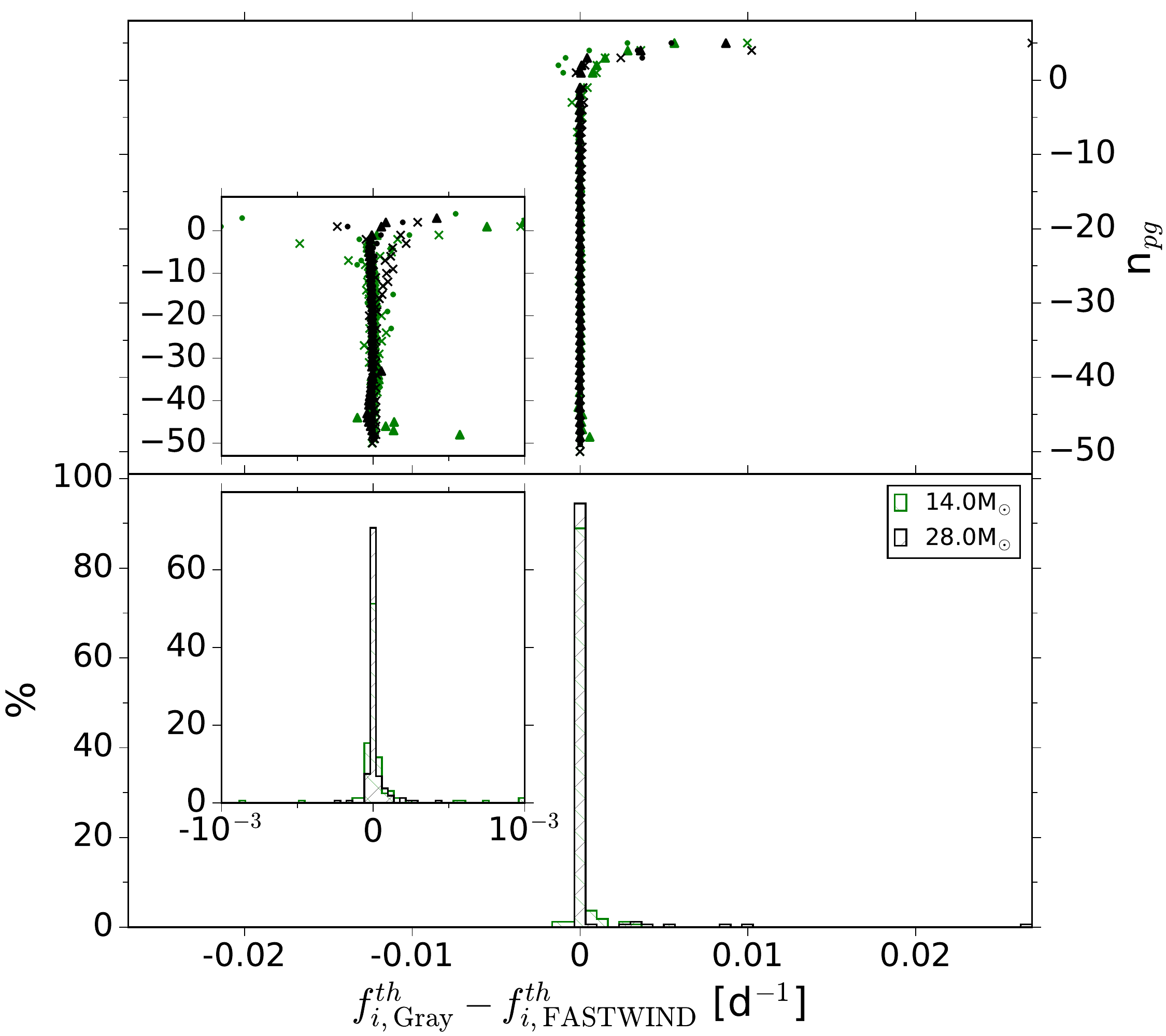}
 \caption{\label{mass-loss} Same as Fig.\,\protect\ref{TA-nonTA} but for a gray
   versus Paczynski static atmosphere model as boundary condition for
   MESA (left) and for a gray atmosphere versus FASTWIND dynamical atmosphere
   model for the two highest masses (right).  }
\end{center}
\end{figure}

Stellar evolution computations treat the stellar wind in a simplistic way in the
sense that a mass-loss rate from a recipe is adopted and the appropriate mass is
peeled off the outer layers of a star at each time step in the evolution, taking
into account the accompanying angular momentum loss and assuming that
hydrostatic equilibrium holds at every time step.  This is a crude
approximation, because the driving force of the wind is ignored in the equation
of motion and a static rather than dynamical atmosphere is considered as
boundary condition to solve the equations of stellar structure.  MESA and GYRE
offer various static atmosphere models from precomputed tables as choices for
these boundary conditions.  The effect of different choices of the atmosphere
model on the frequency values of gravity mode is illustrated in the left panel
of Fig.\,\ref{mass-loss}, where we show the frequency differences obtained for a
gray versus Paczynski atmosphere model.  It can be seen that frequencies are
only modestly affected and that the majority of frequency differences are
within the resolving power of a typical data set, 
except for the lowest
considered mass $M=1.4\,$M$_\odot$ at evolved stages.
The latter models have a thin convective
outer envelope and the interplay amongst convection, pulsation, and atmospheric
properties becomes important for the modes.
For stars without a (thin) convective outer envelope,
the range of the frequency differences is always below 0.01\,d$^{-1}$
(Fig.\,\ref{mass-loss}), which justifies that we neglect this effect compared to other
more important ones.

\subsubsection{Effect of mass loss due to a radiation-driven wind}

So far, the mass-loss rate was not considered as a parameter to estimate 
in forward seismic modeling
for core-hydrogen burning
stars, given that one is already dealing with a +7D optimization problem.
Cluster ensemble asteroseismology from {\it Kepler\/} low-mass red giants did
allow a rough evaluation of the upper limit of mass-loss rate $\dot{\rm M}$ due
to a dust-driven wind on the red giant branch \citep{Miglio2012}. However, the 
potential to estimate $\dot{\rm M}$ seismically and take into account
angular momentum loss during stellar evolution is appealing in view of the
core rotation rates found for intermediate-mass stars \citep{Aerts2017a}. 

We test here how the parameter $\dot{\rm M}$ affects the gravity-mode frequency
predictions.  As shown in the left panel of Fig.\,\ref{mass-loss}, forward
seismic modeling of gravity modes is not affected by the choice of the
atmosphere in the case of static models to compute the boundary conditions,
because such modes are determined by the properties of the deep stellar
interior. Nevertheless, the mode cavities are affected by the interplay between
the stellar wind and for a dynamical atmosphere, energy wave leakage and angular
momentum loss are predicted to occur \citep{Townsend2000a,Townsend2000b}.  We
considered a simplified test by comparing the difference in adiabatic
gravity-mode frequencies for a static gray atmosphere versus a dynamical
atmosphere for particular values of the mass-loss rate as indicated in
Table\,\ref{baseline}.  For each of the parameters of those benchmark models in
Table\,\ref{baseline}, we took the outcome of the MESA model for the luminosity,
radius, and effective temperature in the case of a gray atmosphere.  Along with
the mass and mass-loss rate, these values were then used for the calculation of
a radiation-driven stellar wind structure with the NLTE radiative transfer code
FASTWIND \citep{Puls2005}.  In this code the wind structure is derived from the
mass continuity equation and a pre-described velocity field derived from
standard radiation line-driven wind theory,
$v(r) = v_\infty (1-R_\star/r)^{0.8}$, with the stellar radius 
$R_\star$ and the terminal wind
speed $v_\infty$ as given by MESA.  NLTE Hopf functions are used for the
temperature stratification using an exact $T(\tau_R)$ relation from a converged
plane-parallel NLTE model (with $\tau_R$ the Rosseland optical depth) and taking
into account sphericity effects \citep[see][for an extensive
description]{Santolaya1997}.  This atmosphere and wind structure was then
stitched to the interior structure from MESA. The point of connection was chosen
as the one where the opacities of both models match best at $\tau_R=2/3$. A
smooth and continuous connection was enforced by linear interpolation of the
various quantities in log space around the point of connection.

The comparison between the frequencies of the modes for a static gray atmosphere
versus a dynamical FASTWIND atmosphere is shown in the right panel of
Fig.\,\ref{mass-loss} for the highest two considered masses (for the lower
masses, computations with FASTWIND are not meaningful).  
We find it justified to
ignore the wind for gravity modes, 
but not for pressure modes.  However, we note that our simple test
here only accounts for the wind outflow indirectly through a modified
atmospheric structure, i.e., we do not explicitly account for potential wave
leakage when solving the pulsation equations within GYRE. As such, the effects
seen in Fig.\,\ref{mass-loss} quite likely constitute lower limits.

Any radiation-driven wind effect was so far ignored in
seismic modeling of $\beta\,$Cep stars based on ground-based data.  Given that
stars born with a mass above some 24\,$M_\odot$ experience strong mass loss
throughout their entire life and stars with $M\gtrsim 3\,M_\odot$ as of the blue
supergiant phase, $\dot{\rm M}$ should be considered as an additional parameter
to estimate from forward seismic modeling of such stars. Unfortunately, seismic
data sets to achieve this are currently not yet available for stars with mass
above some 24\,M$_\odot$.  Major progress on this front is within reach of the
TESS Continuous Viewing Zones \citep{Ricker2016} and the long pointings of the
PLATO mission \citep{Rauer2014}.

\subsubsection{Treatment of the pre-main sequence evolution}

Solving the stellar structure equations not only requires boundary conditions at
the centre and the surface of the star, but also initial conditions. In essence,
an initial ``birth model'' is required to ``start'' the evolution from
core-hydrogen burning in full equilibrium and the age of the star is determined
as of that so-called Zero Age Main-Sequence (ZAMS) model. However, partial
nuclear burning already occurs during the contraction phase of the protostar, so
there is no one-to-one equivalence between the stellar age and $X_c$ at the
ZAMS.  Moreover, the physical processes during star formation, including
rotation, magnetism, and accretion during contraction from the Hayashi
  track towards the start of core
hydrogen burning in equilibrium at the ZAMS, 
remain largely unknown.  In view of this, one
tends to circumvent pre-main-sequence evolutionary computations.
MESA comes with a library of pre-computed ZAMS models
for $Z = 0.02$ and with masses between 0.08 and 100\,M$_\odot$ 
\citep{Paxton2011}.
An option to start the evolutionary computations offered to users is to interpolate 
between these pre-computed ZAMS models, in order to achieve a ZAMS model 
for the chosen mass.
We have tested this procedure against the more time-consuming 
computation of contraction models for the chosen $(M,X,Z)$ from the
Hayashi track to the ZAMS to end up with a self-consistent ZAMS model
  according to the chosen $(M,X,Z)$ and frozen input physics as initial conditions.
\begin{figure}
\begin{center}
 \includegraphics[width=3.5in]{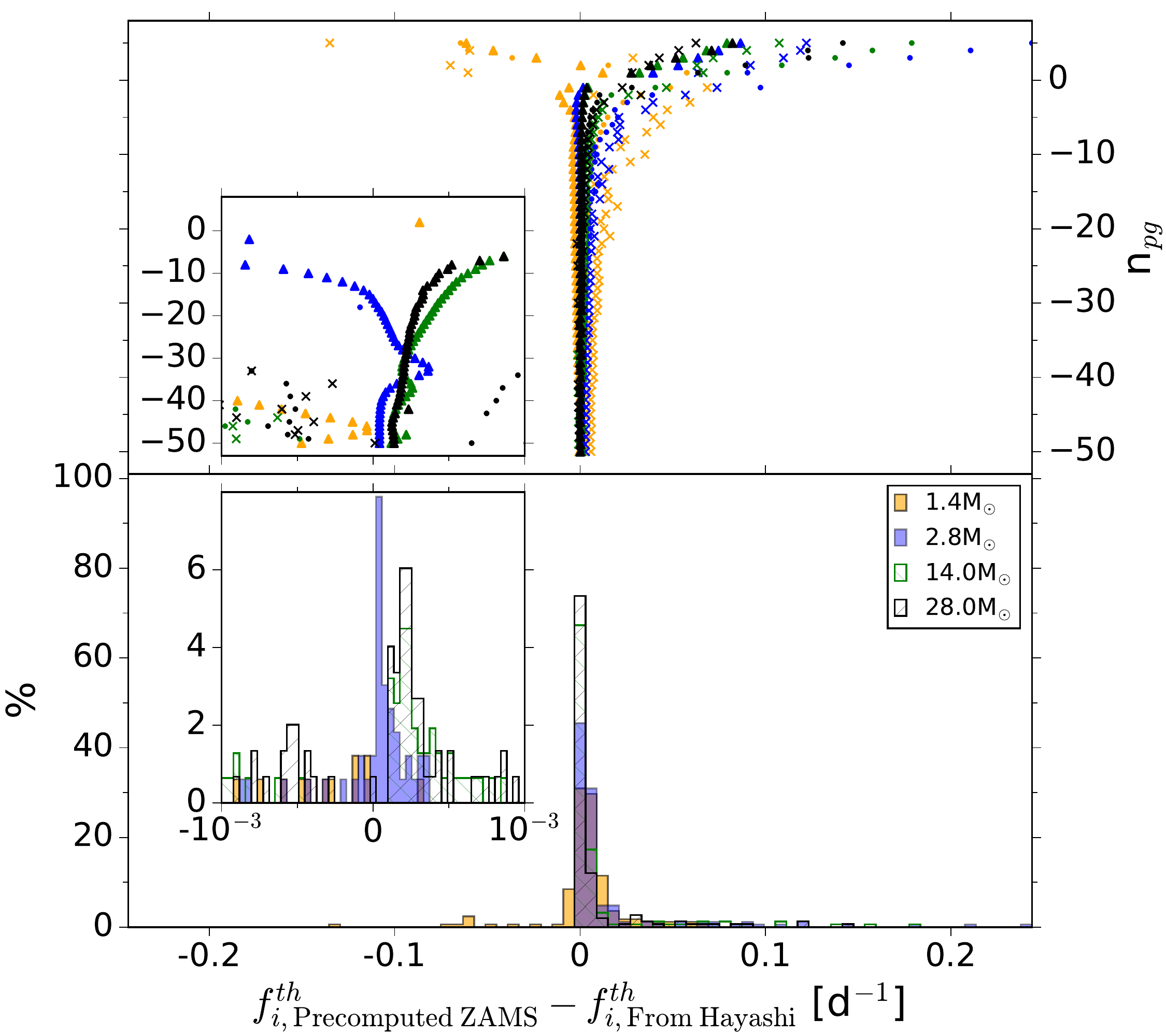}
 \caption{\label{preMS} Same as Fig.\,\protect\ref{TA-nonTA} but comparing 
frequencies for models computed from the Hayashi track versus models based on an
interpolation within a pre-computed grid of ZAMS models.}
\end{center}
\end{figure}
Comparing the frequencies between those two cases is non-trivial as one in
  fact evolves two slightly different stars as of the ZAMS. For this comparison,
Fig.\,\ref{preMS} shows the difference in the oscillation frequencies of
  models at equal values of $X_c$ to within 0.001. This figure
demonstrates that taking an interpolated ZAMS model from
MESA's pre-computed
libraries versus a ZAMS model resulting from contraction starting on the
Hayashi track with consistent input physics for the chosen $(M,X,Z)$
changes the frequencies more than the 
typical frequency error derived from a  half- to one-year
   light curve
for the majority of modes.  It is therefore essential to compute 
models from the Hayashi track to end up with proper
initial conditions
in forward seismic modeling.

Unlike the interior rotation, core overshooting and envelope mixing, we cannot
treat the ``global characteristics'' of the ZAMS model as a simple free
parameter to estimate during the seismic modeling. This aspect is rather a
typical case where various choices for the input physics during the
pre-main-sequence phase each constitute a different stellar model 
$\calm (\bftheta, \bfpsi)$.  The
dependence of the frequency values on the pre-main-sequence structure is good
news for future forward seismic modeling of pre-main-sequence pulsators. While
applications of seismic modeling to this early phase of stellar evolution must
await suitable photometry from space, the same methodology than described here
can be used for these early evolutionary stages once sufficiently identified modes
can be measured \citep[e.g.,][]{Zwintz2014,Zwintz2017}.

\subsection{\label{microphysics}Errors due to choices of microphysics}

\begin{figure}
\begin{center}
 \includegraphics[width=3.5in]{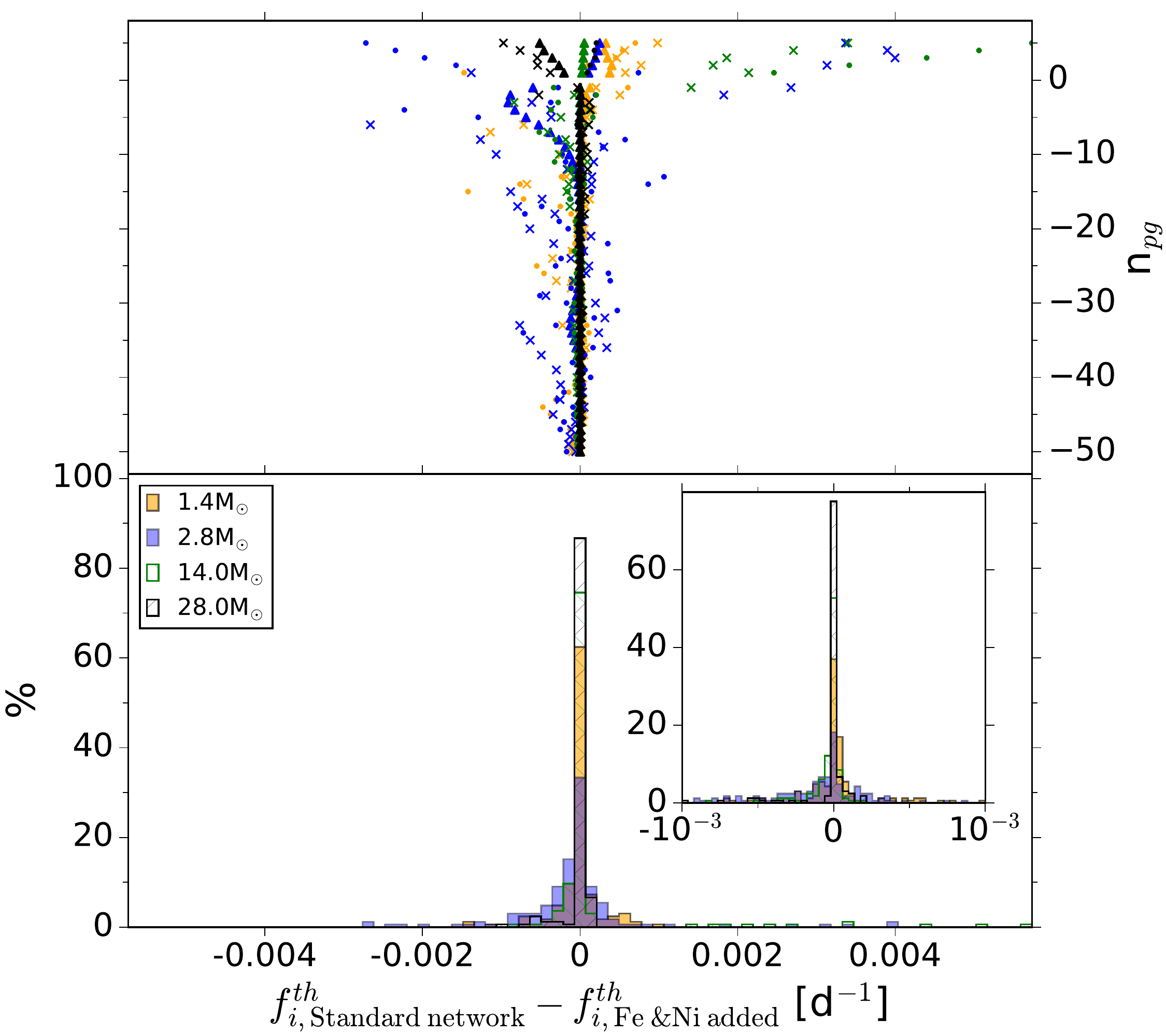}
 \caption{\label{network} Same as Fig.\,\protect\ref{TA-nonTA} but comparing 
frequencies for models computed with the basic network versus a network with 
$^{56}$Fe and $^{58}$Ni added.}
\end{center}
\end{figure}

Aside from the aspects of the macrophysics discussed above, a myriad of choices
for the microphysics can also be made to compute the stellar models. It
concerns, e.g., the Equation-of-State (EOS), thermonuclear and weak reactions,
nuclear reaction networks, opacities, etc.  Here, we focus on the core-hydrogen
burning phase for stars that do not undergo electron degeneracy. Such cases are
the best understood ones in terms of the EOS (an ideal gas with radiation is
appropriate). Moreover, the choices of the nuclear reactions and of the nuclear
reaction networks among a variety of sets affect the seismic properties least
during this simplest evolutionary phase, but may appreciably alter the more
advanced phases of stellar evolution.  For this work, we used the standard MESA
EOS, the NACRE rates \citep{Angulo1999} and the MESA basic rate network based on
eight isotopes: $^1$H, $^3$He, $^4$He, $^{12}$C, $^{14}$N, $^{16}$O, $^{20}$Ne,
and $^{24}$Mg \citep{Paxton2011}.  As an illustration of the effect of such
a choice, we also computed the oscillation frequencies in the case where the
$^{56}$Fe and $^{58}$Ni isotopes were added to the network and provide the
comparison in Table\,\ref{percentages} and Fig.\,\ref{network}. 
It can be seen that, as expected, this only has a minor
effect compared to other aspects of changing the input physics. 

We now turn our attention to some of the more critical aspects of the
microphysics that affect the oscillation frequencies appreciably.

\subsubsection{Opacities and chemical mixtures}

It is well known that the choice of opacities affects the oscillation properties
of B-type pulsators appreciably
\citep[e.g.][]{Daszynska2010}.  In their extensive study of
KIC\,10526294, \citet{Moravveji2015} quantified the effect of choosing different
opacity tables and mixtures for the equilibrium models on the resulting
frequency predictions for gravity modes (see their Fig.\,3). Two different sets
of opacities -- OPAL and OP -- were combined with three different chemical
mixtures to end up with six different models $\calm(\bftheta,\bfpsi)$, 
adopting the MESA
EOS. The six corresponding dipole zonal frequencies for the mode with radial
order $n_{pg}=-32$ differ up to $\simeq 0.004$\,d$^{-1}$, while the frequency
differences for the mode with radial order $n_{pg}=-14$ amount to
$\simeq 0.008$\,d$^{-1}$. This already demonstrated that the theoretical
frequency uncertainty due to different choices of opacity tables cannot be ignored.
\begin{figure}
\begin{center}
 \includegraphics[width=3.5in]{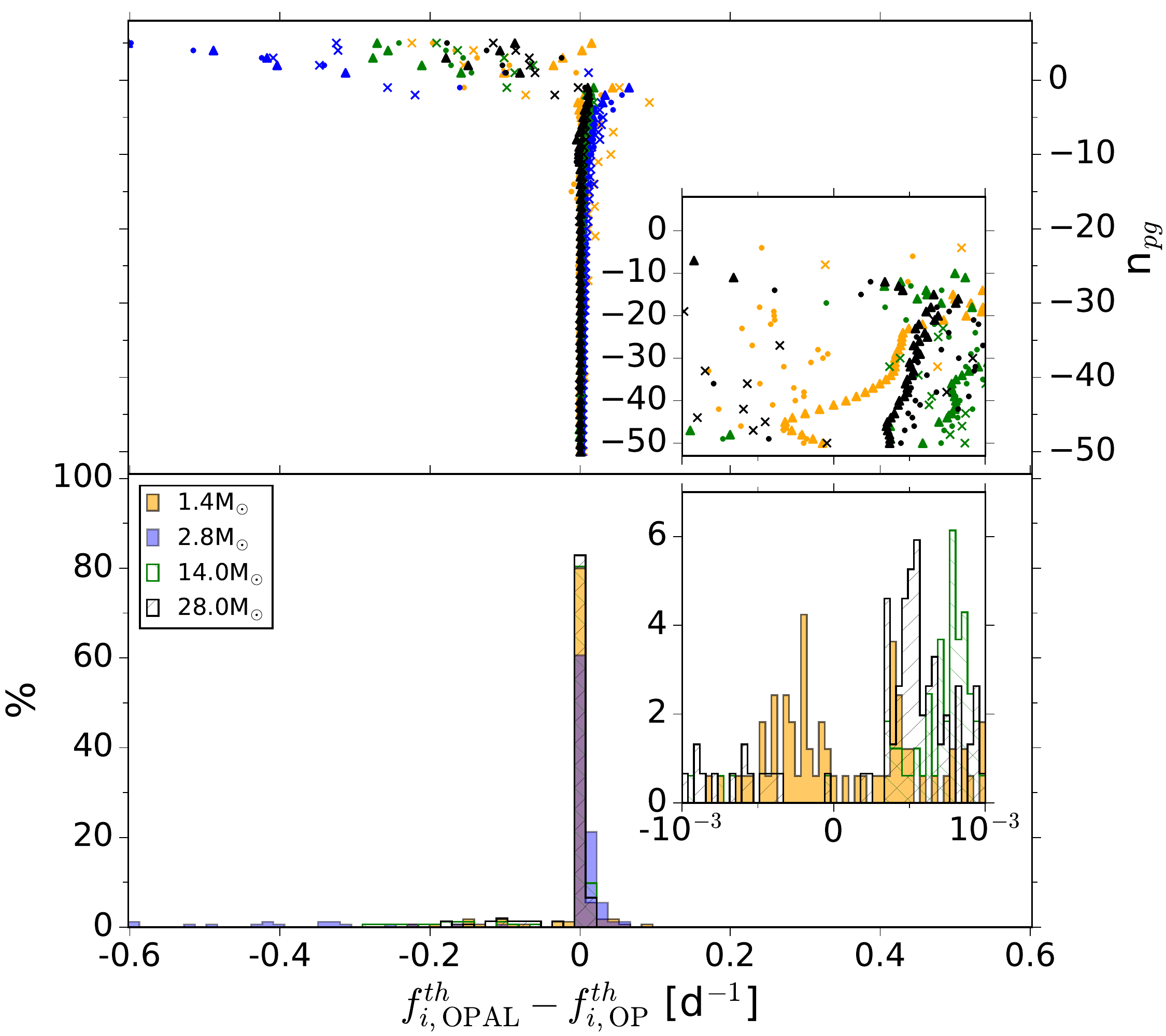}
 \includegraphics[width=3.5in]{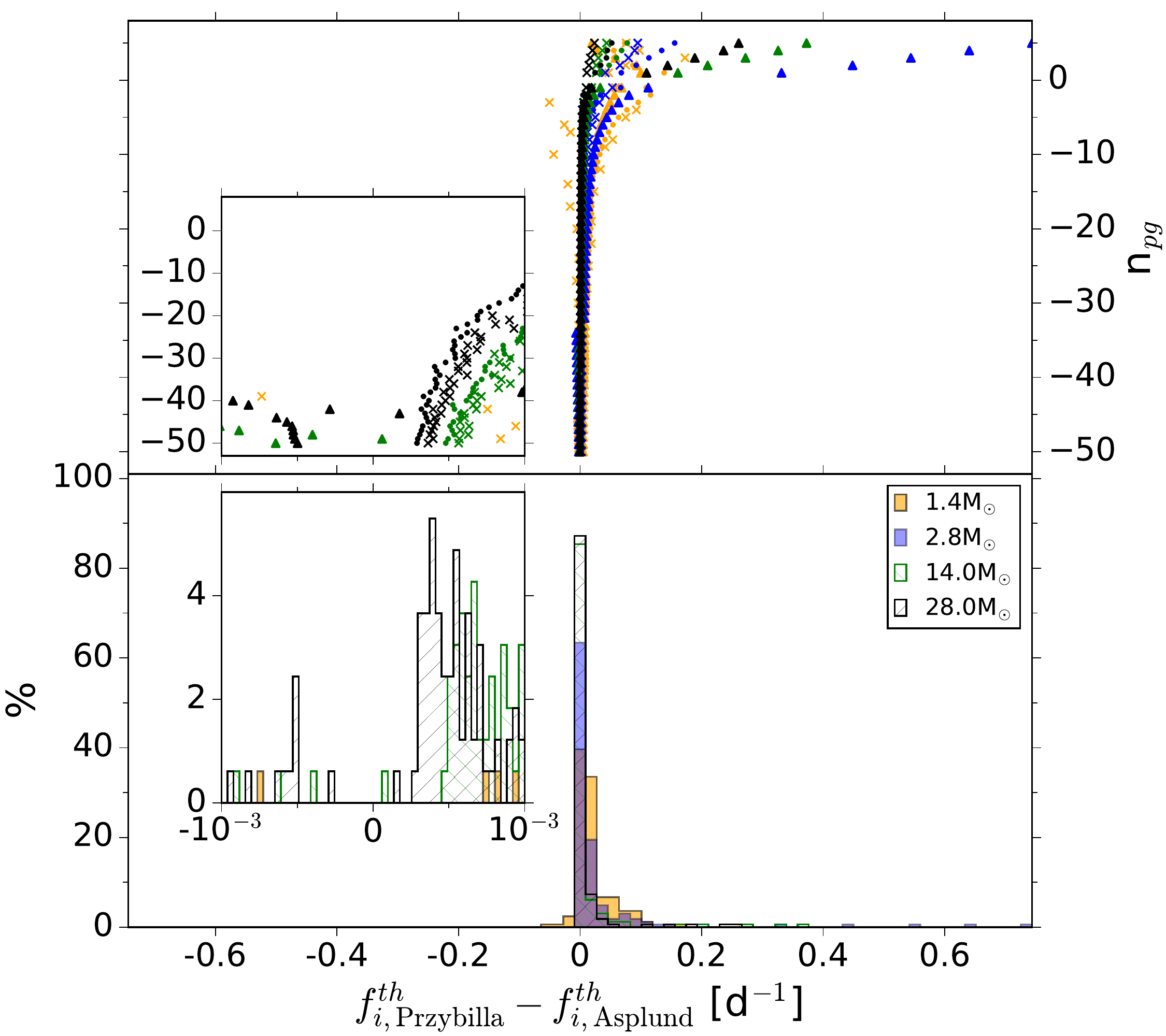}
 \caption{\label{opal-op} Same as Fig.\,\protect\ref{TA-nonTA} but comparing
   frequencies for models based on OPAL versus OP opacities (left) and for
   models based on the solar mixture by \citet{Asplund2009} versus the standard
   mixture of OB stars in the solar neighbourhood \citet{Przybilla2013},
   adopting OPAL opacities for both mixtures (right). }
\end{center}
\end{figure}

Figure\,\ref{opal-op} shows the deviations between the theoretical frequencies
for the benchmark models in Table\,\ref{baseline} due to a different choice of
opacities (left) and chemical mixture (right). Both these choices have a major
effect on the frequency values in the sense that the differences are larger than
the frequency error of a typical half- to one-year light curve. 
Both the pressure-mode and
gravity-mode frequencies get shifted with a systematic effect that cannot be
ignored.  
We
conclude that, with the current poor knowledge of opacities, 
asteroseismology in the mass regime considered here
should be based on various stellar models $\calm (\bftheta,\bfpsi)$ 
relying on different opacity
tables and chemical mixtures.

\subsubsection{Microscopic atomic diffusion}

In addition to the aforementioned parameter $D_{\rm ext}$ to be estimated in
the radiative envelope, composition changes may also occur due to transport
processes following microscopic atomic diffusion in the radiative layers of the
star.  The computation of such microscopic processes requires detailed
evaluation of frequency-dependent absorption coefficients and re-evaluation of
opacities from all the relevant ions involved in the various layers of the star,
for each time step along the evolution. This obviously constitutes an immense
challenge for the evolutionary computations (CPU-wise) and is usually omitted
for massive stars as these evolve on a timescale of millions of years.  For cool
stars with an extensive convective envelope, it is necessary to consider models
with gravitational settling for evolutionary computations
\citep[e.g.,][]{Michaud2004} and for proper chemical tagging for galactic
archeology \citep{Dotter2017}. Here, we investigate the effect of atomic
diffusion for intermediate- and high-mass stars, which also requires the
  inclusion of radiative levitation \citep[e.g.][]{Deal2016,Deal2017}.

Within MESA, four components of atomic diffusion can be included, each for
various chemical elements: gravitational settling, concentration diffusion,
thermal diffusion, and radiative levitation \citep{Hu2011,Paxton2015}.
Including atomic diffusion in MESA requires any of these
phenomena to be turned ``on'' or ``off'', i.e., the activation of
any of these particular microscopic effects can be included or not.
Moreover, each of the phenomena can be included with or without a
parameter to be estimated.  In our methodological set up, including atomic
diffusion for each different chemical element would imply for each a different
stellar model $\calm (\bftheta, \bfpsi)$.  
For low-mass stars \citep[e.g.,][]{Verma2017} and for
white dwarfs \citep[e.g.,][]{Romero2017}, gravitational settling is an important
ingredient for asteroseismic modeling.  On the other hand, radiative levitation
acting upon Fe and Ni is critical to take into account when estimating
oscillation frequencies for hot subdwarfs in core helium burning
\citep[e.g.,][]{Charpinet2011,Bloemen2014}.
\begin{figure}
\begin{center}
 \includegraphics[width=3.5in]{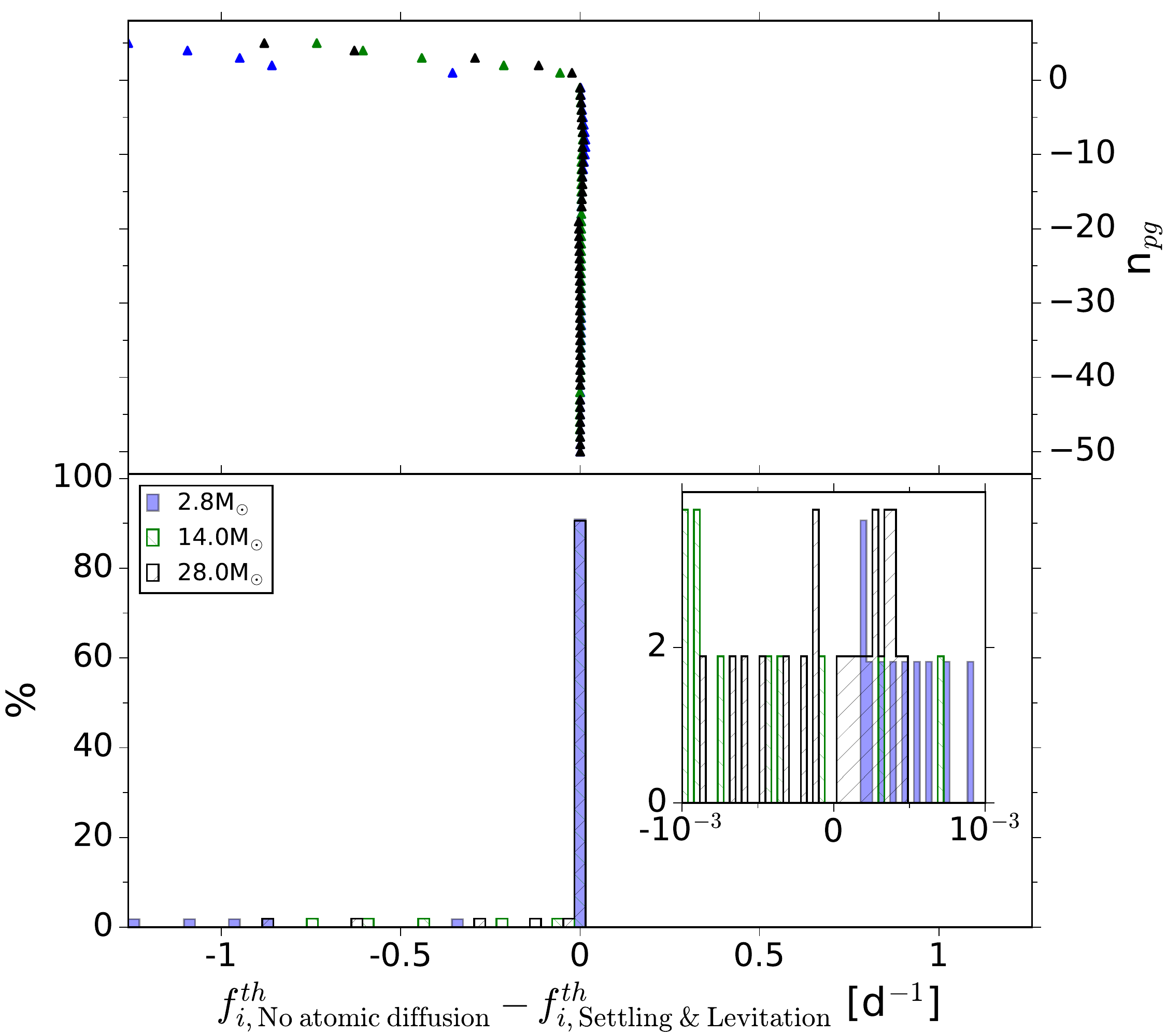}
 \caption{\label{diffusion} Same as Fig.\,\protect\ref{TA-nonTA} but comparing
   models without and with atomic diffusion, taking into account gravitational
   settling, concentration diffusion, thermal diffusion, and
radiative levitation. This comparison is limited to the case of $X_c=0.7$  
for computational reasons. For all these cases, $^{56}$Fe and $^{58}$Ni 
were added to
the network.}
\end{center}
\end{figure}

\begin{figure}
\begin{center}
\includegraphics[width=3.5in]{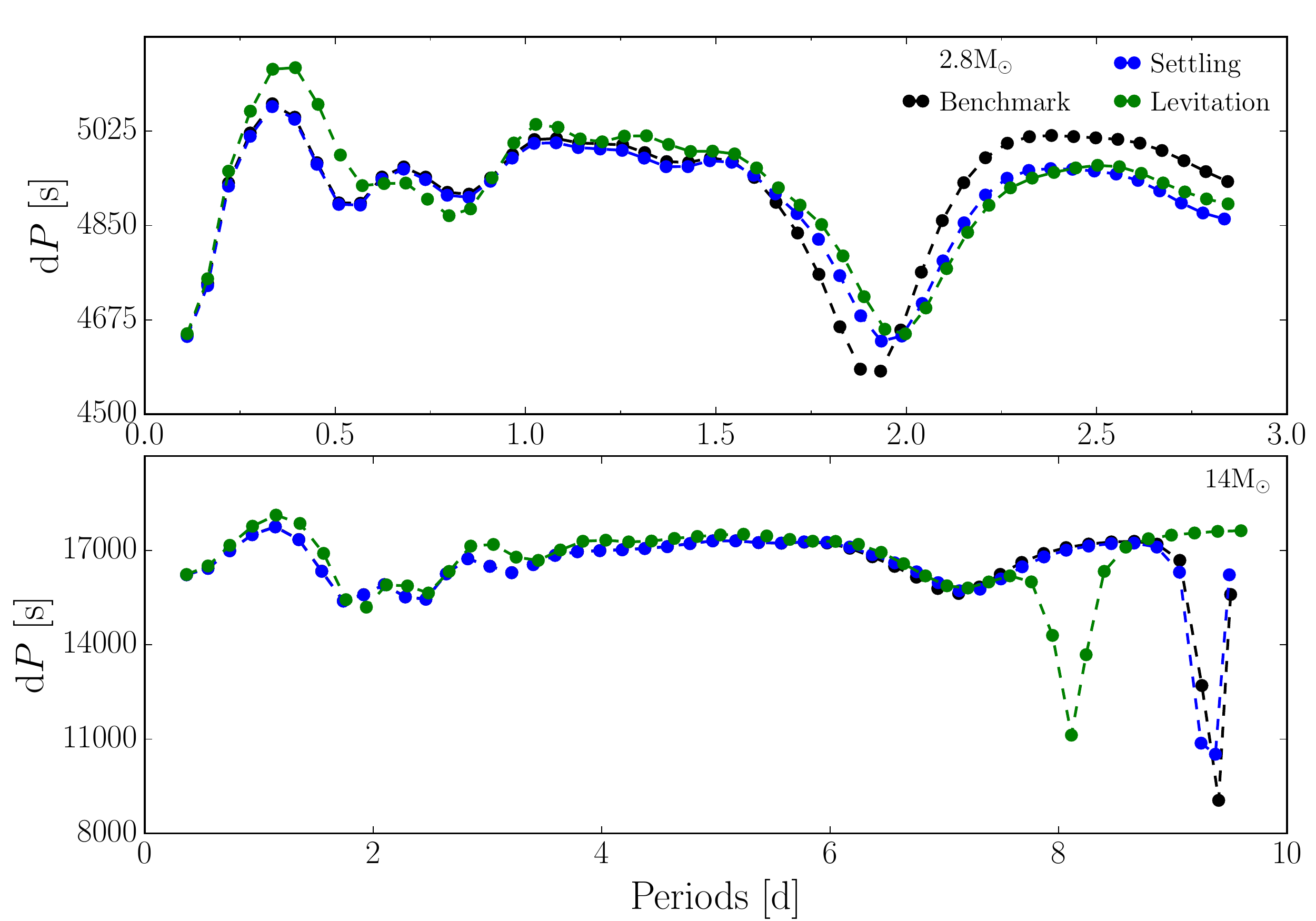}
 \caption{\label{DeltaP-Levitation} Change in the 
gravity-mode period spacing patterns for the 
2.8 and  14\,M$_\odot$ models at $X_c= 0.7$ when settling alone and levitation
in addition is considered.} 
\end{center}
\end{figure}

For the stars under study here, we show in Fig.\,\ref{diffusion} the comparison
between the frequencies due to models without and with atomic diffusion.  For
this comparison, we used the treatment by \citet{Hu2011}, as implemented in
MESA, taking into account gravitational settling, concentration diffusion,
thermal diffusion, and radiative levitation for the elements H, He, C, N, O, Ne,
Mg, Fe and Ni.  For this comparison we limited to the models from the ZAMS until
$X_c=0.7$ for CPU reasons, where we adopted the most recent MESA implementation
\citep{Paxton2018}.  In this comparison, we omit the models with
  $M=1.4\,$M$_\odot$, because 
settling is the dominant effect for those and this
  requires some level of turbulent mixing (combined or not with envelope
  undershooting) to be included in the model computations, in order to 
end up with appropriate surface abundances of metals
\citep[see, e.g., ][for pioneering work]{Turcotte1998}.
One usually takes such ad-hoc turbulent mixing to be a
  function of density
  \citep[e.g.,][]{Richer2000,Deal2016,Dotter2017}.  Since our
  baseline models do not have such density-dependent turbulence, nor
  undershooting, the comparison with the baseline models indicated in
  Table\,\ref{percentages} is not meaningful for this mass.

Figure\,\ref{diffusion} and Table\,\ref{percentages} reveal that atomic
  diffusion has a large effect on the frequencies for all the modes. We find
  that this is mainly due to the radiative levitation for the three masses we
  considered here.  This is illustrated for the period spacing patterns of the
  2.8 and 14\,M$_\odot$ models in Fig.\,\ref{DeltaP-Levitation}, where it is
  seen that the mode trapping changes when radiative levitation is
  taken into account, because it modifies 
the $\mu$-gradients and hence the  convection zones.
The importance of radiative levitation was already
  stressed by \citet{Deal2017} in their study of the solar-like oscillations of
  the F-type exoplanet host star 94\,Ceti.  Radiative levitation has so far
  hardly been included in forward seismic modeling of gravity modes, mainly
  because it is too demanding in computation time. Our results show it to be
  worthwhile to undertake future studies on how to include the combined effect
  of settling and levitation in the forward modeling of intermediate-mass and
  high-mass stars.

As a side remark in this section, we stress that atomic diffusion may be of
importance locally in and near the driving layer in the envelope for mode
excitation of opacity-driven modes during the main sequence, since more modes
are detected than predicted to be excited in massive stars
\citep[e.g.,][]{Moravveji2016,Daszynska2017}.  For this reason, mode excitation
is not part of our forward seismic modeling scheme but is treated as {\it a
  posteriori\/} evaluation of how good the best seismic models are in terms of
explaining overall stellar properties, along with other observables such as a
spectroscopic $T_{\rm eff}$ and $\log\,g$, surface abundances, and
luminosities deduced from Gaia astrometry.

In general, our statistical framework outlined in the following sections is only
based on gravity-mode frequencies and does {\it not\/} rely on surface
properties from spectroscopy, nor on astrometry or interferometry. The reason is
that the gravity modes provide direct observational information on the deep
stellar interior, at the level of $\varepsilon_i^\ast/f_i^\ast<\!\!<\,0.1\%$,
while the relative precisions for $T_{\rm eff}$, $\log\,g$ and abundances from
spectroscopy of stars in the mass range considered here are far worse, often
above 10\% \citep[e.g.][]{Morel2008,Martins2012a,Martins2012b,VanReeth2015}.
Moreover, our aim is to tune the near-core interior properties of the
stars. After deducing the best seismic models, we will confront their
predictions with measured surface properties, to assess the physical conditions
in the outer envelopes of the stars. In this respect, our set-up and application
is quite different than the case of low-mass stars with pressure modes.  Of
course, whenever additional spectroscopic, interferometric or astrometric data
would become available at precise enough level for our scientific aims, it can
also be used as {\it input\/} in the forward modeling, together with the
oscillation modes. Currently, this is not yet the case for the mass regime we
treat here.

\section{\label{MLE} 
Statistical model formulation and parameter  estimation}

We now place the principles outlined in the previous
Sections\,\ref{ingredients}, \ref{pulsation-error}, and \ref{models-error} into
a formal and appropriate statistical framework.  We assume that we have various
stellar model grids based on different stellar models $\calm(\bftheta,\bfpsi)$,
along with their pulsational frequencies, at our disposal.  We want to reach two
major goals:
\begin{enumerate}
\item to fit as closely as possible the observed and identified gravity-mode
  oscillation frequencies $f_i^\ast$ ($i=1,\dots,n$) by model (theoretical)
  values $f_i^{\rm th}$.  As outlined above, we are in the situation that some
  of the uncertainties on the theoretical frequencies are smaller than the
  errors of the observed frequencies, while others are larger or of the same
  order of magnitude.  The overall
  theoretical errors (for $f_i^{\rm th}$) are unknown, because additional factors than
  those shown in Figs\,\ref{TA-nonTA} -- \ref{diffusion} are at play, such
  as the limitation to 1D stellar structure models, the absence of certain forces, missing
  input physics, etc. Hence, heteroscedasticity must be included in the statistical 
model formulation.
\item to select the most likely physical model $\calm(\bftheta,\bfpsi)$ for a
  single star or for an ensemble of stars, without introducing {\it a priori\/}
  bias about the unknown input physics, and in particular keeping in mind that we
  are dealing with a different physical regime compared to Sun-like stars with a
  radiative core: here, the Coriolis force, core overshooting and transport of
  elements in the radiative envelope are dominant phenomena that cannot be
  ignored.
\end{enumerate}
This setting is different from many statistical problems of parameter
estimation, where a single parameter vector is estimated from a set of
repetitions.  Here, we are {\it not dealing with replicates\/} of one and the
same mode frequency but rather with a single measurement for a set of observed
frequencies $f_i^\ast$ ($i=1,\dots,n$) belonging to modes that each have
their own mode cavity and probing power for the interior physics.  
Each of the measurements, $f_i^\ast$, is accompanied by its own single error
measurement $\varepsilon_i^\ast$ for which we assume 
$\varepsilon_i^\ast\sim N(0,\lambda_i)$.

We now introduce convenient vectorial notations to outline the above aims into
suitable statistical frameworks. An observation takes the form of a vector
$\BY^\ast$, consisting of the observed frequency values,
$Y^\ast_i=f_i^\ast$ with $i=1,\dots,n$, where the length of the vector,
$n$, is star specific (the number of identified gravity modes).  Let us denote
the measured frequency precisions as $\Lambda$, a diagonal matrix with elements
$\lambda_i$.  Given a physical model $\calm(\bftheta,\bfpsi)$, a vector
$\bftheta$ of stellar parameters, and a vector of fixed properties $\bfpsi$ that
is assumed known, the model predicts the observation: $\BY(\bftheta,\bfpsi)$,
where $\BY$ is a vector of $n$ predicted frequencies $f_i^{\rm th}$ from model
$\calm(\bftheta,\bfpsi)$ for the identified frequencies corresponding with
$Y^\ast_i$.  Denote by $p$ the length of vector $\bftheta$. For convenience,
$\bfpsi$ will sometimes be dropped from notation:
$\BY(\bftheta)\equiv\BY(\bftheta,\bfpsi)$.  The inclusion of a parameter
$\bfpsi$ in the statistical model allows to separate key parameters to be
optimized (grouped in $\bftheta$) from the ones that are considered fixed for
the fit. This is necessary when the overall parameter vector (grouping
$\bftheta$ and $\bfpsi$) is high-dimensional as in our problem of forward
seismic modeling.

In our application, the number of detected gravity modes with well-determined
frequency,  $n$, 
is typically between 10 and 40 when dealing with nominal
{\it Kepler\/} data but smaller for other data sets, while $p$ is at least
seven: aside from the four standard input parameters for the computation of
stellar evolution, $(M,X,Z,X_c)$, we need to estimate as well the near-core
rotation with period $P_{\rm rot}=1/f_{\rm rot}=2\pi/\Omega_{\rm rot}$, at least
one parameter connected with the core overshooting, $D_{\rm ov}$, and one
parameter describing the mixing in the radiative envelope responsible for the
mode trapping.  Ideally, however, $p$ is larger, as one also wants to estimate
the {\it profiles\/} of the core overshoot, envelope mixing, and interior
rotation.  As a concrete example of this approach in the literature, $\bftheta$
was taken to be $(M,X,Z,X_c,D_{\rm ov},D_{\rm ext})$ (hence $p=6$), while
$\bfpsi$ stood for fixed values of $\alpha_{\rm MLT}=1.8$,
$\alpha_{\rm sc}=0.01$, $\dot{M}=0.0$, along with the fixed choices of the MESA
EOS, the chemical mixture by \citet{Przybilla2013}, and OPAL opacities, in
``Model\,4'' of the study of the ultra-slow rotator KIC\,10526294 by
\citet{Moravveji2015}. The latter authors tested the performance of this
Model\,4 with respect to a simpler stellar model having no extra diffusive
mixing in the envelope, i.e., a stellar model with
${\bftheta}=(M,X,Z,X_c,D_{\rm ov})$ ($p=5$) and $\bfpsi$ defined by
$\alpha_{\rm MLT}=1.8$, $D_{\rm ext}=0.0$, $\alpha_{\rm sc}=0.01$,
$\dot{\rm M}=0.0$ with the same fixed choices of the MESA EOS, chemical mixture,
and opacities \citep[``Model\,1'' in][]{Moravveji2015}. The result was that
Model\,4 gave a better representation of the detected zonal dipole gravity mode
frequencies than Model\,1, hence the conclusion that this star undergoes
envelope mixing in addition to core overshooting. In such an experiment,
Models\,1 and 4 are called {\it nested\/} models, because Model\,4 has one more
parameter to estimate than Model\,1 but is otherwise the same.

Commonly, the estimation of $\bftheta$ is evaluated in all points of a grid.
This grid can result from a cartesian product in the components of $\bftheta$,
but this is not a necessary requirement. Also, in practical applications, the
grid will be finer in one component than in another. 
Let us index the grid points by a single index $j=1,\dots,q$. This leads to a
collection of $q$ theoretically-generated vectors in each grid point $j$,
denoted as $\BY_j\equiv \BY_j(\bftheta_j,\bfpsi)$.  These vectors $\BY_j$ each have
length $n$. They result from the theoretical model $\calm(\bftheta,\bfpsi)$ that
lies at the basis of the grid. We consider and distinguish between the following
four problems to solve:
\begin{description} 

\item[\bfseries Problem 1] Given one set of observed identified frequencies
  $\BY^\ast$ and a theoretical model $\calm(\bftheta,\bfpsi)$ used to compute a
  grid with $j=1,\ldots,q$ grid points, find the most plausible value for
  $\bftheta$ from the grid, i.e., select $\bftheta_j$ for which
  $\BY_j(\bftheta_j,\bfpsi)$ comes
  closest to $\BY^\ast$. By extension, a small set of candidate $\bftheta_j$ can
  be selected. Assess the errors of the selected $\bftheta_j$.

\item[\bfseries Problem 2] We wish to compare the capacity of various
  theoretical models to describe the data.  In order to tackle this, it is
  convenient to compare the estimates among two theoretical models,
  $\calm(\bftheta^{(m)},\bfpsi^{(m)})$ and
  $\caln(\bftheta^{(n)},\bpsi^{(n)})$. The problem is thus for a given value of
  the vector $\bftheta^{(m)}$ from a benchmark model of grid $(m)$, to find
  candidates $\bftheta^{(n)}$ from grid $(n)$ that provide the best match.

\item[\bfseries Problem 3] Search for a statistical model to replace theoretical
  model $\calm(\bftheta,\bfpsi)$, delivering an outcome vector sufficiently
  close to $\BY$ but much faster to evaluate and lending itself for error
  estimation of $\bftheta$ and for statistical model selection.

\item[\bfseries Problem 4] Find the most adequate theory
  $\calm(\bftheta,\bfpsi)$, among a collection of competing theories, using a
  set of $N$ stars with $\BY^\ast_t$ (where $t=1,\dots, N)$.
\end{description} 
We provide methodology to treat each of these four problems in the following
subsections, adopting vectorial notation as is common in statistics.

\subsection{Problem 1: A Mahalanobis distance based solution to 
match the identified gravity modes of one star}

The problem is to find the value for $\bftheta$ that best predicts the stellar
observation $\BY^\ast$, $\bftheta_0$ say, with corresponding value
$\BY_0\equiv\BY_0(\bftheta_0,\bfpsi)$. We need to select $\bftheta_0$ such that
the distance between $\BY^\ast$ and $\BY_0$ is smallest, keeping in mind the
measured frequency precisions $\Lambda$.  One should accommodate two
features. The first is that the variability in one component of $\bftheta$ may
be larger than in another; in addition, the units may vary from component to
component. This implies that standardization is necessary. The second feature is
that correlations occur between the components of $\bftheta$ \citep[cf.\ Fig.\,5
in][]{Moravveji2015,Pedersen2018}.  The stronger a component correlates with the other
components, the less independent information it contributes.  An appropriate
solution to this problem is offered by the Mahalanobis distance \citep{JW02}:
\begin{equation}
\label{maldist}
\bftheta_0=\arg\min\left\{ (\BY(\bftheta)-\BY^\ast)^\top V^{-1}(\BY(\bftheta)-\BY^\ast)\right\},
\end{equation}
where $V=\mbox{var}(\Y)$, the variance-covariance matrix 
of the vector $\Y (\bftheta,\bfpsi)$.  
The Mahalanobis distance naturally takes into account correlations among the 
components of $\bftheta$ and reduces to the Euclidean distance in absence of
any such correlations, i.e., when $V$ is a diagonal matrix.

To evaluate Eq.\,(\ref{maldist}), we need an expression for $V$. This matrix can be
seen as the intrinsic variability in $\BY$ across a relevant range of $\Y$. The
grid based on $\calm(\bftheta,\bfpsi)$,
especially when sufficiently fine, will provide an approximation of
this quantity when computed over a reasonable range of $\bftheta$. 
This implies that it can
be estimated from the grid. Considering all grid points gives:
\begin{equation}
\label{vmat}
\widehat{V}=\frac{1}{q-1}\sum_{j=1}^q(\BY_j-\overline{\Y})(\BY_j-\overline{\Y})^\top,
\end{equation}
where the average is
$$\overline{\BY}=\frac{1}{q}\sum_{j=1}^q\BY_j.$$
In practice, we will typically evaluate Eq.\,(\ref{maldist}) using all grid points: 
\begin{equation}
\label{maldistgrid}
\bftheta_0=\arg\min_{j=1}^q\left\{ (\BY_j-\BY^\ast)^\top V^{-1}(\BY_j-\BY^\ast)\right\}.
\end{equation}
Taking the measurement errors of the frequencies $\BY^\ast$ into account in the
computation of the Mahalanobis distance is done by replacing $\widehat{V}$ by
$\widehat{V}+\Lambda$ in Eq.\,(\ref{maldistgrid}).
Note that the inverse $V^{-1}$ may need to be replaced with the Moore-Penrose
inverse \citep{Penrose1955} for models that are not of full rank (see the
statistical models treated in Problem\,3).  Expression\,(\ref{maldistgrid})
provides a more appropriate measure than the often used $\chi^2$ based on an
Euclidian distance, which does not
take into account the correlated nature of the components of the vector
$\bftheta$.

As for a first estimate of 
the error of $\bftheta_0$, we consider the Mahalanobis distance to comply
with a normal distribution. In that case, it follows a $\chi^2_p$
distribution. Let the critical point be $c_p$, i.e.,
$P(\chi^2_p\ge c_p)=0.05$.  We then consider the empirical prediction
ellipsoid across a relevant part of the grid:
\begin{equation}
\label{predictionellipsoid}
\calc=\left\{
\bftheta_j|j=1,\dots,q; (\BY_j(\bftheta_j,\bfpsi)-
\BY^\ast)^\top \widehat{V}^{-1}(\BY_j(\bftheta_j)-\BY^\ast)\le c_p\right\}.
\end{equation}
The theoretical ellipsoid would range over all $\bftheta$ in continuous space,
which is impractical, of course. As a consequence, the prediction ellipsoid
depends on the coarseness of the grid, so it is essential to have it
sufficiently fine yet covering an approriately broad range for $\bftheta$.  We
can circumvent this below when treating Problem\,3, where $\bftheta$ is
determined from a statistical model rather than a theoretical model. In the case that
this can be done successfully, we improve the error estimation from
Eq.\,(\ref{predictionellipsoid}) by relying on the errors of the parameters of
the statistical model to determine the errors of $\bftheta$.

\begin{figure}
\centering
  \includegraphics[width=0.95\linewidth]{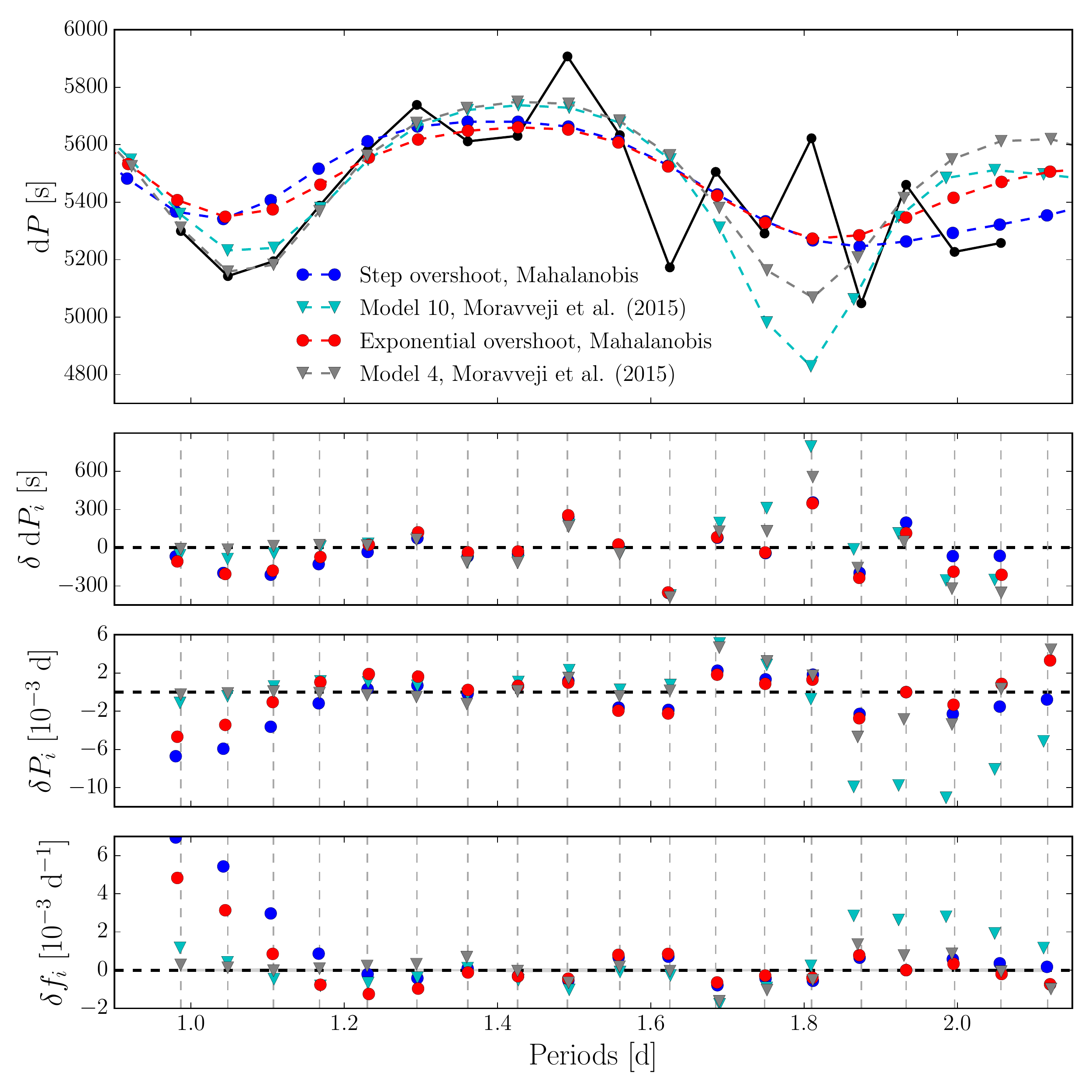}
  \caption{Comparison between observed and theoretical period spacing patterns
    (upper panel) and their difference (2nd panel), the corresponding mode
    period (3rd panel) and mode frequency (lowest panel) differences between the
    models and observations.  The gray dashed lines indicate the observed
    frequencies listed in Table\,1 of \citet{Moravveji2015}. The errors of the
    measured mode periods are represented in Fig.\,\protect\ref{KIC-Var-Covar}.}
\label{KIC-figure}
\end{figure}
\begin{figure}
\centering
\includegraphics[width=0.45\linewidth]{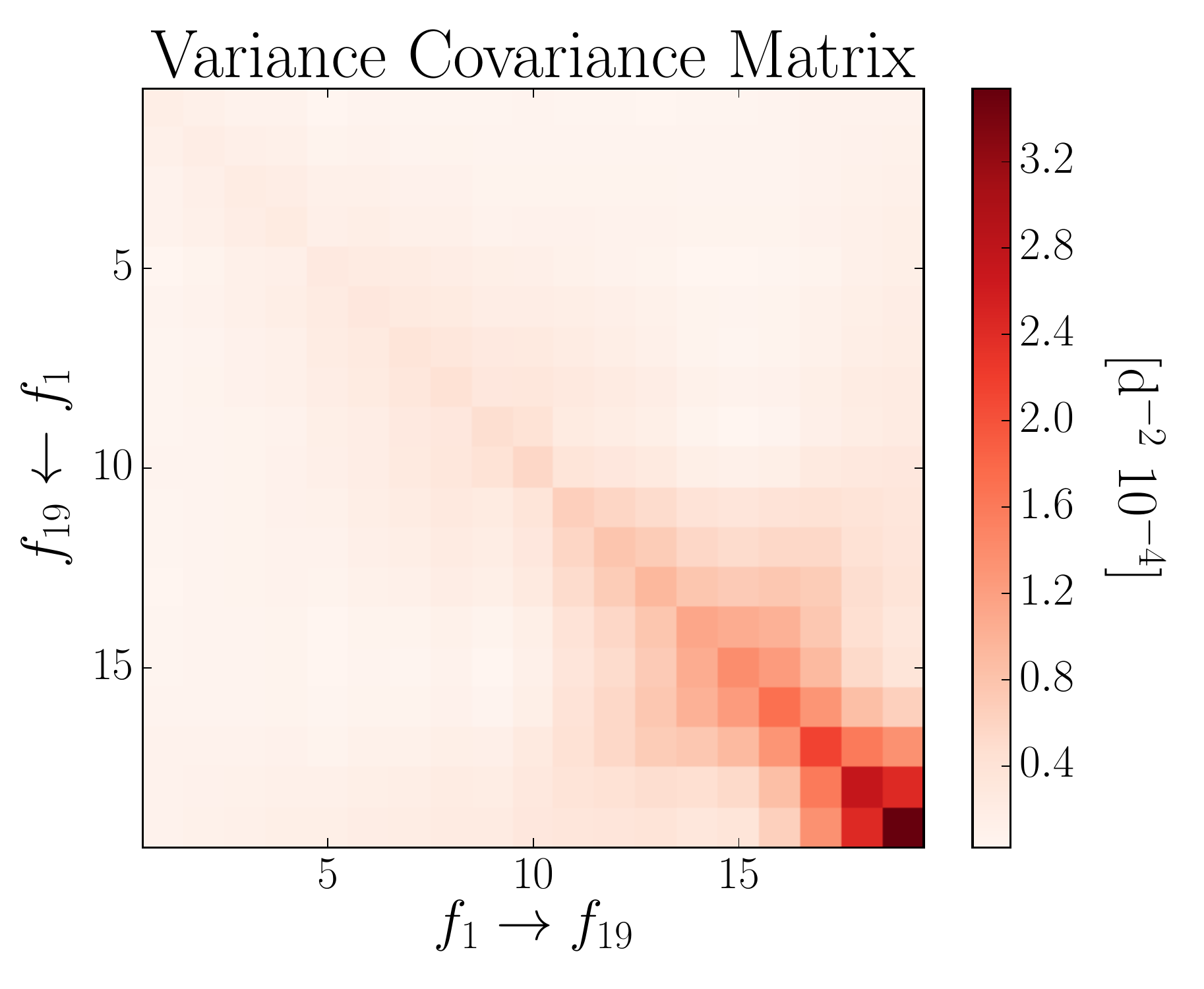}
\includegraphics[width=0.45\linewidth]{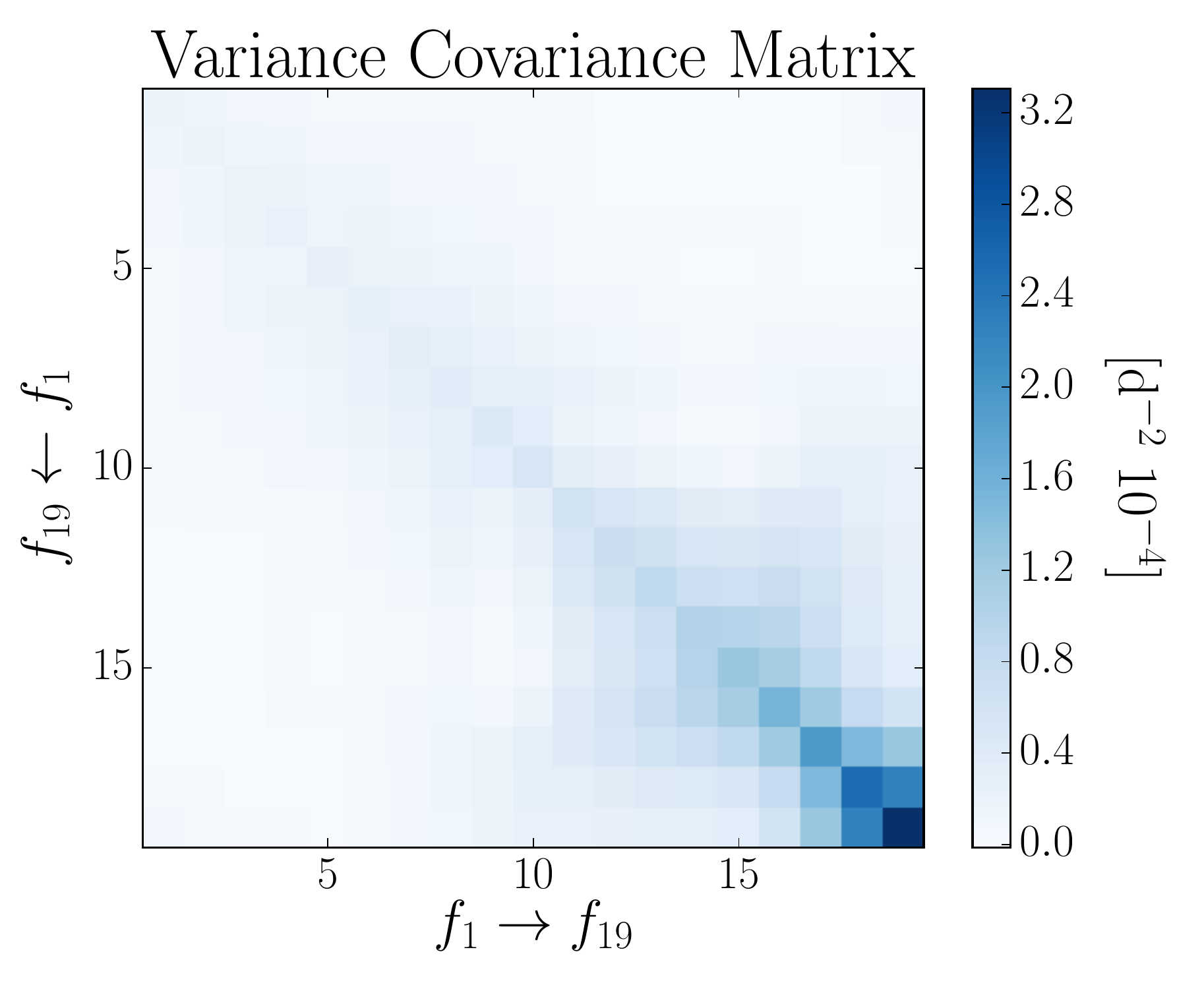}
\caption{Variance-covariance matrix $\hat{V} + \Lambda$ of dimension 19 $\times$
  19 representing the 19 detected zonal mode frequencies of KIC\,10526294, for 
two considered dense grids of models, including the observational errors 
(left: for the 
exponential overshoot grid whose best model is
represented by the red dots in Fig.\,\ref{KIC-figure}; right: for the step
overshoot grid whose best model is indicated in blue in Fig.\,\ref{KIC-figure}).} 
\label{KIC-Var-Covar}
\end{figure}
\begin{table}
\caption{\label{KIC-result}
Parameter estimation
based on a $\chi^2$ statistic \citep{Moravveji2015} and on the
Mahalanobis distance $\bftheta_0$ from the 19 dipole zonal modes of KIC\,10526294.}
\begin{tabular}{r|ccccc}
\hline
Statistic & $M (M_\odot)$ & $Z$ & $X_c$ & $D_{\rm ov}$ 
& $D_{\rm  mix}$\,(cm$^2$\,s$^{-1}$)\\  \hline  
$\chi^2$ & 3.25 & 0.014 &  0.627 & $f_{\rm ov}=0.017$ &  56.2 \\
$\bftheta_0$ & 3.24 & 0.017 &  0.644 & $f_{\rm ov}=0.018$ &  177.8 \\
\hline
$\chi^2$ & 3.19 & 0.019 &  0.628 & $\alpha_{\rm ov}=0.21$ &  56.2 \\
$\bftheta_0$ & 3.23 & 0.017 &  0.640 & $\alpha_{\rm ov}=0.20$ &  177.8 \\
\hline
\end{tabular}
\end{table}
As an illustration of the benefit of $\bftheta_0$ with respect to a $\chi^2$
statistic, we rely on two very dense seismic model grids computed to explain the
19 identified dipole zonal modes of the slowly pulsating B star 
KIC\,10526294 by \citet[see Table\,1
in][for the frequencies and Table\,3 for the grids and selection of the best
models 4 and 10 from a $\chi^2$ statistic]{Moravveji2015}.  Both these grids
have $p=5$ and their input physics only differs in the treatment of the core
overshooting, with one grid using exponential overshoot described by the
parameter $f_{\rm ov}$ and the other grid relying on a step overshoot with
parameter $\alpha_{\rm ov}$ (both have the radiative temperature gradient in the
overshoot zone).  This star rotates at less than 1\%$v_{\rm crit}$ and it revealed
zonal modes with unambiguous identification from rotational splitting,
so rotation can be ignored for the matching of the mode frequencies.

\begin{figure}
\centering
\includegraphics[width=0.45\linewidth]{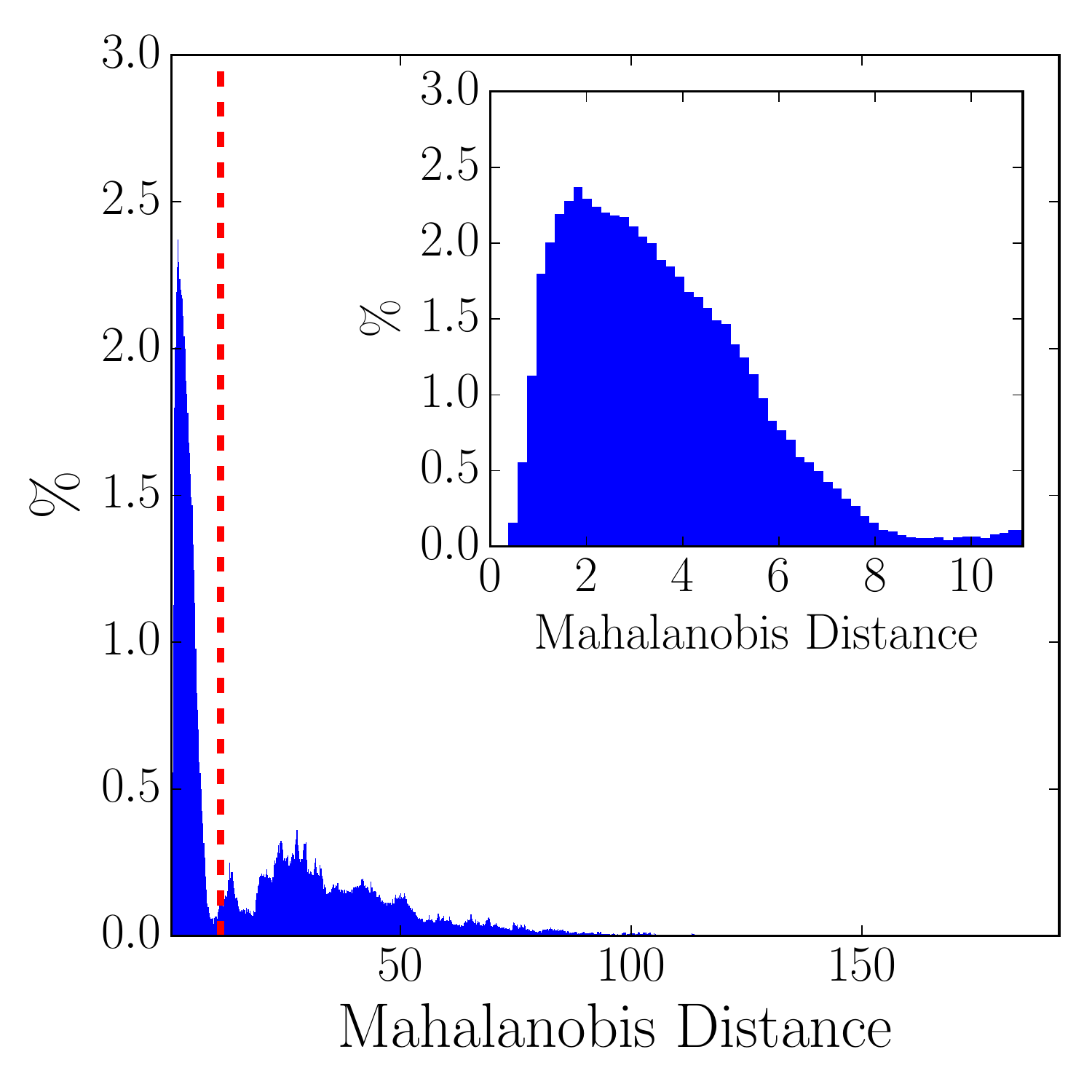}
\includegraphics[width=0.45\linewidth]{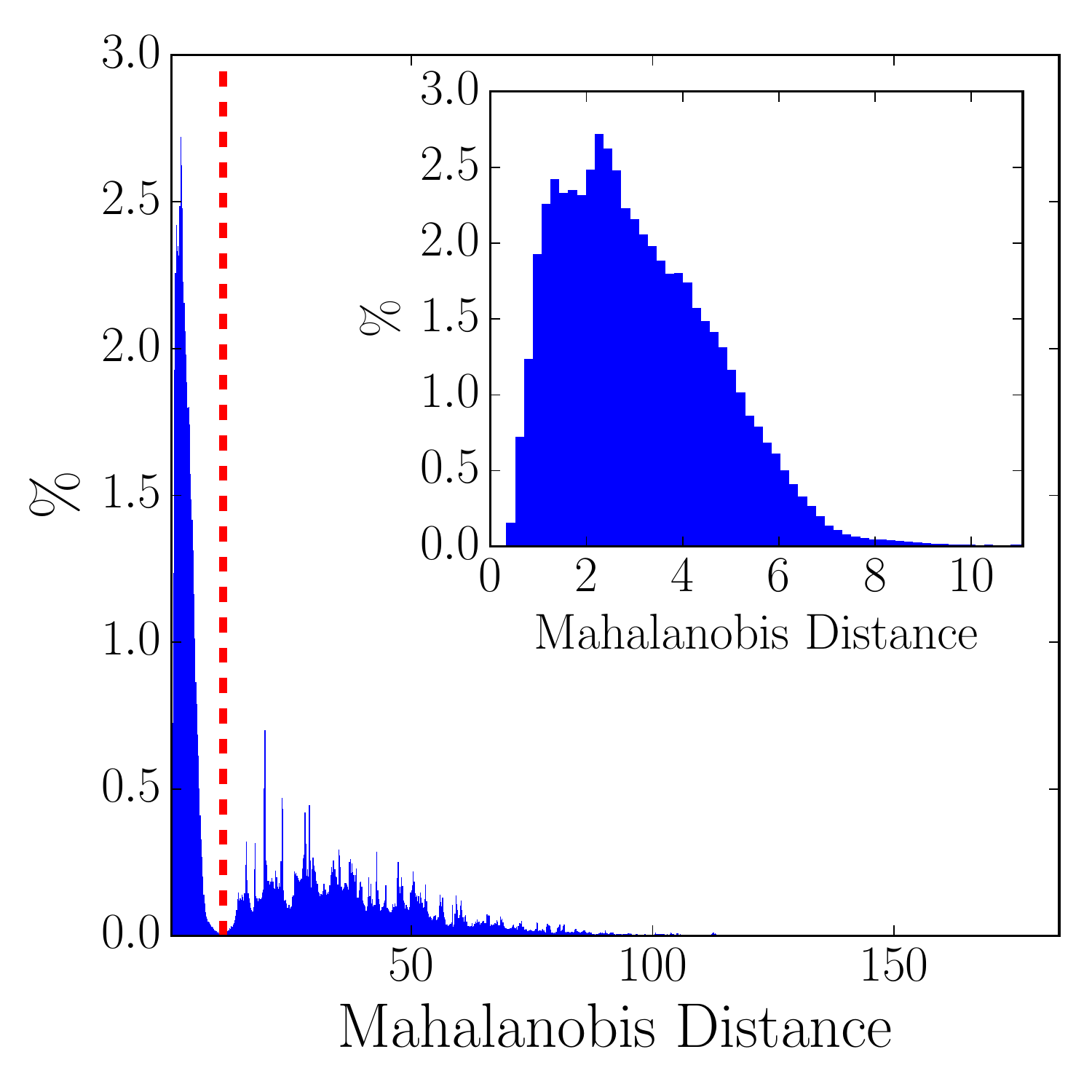}
\caption{Histograms of the Mahalanobis distances where the overall range was
  divided over 1,000 bins, for each of the two considered dense grids of models
  for KIC\,10526294 (left: exponential overshoot; right: step overshoot). The
  vertical dashed line indicates value 11.07 ($P(\chi^2_5\ge 11.07)=0.05$) and
  the inset shows the distance over that interval.}
\label{KIC-error}
\end{figure}
With our description based on the Mahalanobis distance in
Eq.\,(\ref{maldistgrid}), we improve the modeling compared to
\citep{Moravveji2015} by taking into account the correlations among the
parameters $\bftheta=(M,Z,X_c,D_{\rm ov},D_{\rm ext})$, as well as allow
non-linear dependence of the observed frequencies on those five parameters. 
The
parameter  estimation from the two grids based on $\chi^2$ and $\bftheta_0$ are
listed in Table\,\ref{KIC-result}. The period spacing patterns of the best
solutions are compared with the observed ones in Fig.\,\ref{KIC-figure},
while the two variance-covariance matrices for the two grids, $\hat{V} +
  \Lambda$ (hence  taking into
  account the measured frequency errors) are shown in
  Fig.\,\ref{KIC-Var-Covar}. The observational errors decrease from 
0.000057\,d$^{-1}$ for $f_1=0.472220$\,d$^{-1}$ to 0.000036\,d$^{-1}$ 
for $f_{19}=1.013415$\,d$^{-1}$. Hence, almost all the variance represented in 
Fig.\,\ref{KIC-Var-Covar} comes from the variability in the 
model grid predictions of the
frequencies.  

It is
found that the estimated parameter values for $\bftheta$ are not 
exactly the same, yet very
similar in value.  The $\chi^2$ selection by \citet{Moravveji2015} 
focused on the best agreement
for the lowest-order modes, while the Mahalanobis distance selects models that
give the overall best agreement.  From the middle two panels of the figure, we
deduce that the largest discrepancies between observed and theoretically
predicted modes for the $\chi^2$ prediction occur for the highest-order
modes and reach to about 900\,s, while the Mahalanobis prediction has the
poorest agreement for the lowest-order modes at a level of about 600\,s. All
differences between the frequencies of the four considered models and
KIC\,10526294 translate into $|f_i^\ast-f_i^{\rm th}|/f_i^\ast < 1\%$.
Moreover, the discrepancy between the model and observed frequencies in the
bottom panel of Fig.\,\ref{KIC-figure} are smaller than most of the
theoretical uncertainties found in Sections\,\ref{pulsation-error} and
\ref{models-error} (see upper left insets of Figs\,\ref{TA-nonTA} -- \ref{diffusion}).
Hence, with the current knowledge of stellar interiors, this
is an excellent overall agreement (we have stretched the ordinate axes in
Fig.\,\ref{KIC-figure} for optimal visibility of the small discrepancies).  

Given that these two dense grids resulted from a zoom-in based on an initial
screening of a much sparser one, optimally tuned to comply with the data, we
expect the error estimation according to Eq.\,(\ref{predictionellipsoid}) to
encapsulate these entire dense grids. Their ranges are given in Table\,2 of
\citet[][Mixing Grid \& Step Overshoot Grid]{Moravveji2015} and the distribution
of the Mahalanobis distances is shown in Fig.\,\ref{KIC-error}. The dashed
vertical line is the $\chi^2_5=11.07$ cutoff value. About half of the models in
the two grids have a lower distance and as expected, the two prediction
ellipsoids for $\bftheta$ span the entire grids in both cases. As a final note,
we point out that the best models selected by the Mahalanobis distance do not
change the inversion results for the rotation profile of the star determined in
\citet{Triana2015}. Indeed, it was already shown in that paper that this profile
was hardly dependent on the best forward model selection as long as it provided
a reasonable match to the zonal mode frequencies.

\subsection{Problem 2: A Mahalanobis distance based solution to 
compare two stellar models}

A pertinent question in seismic modeling of stars with a convective core is
whether the core overshooting is better described either by a step function or
by an exponentially decaying one. \citet{Pedersen2018} assessed the capacity of
the nominal {\it Kepler\/} data to achieve such a comparison based on gravity
modes. Another case of comparing the capacity of two stellar models was already
mentioned above, namely with and without chemical transport in the radiative
envelope \citep{Moravveji2015,Pedersen2018}.  Such assessments require
comparison between two theoretical models, $\calm(\bftheta^{(m)},\bfpsi^{(m)})$
and $\caln(\bftheta^{(n)},\bpsi^{(n)})$.  We assume, of course, that both models
produce observation vectors $\Y$ and $\Z$, respectively, of the same length $n$
and having the same meaning for all of their components. That said, the length
of the vector and the meaning of the components of $\bftheta^{(m)}$ and
$\bftheta^{(n)}$ may be different.

A Mahalanobis-based solution, similar to that of Problem 1, is 
considered. It should be clear that there is some asymmetry to the problem, as
each of the two grids has its own metric. Denote the grids by:
\begin{itemize}
\item For $\calm(\bftheta^{(m)},\bfpsi^{(m)})$, we have resulting
grid points $j=1,\dots,q$, with 
$\BY_j(\bftheta_j^{(m)},\bfpsi^{(m)})$;
\item For $\caln(\bftheta^{(n)},\bfpsi^{(n)})$, we have resulting 
grid points $s=1,\dots,r$, with 
$\BZ_s(\bftheta_s^{(n)},\bfpsi^{(n)})$.
\end{itemize}
Let the variance-covariance matrix $V_m$, like in Eq.\,(\ref{vmat}) correspond to the
first grid, with the equivalent quantity $V_n$ corresponding to the second
grid. 
Suppose that a benchmark model computed from 
the first theory $\calm(\bftheta_0^{(m)},\bfpsi^{(m)})$
with parameters $\bftheta_0^{(m)}$ results in frequencies
$\BY_0$, then we
need to find 
\begin{equation}
\label{maldist1}
\bftheta^{(n)}_0=\arg\min_{s=1}^r\left\{ (\BZ_s-\BY_0)^\top V_n^{-1}(\BZ_s-\BY_0)\right\}.
\end{equation}
In the other direction, 
we pick 
 a benchmark model computed from 
the second theory $\caln(\bftheta_0^{(n)},\bfpsi^{(n)})$
with parameters $\bftheta_0^{(n)}$ 
resulting in frequencies
$\BZ_0$,
and we
need to 
we solve:
\begin{equation}
\label{maldist2}
\bftheta^{(m)}_0=\arg\min_{j=1}^q\left\{ (\BY_j-\BZ_0)^\top V_m^{-1}(\BY_j-\BZ_0)\right\}.
\end{equation}
\begin{figure}
\centering
\includegraphics[width=0.45\linewidth]{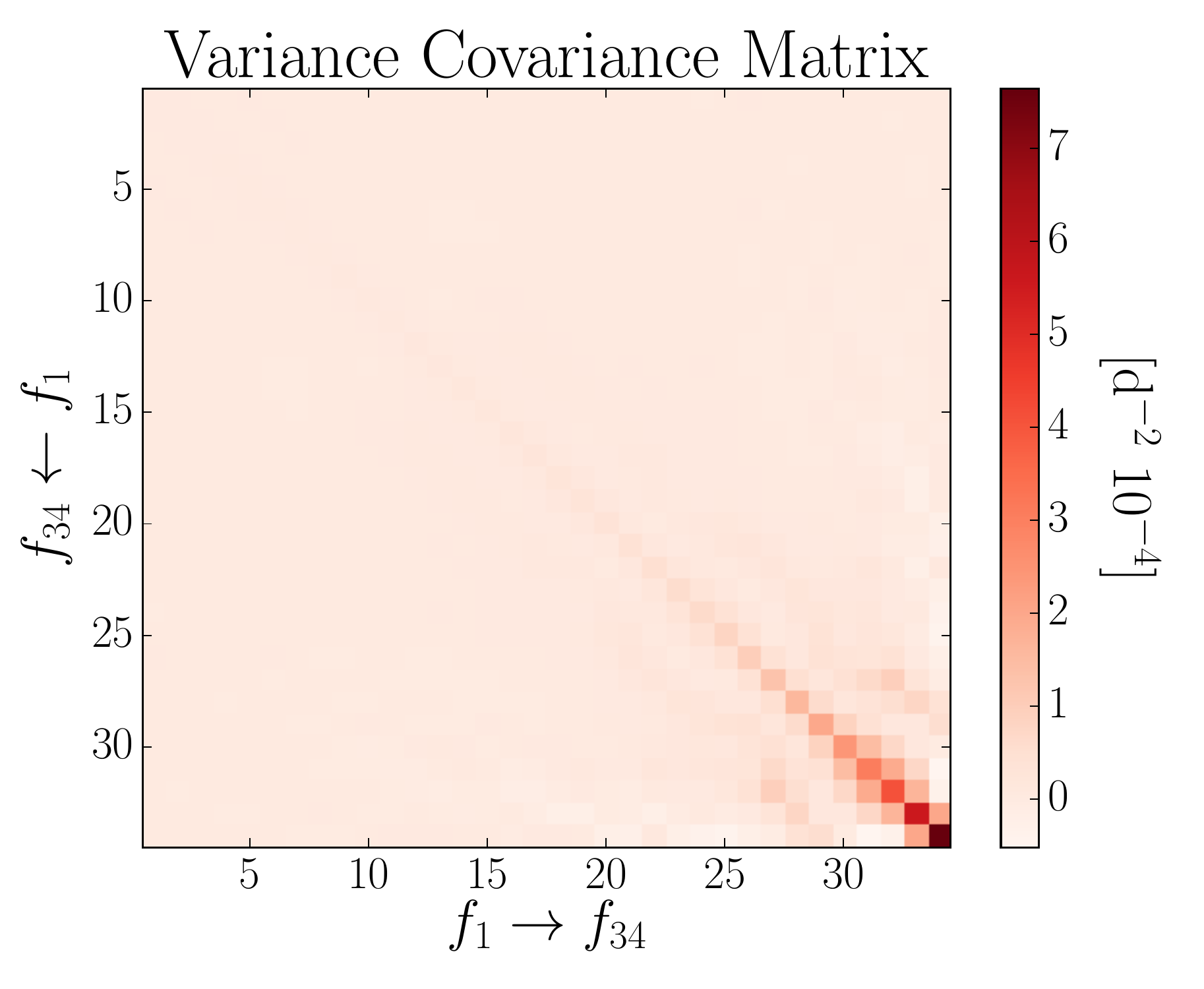}
\includegraphics[width=0.45\linewidth]{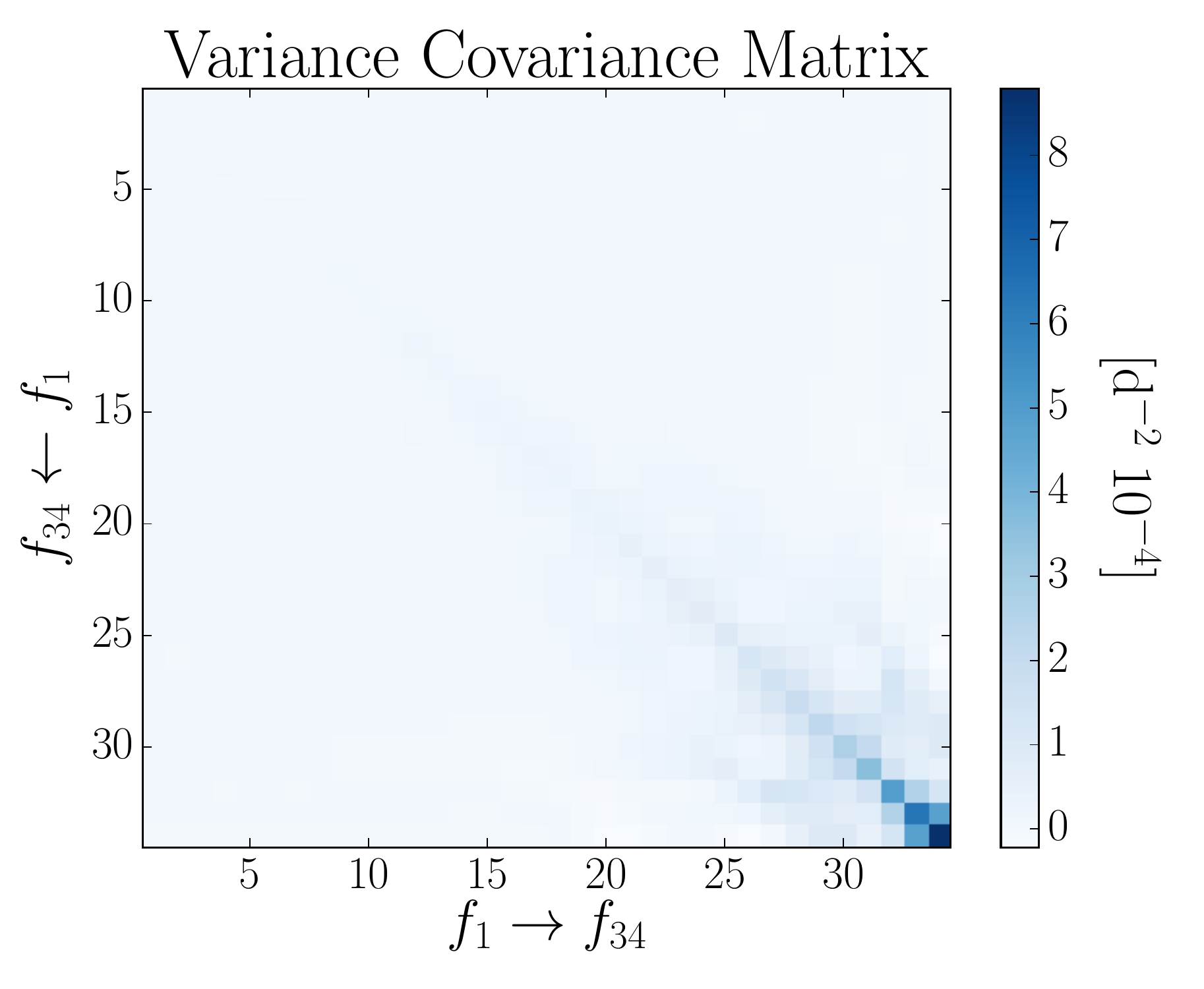}
\caption{Variance-covariance matrix $\hat{V}$ of dimension 34 $\times$
  34 representing 34 prograde dipole mode frequencies of benchmark model
$\bftheta_0^{(m)}=(M,X,X_c,\alpha_{\rm ov})$=$(3.25, 0.71, 0.5, 0.20)$
for a dense grid of models with exponential overshooting
(left) and 
of benchmark model
$\bftheta_0^{(n)}=(M,X,X_c,f_{\rm ov})$=$(3.25, 0.71, 0.5, 0.015)$
for a dense grid of models with step overshooting
(right).}
\label{Bench-Var-Covar}
\end{figure}
\begin{figure}
\centering
\includegraphics[width=0.30\linewidth]{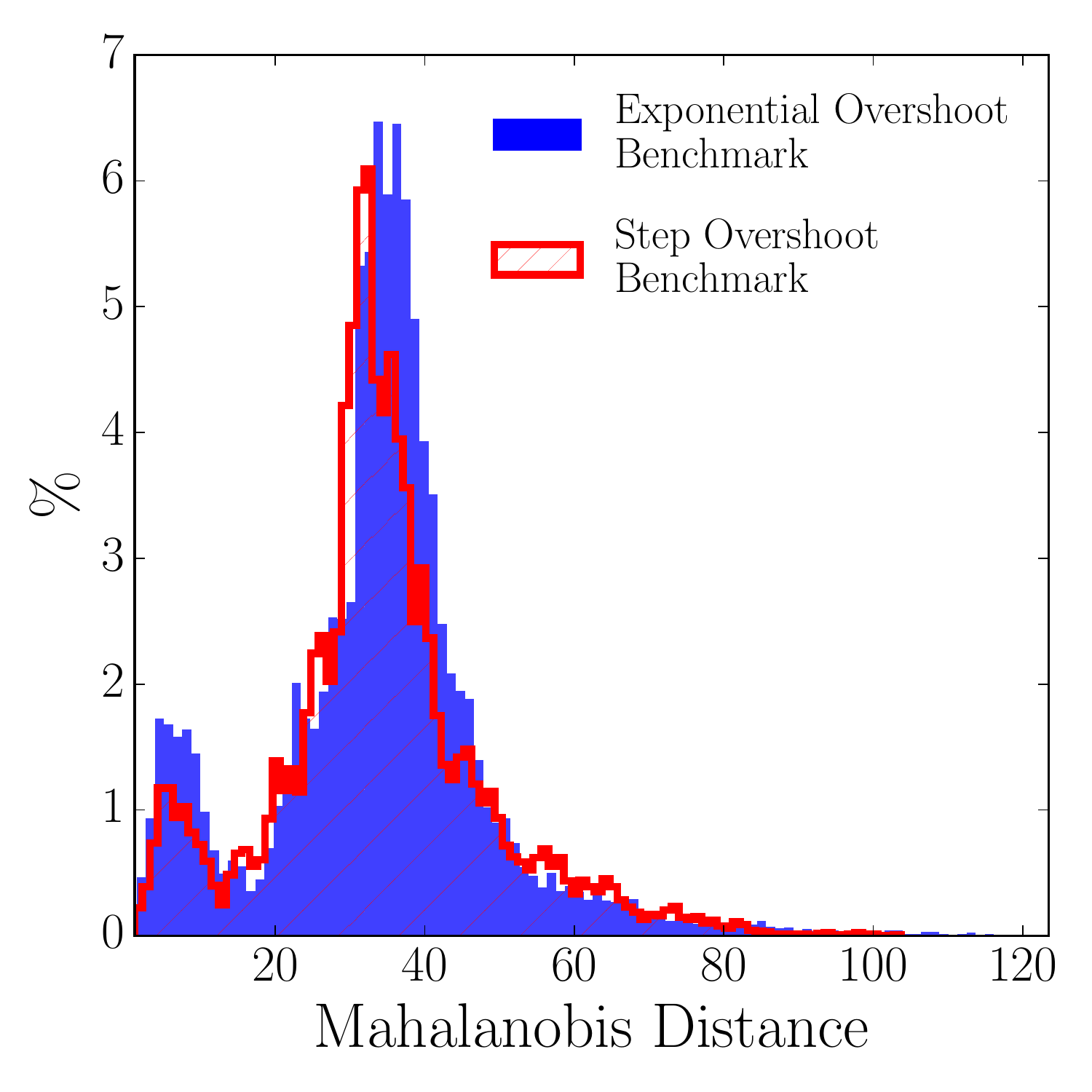}
\includegraphics[width=0.30\linewidth]{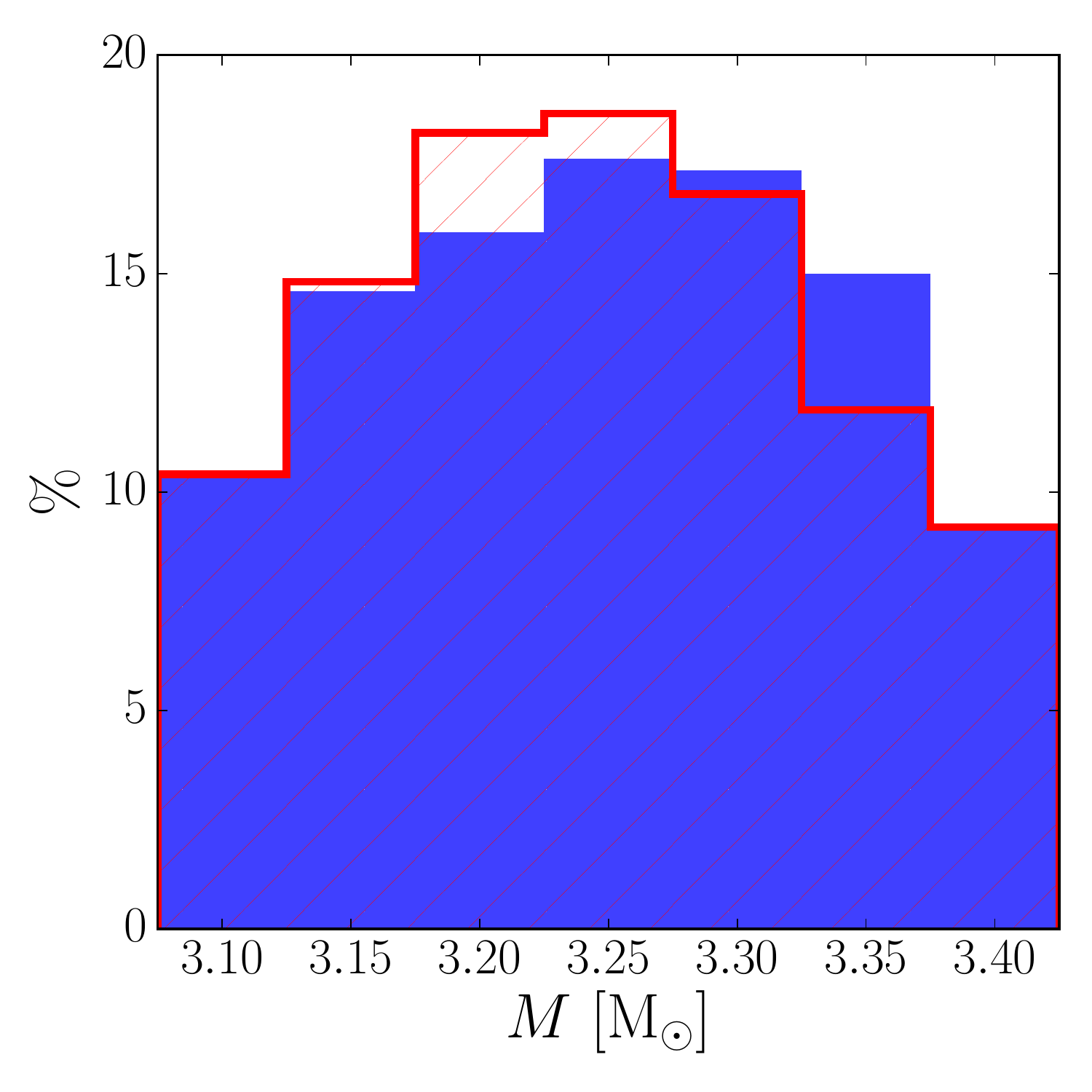}
\includegraphics[width=0.30\linewidth]{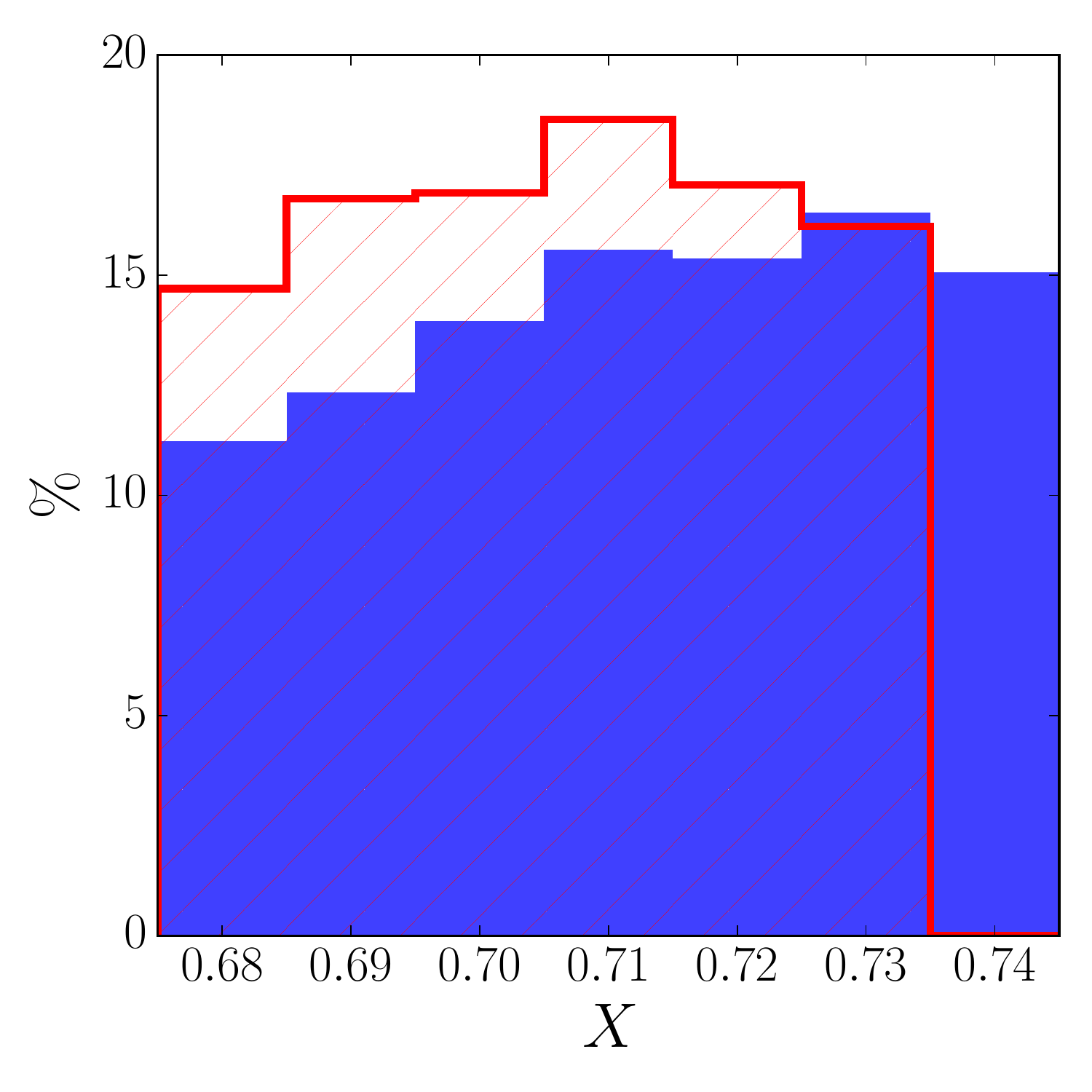}\\
\includegraphics[width=0.33\linewidth]{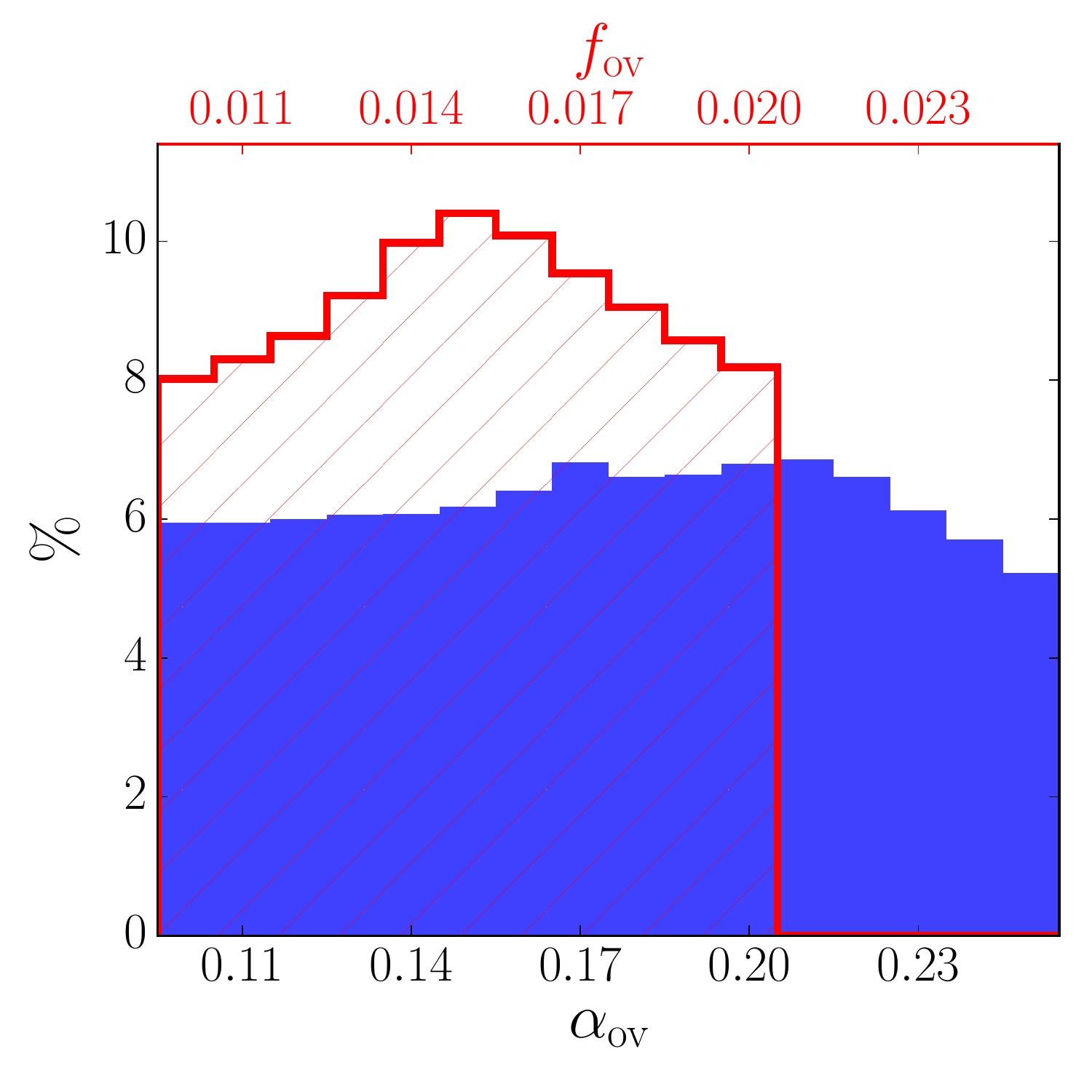}
\includegraphics[width=0.30\linewidth]{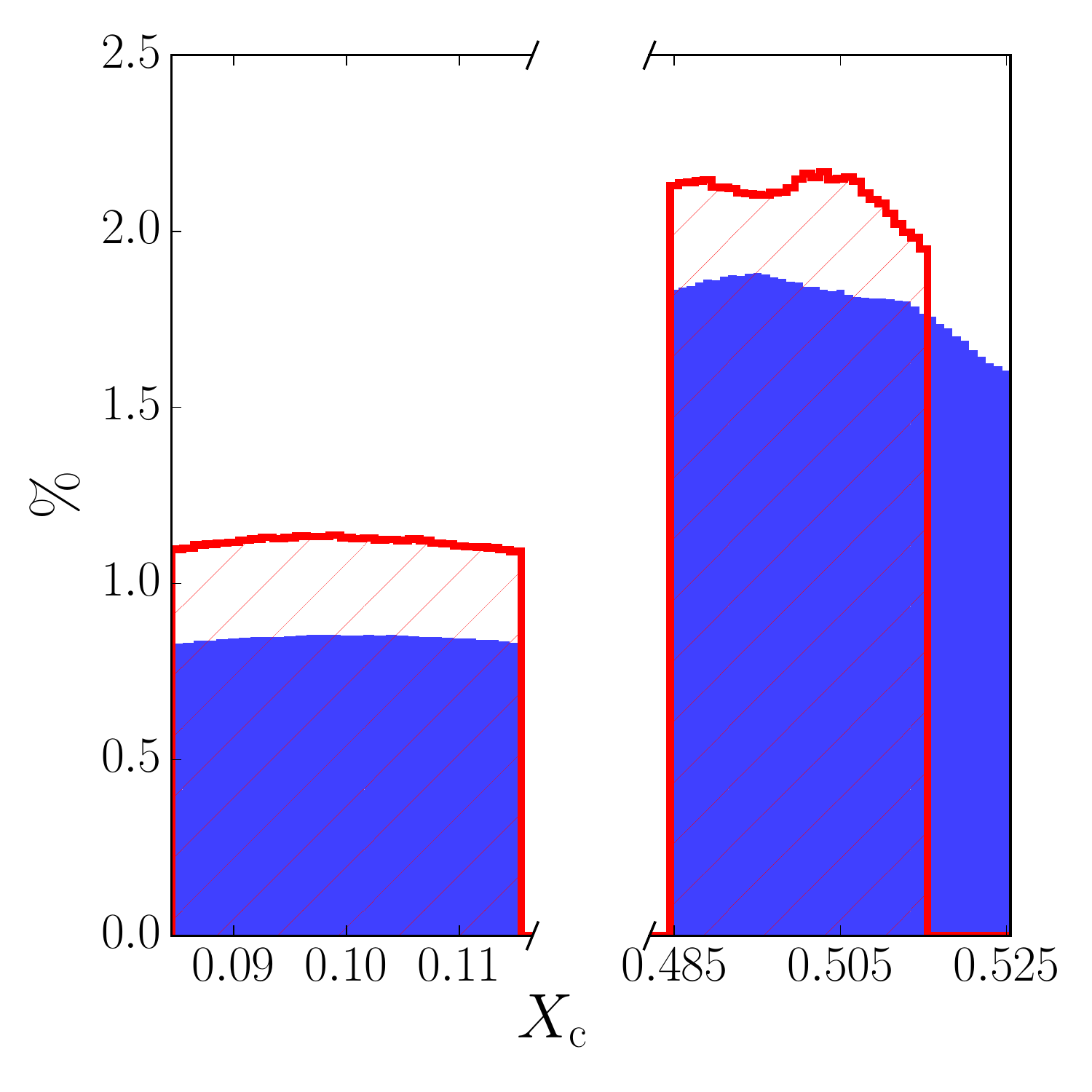}
\caption{Distributions of the Mahalanobis distance and of the four parameter
  predictions for $\bftheta$
  resulting from the comparison of 34 zonal dipole mode frequencies of a
  benchmark model with $(M,X,X_c,\alpha_{\rm ov})$=$(3.25, 0.71, 0.5, 0.20)$ as
  found from a grid of models with exponential overshoot (red) and of a
  benchmark model with $(M,X,X_c,f_{\rm ov})$=$(3.25, 0.71, 0.5, 0.015)$ 
compared to a grid of step overshoot models (blue).}
\label{BenchStep-Exp}
\end{figure}
As an illustration of this Problem\,2, we rely on the 4D grids of stellar models
computed by \citet[][Table\,2]{Pedersen2018}
and their high-order dipole modes with periods ranging from 0.8 to 3\,d.
For computational reasons, these
grids were centered around two evolutionary stages ($X_c$ near 0.5 and 0.1).  We
consider the $n=34$ frequencies of the dipole prograde gravity modes as
$\BY$ to be those of the benchmark model from the step overshoot grid with
$\bftheta_0^{(m)}=(M,X,X_c,\alpha_{\rm ov})$=$(3.25, 0.71, 0.5, 0.20)$ and we
search for the best model in the grid computed for exponential overshooting with
$f_{\rm ov}$ instead of $\alpha_{\rm ov}$ to ascertain which of the
exponential overshoot models explains the one with the step overshooting of the
benchmark model best. We refer to \citet{Pedersen2018} for the details on the
fixed input physics $\bfpsi$, which is the same for both grids except for the
mixing efficiency in the overshoot region.

We show the variance-covariance matrix $\hat{V}$ in the left panel of
Fig.\,\ref{Bench-Var-Covar}, revealing once again that the 
dominant variability among the models in the grid occurs for modes in
the high-frequency domain.
The Mahalanobis distance leads to the parameters estimation given by
$\bftheta_0^{(n)}$=$(M,X,X_c,f_{\rm ov})$=$(3.25,0.71,0.505,0.015)$ from
Eq.\,(\ref{maldist1}). The distribution of the best solutions for the parameters
$\bftheta$ are shown, along with the Mahalanobis distance distribution, in
Fig.\,\ref{BenchStep-Exp} (red). The parameter distributions were constructed
by giving every grid point $s=1, \ldots,r$ 
a weight according to the minimum Mahalanobis
distance of the entire grid divided by the Mahalanobis distance at this grid
point $s$.

Conversely, we select the dipole prograde gravity-mode frequencies $\BZ$ from a
benchmark model with exponential overshoot
$\bftheta_0^{(n)}$=$(M,X,X_c,f_{\rm ov})$=$(3.25, 0.71, 0.5, 0.015)$ and we
search for the best model in the grid computed for step overshooting with
parameter $\alpha_{\rm ov}$ to find
$\bftheta_0^{(m)}$=$(M,X,X_c,\alpha_{\rm ov})$=$(3.25,0.71,0.495,0.20)$ from
Eq.\,(\ref{maldist2}) as represented in Fig.\,\ref{BenchStep-Exp} (blue),
  with the variance-covariance matrix $\hat{V}$ given in the right panel of
Fig.\,\ref{Bench-Var-Covar}.
It is
visually clear that this problem set is asymmetric, as demonstrated by the
asymmetry of the distributions of the parameters $\bftheta^{(n)}$ (red) and
$\bftheta^{(m)}$ (blue). This reflects the fact that the parameters
$\alpha_{\rm ov}$ and $f_{\rm ov}$ correlate differently with $(M,X,X_c)$. Such
a different correlation structure was also hinted at from modeling of close binary
companions with a convective core \citep{ClaretTorres2017} but was not yet taken
up coherently in treating determinations of core masses in such objects from
isochrone fitting.
The bottom right panel of Fig.\,\ref{BenchStep-Exp} shows the power of the
Mahalanobis distance to perform stellar aging from gravity modes as there is no
confusion between young and old models. Finally, the mass and initial hydrogen
distributions  in Fig.\,\ref{BenchStep-Exp} show the opportunity to determine
stellar masses 
with high precision from gravity modes, typically better than 5\%,
while the chemical composition is harder
to derive as it suffers more from the degeneracies among the parameters. 

\subsection{Problem 3: Fitting a statistical model to frequencies 
generated from a physical model}
Given that evaluating theoretical models and their frequencies is
computationally expensive, one might aim to replace the $\BY$ derived from a
theoretical model $\calm(\bftheta,\bfpsi)$ in the $q$ grid points 
with a statistical model,
sufficiently close in outcome but easier to evaluate. Such a situation typically
occurs when grids of stellar evolution and pulsation computations turn out to be
insufficiently fine for forward seismic modeling and one wants to avoid having to
re-compute models with a too extensive number of extra grid points
\citep[cf.,][]{Moravveji2016}.  Therefore, we write such a statistical model as:
\begin{equation}
\BY_j\sim N\left(\bfmu_j=\bfmu(\bfbeta;\bftheta_j,\bfpsi),\Sigma\right).
\label{modelone}
\end{equation}
In other words, we treat the vectors with 
$n$ frequencies resulting from a theoretical model 
$\calm(\bftheta,\bfpsi)$ as
outcome variables $\BY_j$ and we want to describe these outcome variables by a
function $\bfmu_j=\bfmu(\bfbeta,\bftheta_j,\bfpsi)$. 
The latter can be linear or non-linear.
The statistical model will typically contain unknown 
parameters to be estimated, grouped in a vector $\bfbeta$.  Because a perfect
match between the frequencies resulting from the theoretical stellar model 
$\calm(\bftheta,\bfpsi)$ 
and from the statistical approximate
model never occurs, we introduce an error term for each grid point, $\beps_j$,
with variance-covariance matrix $\Sigma$.
Hence, we rewrite the statistical model~(\ref{modelone}) as:
\begin{equation}
\label{modelonebis}
\BY_j=\bfmu(\bfbeta;\bftheta_j,\bfpsi)+\beps_j,
\end{equation}
where $\beps_j$ is the 
error of the estimated frequency vector in the grid point
$j$ resulting from the statistical model.
At the level of the individual frequencies labeled with $i=1,\ldots,n$, 
Eq.\,(\ref{modelonebis})
can be written as:
\begin{equation}
\label{modelonebisj}
Y_{ji}=\mu_i(\bfbeta_i;\bftheta_j,\bfpsi)+\varepsilon_{ji},
\end{equation}
with $\mbox{var}(\varepsilon_{ji})=\sigma_{ii}$, 
the $i$th diagonal element of $\Sigma$. 
Also, $\mbox{cov}(\varepsilon_{ji},\varepsilon_{jk})=\sigma_{ik}$, the
covariance, and $(i,k)$ the off-diagonal element of $\Sigma$.\footnote{
  Note that we use the notation $\sigma_{ik}$ for the generic element of a
  variance-covariance matrix $\Sigma$, as is commonly done in the statistical
  literature.   The elements correspond to variances if
  $i=k$ and to covariances if $i\ne k$ \citep{JohnsonWichern2000}.
The
  corresponding standard deviations are then $\sqrt{\sigma_{ii}}$.}
So, in general, each regression component $i$ is allowed its own variance
$\sigma_{ii}$ for $\varepsilon_{ji}$, implying heteroscedasticity, as required
following Sections\,\ref{pulsation-error} and 
\ref{models-error}.

With this setup, we assume that the normal distribution offers a sensible
statistical working model to represent the theoretical models
$\calm(\bftheta,\bfpsi)$.  These theoretical models lead to uncertainties for
the theoretically predicted frequencies as illustrated in Figs\,\ref{TA-nonTA}
-- \ref{diffusion}. Several of these have symmetric distributions, while others
have skewed distributions due to systematic uncertainties.  Nevertheless, there
are several arguments supporting the choice of a normal statistical theory for
$\BY_j$ in Eq.\,(\ref{modelonebis}).  First, apart from computational
convenience, this rests upon the fact that, for inferences to be valid, first-
and second-order moment assumptions suffice, even if normality is not satisfied
\citep{JW02}.  This also explains why normal regression and linear regression
lead to the same results. Second, \cite{White1982} provides a theory for
misspecified statistical models, indicating that fitting misspecified models
with a normal distribution still leads to the best possible description given a
class of statistical models considered. Third, transformations can be applied,
if deemed appropriate, to improve the quality of the normal
approximation. Fourth, the fact that $\bfmu(\bfbeta;\bftheta_j,\bfpsi)$ is
allowed to be highly non-linear brings appropriate flexibility to the model
formulation. For all these reasons, the normal distribution is a very defensible
model hypothesis to make \citep[see also][for similar arguments in the case of
solar-like oscillations]{Gruberbauer2013,Appourchaux2014}.

\subsubsection{Maximum likelihood estimation}
One convenient way to estimate parameters is via maximum likelihood
\citep[MLE,][]{Welsh96}. When each component of $Y_{ji}$ has its own regression
parameters, the overall parameter vector partitions as
$\bfbeta=(\bfbeta_1^\top,\dots,\bfbeta_n^\top)^\top$, and we can separately
solve each of the $n$ problems. In this case, we can write
$Y_{ji}=\mu_{ji}(\bfbeta_i;\bftheta_j,\bfpsi)+\varepsilon_{ji}$, where the length
of the entire vector $\bfbeta$ then partitions as $u=u_1+\dots+u_n$, with $u_i$
the length of the sub-vector pertaining to the regression for the $i$th
frequency component. The normal likelihood for component $i$ is then written as:  
\begin{equation}
\label{likelihood}
L(\bfbeta_i,\sigma_{ii})=\prod_{j=1}^q\frac{1}{\sqrt{\sigma_{ii}}\sqrt{2\pi}}\exp\left\{
-\frac{1}{2}\frac{(Y_{ji}-\mu_{ji})^2}{\sigma_{ii}}
\right\},
\end{equation}
where $\mu_{ji}\equiv\mu_{ji}(\bfbeta_i;\bftheta_j,\bfpsi)$ is used for
shorthand. 
One proceeds by writing the corresponding log-likelihood:
$$
\ell(\bfbeta_i,\sigma_{ii})=-\frac{q}{2}\ln(\sigma_{ii})-\frac{q}{2}\ln(2\pi)-\frac{1}{2}\sum_{j=1}^q\frac{(y_{ji}-\mu_{ji})^2}{\sigma_{ii}},
$$
where $y_{ji}$ stands for the realized value of the random variable $Y_{ji}$. 
In practice, it is sufficient to work with the kernel of the log-likelihood:
$$
\widetilde{\ell}(\bfbeta_i,\sigma_{ii})=-\frac{1}{2}\ln(\sigma_{ii})-\frac{1}{2}\sum_{j=1}^q\frac{(y_{ji}-\mu_{ji})^2}{\sigma_{ii}}.
$$
Its first derivative w.r.t.\ $\bfbeta_i$, the so-called score function for $\bfbeta_i$, is:
\begin{equation}
\label{scoretheta}
\BS(\bfbeta_i)=\frac{\partial\ell}{\partial\bfbeta_i}=
\frac{1}{\sigma_{ii}}\sum_{j=1}^q(y_{ji}-\mu_{ji})\cdot\frac{\partial\mu_{ji}}
{\partial\bfbeta_i}.
\end{equation}
The score equation to solve hence is: 
$$
\BS(\bfbeta_i)=
\frac{\partial\ell}{\partial\bfbeta_i}=
\frac{1}{\sigma_{ii}}\sum_{j=1}^q(y_{ji}-\mu_{ji})\frac{\partial\mu_{ji}}{\partial\bfbeta_i}=
\mbox{\bfseries 0}.
$$ 
For non-linear models for $\mu_{ji}$, 
this first derivative will typically not allow for an explicit expression 
and may well be cumbersome \citep{SW03}. 
Note that $\BS(\bfbeta_i)=\mbox{\bfseries 0}$ if and only if
\begin{equation}
\label{scoretwee}
\sum_{j=1}^q(y_{ji}-\mu_{ji})\frac{\partial\mu_{ji}}{\partial\bfbeta_i}=\mbox{\bfseries 0}.
\end{equation}

Evidently, the log-likelihood function also depends on the 
argument $\sigma_{ii}$. The score equation for this parameter is
\begin{equation}
\label{scoredrie}
S(\sigma_{ii})=-\frac{1}{2}\frac{1}{\sigma_{ii}}+
\frac{1}{2}\frac{1}{\sigma_{ii}^2}\sum_{j=1}^q(y_{ji}-\mu_{ji})^2=0.
\end{equation}
Once (\ref{scoretwee}) has been solved, 
the solution to (\ref{scoredrie}) follows as:
\begin{equation}
\label{maxlikjj}
\widehat{\sigma_{ii}}=\frac{1}{q}\sum_{j=1}^q
\left[y_{ji}-\mu_{ji}(\widehat{\bfbeta_i})\right]^2,
\end{equation}
where $\widehat{\bfbeta_i}$ is the solution 
to (\ref{scoretwee}), the maximum likelihood estimator (MLE) for $\bfbeta_i$.

Because of small-sample bias in the MLE for $\sigma_{ii}$, 
one often uses the least squares version: 
\begin{equation}
\label{olsjj}
\widetilde{\sigma_{ii}}=\frac{1}{q-u_i}\sum_{j=1}^q
\left[y_{ji}-\mu_{ji}(\widehat{\bfbeta_i})\right]^2.
\end{equation}
Clearly, when the number of grid points $q$ is large relative to the number 
of parameters considered for optimization, $u_i$, as is the case in any
practical application of forward statistical modeling, the bias becomes negligible. 

To estimate the standard errors of $\bftheta$ and $\sigma^2$, we proceed by
calculating the second derivates of the log-likelihood function, the so-called
Hessian matrix, denoted by $H(\cdot)$. The argument is left unspecified because
we can calculate it for both $\bftheta$ and $\sigma^2$. The negative of the
Hessian is the information matrix, denoted $I(\cdot)$. The inverse of this
produces the variance-covariance matrix of the parameter estimators, denoted by
$V(\cdot)$. We often work with the expected values of both $H(\cdot)$ and
$I(\cdot)$, for ease of computation and because they are asymptotically
equivalent. We will denote these expected values by ${\cal H}(\cdot)$ and ${\cal
  I}(\cdot)$.
At this point, it is useful to note that, in fact, the entire
variance-covariance matrix 
$\Sigma$ can be estimated, by vectorizing (\ref{maxlikjj}) or (\ref{olsjj}):
\begin{equation}
\label{maxliksigma}
\widehat{\Sigma}=\frac{1}{q}\sum_{j=1}^q
\left[\BY_{j}-\bfmu_{j}(\widehat{\bfbeta})\right]
\left[\BY_{j}-\bfmu_{j}(\widehat{\bfbeta})\right]^\top
\end{equation}
and
\begin{equation}
\label{olssigma}
\widetilde{\Sigma}=\frac{1}{q-p}\sum_{j=1}^q
\left[\BY_{j}-\bfmu_{j}(\widehat{\bfbeta})\right]
\left[\BY_{j}-\bfmu_{j}(\widehat{\bfbeta})\right]^\top.
\end{equation}

The Hessian for $\bfbeta_i$ is:
$$
H(\bfbeta_i)=\frac{\partial^2\ell}{\partial\bfbeta_i\partial\bfbeta_i^2}=
\frac{\partial S(\bfbeta_i)}{\partial\bfbeta_i}
=-\frac{1}{\sigma_{ii}}\sum_{j=1}^q \left(\frac{\partial\mu_{ji}}{\partial\bfbeta_i}\right)
\left(\frac{\partial\mu_{ji}}{\partial\bfbeta_i}\right)^\top
+
\frac{1}{\sigma_{ii}}\sum_{j=1}^q
(y_{ji}-\mu_{ji})\frac{\partial^2\mu_{ji}}{\partial\bfbeta_i\partial\bfbeta_i^\top}.
$$
Hence,
$$
{\cal H}(\bfbeta_i)=-\frac{1}{\sigma_{ii}}
\sum_{j=1}^q \left(\frac{\partial\mu_{ji}}{\partial\bfbeta_i}\right)
\left(\frac{\partial\mu_{ji}}{\partial\bfbeta_i}\right)^\top
$$
because $E[y_{ji}-\mu_{ji}]=E[y_{ji}]-\mu_{ji}=\mu_{ji}-\mu_{ji}=0$.
Further, the variance-covariance matrix of $\bfbeta_i$ takes the form:
$$V(\bfbeta_i)=-{\cal H}(\bfbeta_i)^{-1}=\left[\frac{1}{\sigma^2}
\sum_{j=1}^q \left(\frac{\partial\mu_{ji}}{\partial\bfbeta_i}\right)
\left(\frac{\partial\mu_{ji}}{\partial\bfbeta_i}\right)^\top\right]^{-1}.
$$
Turning to $\sigma_{ii}$:
\begin{eqnarray*}
H(\sigma_{ii})&=&
\frac{q}{2}\frac{1}{\sigma_{ii}^2}-\frac{1}{\sigma_{ii}^3}\sum_{j=1}^q(y_{ji}-\mu_{ji})^2,\\
{\cal H}(\sigma_{ii})&=&\frac{q}{2}\frac{1}{\sigma_{ii}^2}-\frac{1}{\sigma_{ii}^3}q\sigma_{ii}=-\frac{q}{2}\frac{1}{\sigma_{ii}^2},\\
V(\widehat{\sigma_{ii}})&=&\frac{2\sigma_{ii}^2}{q}.
\end{eqnarray*}
From the variance-covariance matrices (note that for $\sigma_{ii}$ it is a
scalar), we derive the standard errors. For $\bfbeta_i$, we take the square root
of the diagonal of $V(\bfbeta_i)$. For $\sigma_{ii}$, we find
$\mbox{s.e.}(\sigma_{ii})=\sqrt{2/q}\sigma_{ii}$.
Strictly speaking, we should also consider the covariance between 
the estimated $\bfbeta_i$ and $\sigma_{ii}$. 
However, this is asymptotically zero. Indeed:
$$\frac{\partial S(\bfbeta_i)}{\partial\sigma_{ii}}=
\frac{\partial S(\sigma_{ii})}{\partial\bfbeta_i}=
H(\bfbeta_i,\sigma_{ii})=
-\frac{1}{\sigma_{ii}^2}\sum_{j=1}^q(y_{ji}-\mu_{ji})
\frac{\partial\mu_{ji}}{\partial\bfbeta_i}.$$
Hence,
$$
{\cal H}(\bfbeta_i,\sigma_{ii})=
-\frac{1}{\sigma_{ii}^2}\sum_{j=1}^q E(y_{ji}-\mu_{ji})
\frac{\partial\mu_{ji}}{\partial\bfbeta_i}=0.$$

\noindent\underline{Special Case: Linear Statistical Models\label{linmod1}}\\ 
In linear models, all of
the above estimators and precision estimators have closed forms. Write the
linear model as $\mu_{ji}=\bx_{ji}^\top\bfbeta_i$, where $\bx_{ji}$ is a vector
of predictors, consisting of the components of $\bftheta_j$ and functions
thereof, and create a convenient design matrix and outcome vector:
$$X_i=\left(
\begin{array}{c}
\bx_{1i}^\top\\
\bx_{2i}^\top\\
\vdots\\
\bx_{qi}^\top
\end{array}
\right),\qquad
Y_i=\left(
\begin{array}{c}
y_{1i}^\top\\
y_{2i}^\top\\
\vdots\\
y_{qi}^\top
\end{array}
\right).
$$
\begin{eqnarray*}
S(\bfbeta_i)=0&\Longleftrightarrow&\sum_{j=1}^q(y_{ji}-\bx_{ji}^\top \bfbeta_i)\bx_{ji}=0\\
&\Longleftrightarrow&\sum_{j=1}^q\bx_{ji}
                      y_{ji}=\left(\sum_{j=1}^q\bx_{ji}\bx_{ji}^\top
\right)\bfbeta_i\\
&\Longleftrightarrow&X_i^\top Y_i=\left(X_i^\top X_i\right)\bfbeta_i.
\end{eqnarray*}
This leads to  the classical ordinary least squares (OLS) solution for linear 
regression \citep{NWK90}:
\begin{equation}
\label{beta-OLS}
{\bfbeta_i}=\left(X_i^\top X_i\right)^{-1}X_i^\top Y_i.
\end{equation}
The expected Hessian becomes:
$${\cal H}(\bfbeta_i)=
-\frac{1}{\sigma_{ii}}\sum_{j=1}^q\bx_{ji}\bx_{ji}^\top=-\frac{1}{\sigma_{ii}}X_i^\top X_i$$
and hence
\begin{equation}
\label{variance-OLS}
V({\bfbeta_i})={\sigma_{ii}}\left(X_i^\top X_i\right)^{-1}.
\end{equation}
For $\sigma_{ii}$, all remains the same. 
Note that the space spanned by $\bfmu(\widehat{\bfbeta};\bftheta_j,\bfpsi)$
is at most of dimension length $(\bftheta_j)=t$, even when $t<n$.

\subsubsection{Including a measurement error component}
In the above, Model~(\ref{modelonebis}) is heteroscedastic in the sense that it
allows for a different variance $\sigma_{ii}$ for each of the components
$Y_{ji}$ of the vector $\BY_j$. In concrete applications, one would like to
assess how the difference between the $\BY$ vector derived from the 
statistical model and derived from the ``true'' physical model
$\calm(\bftheta,\bfpsi)$ compares to the measured frequency precision that
applies to $\BY^\ast$. Indeed, in practise, the various grids will have been
computed for a particular star, whose gravity-mode oscillations and their errors
have been measured for its $i=1,\ldots,n$ frequencies.

We recall that the measured frequency precisions
are denoted as $\Lambda_j$ and that its diagonal matrix elements are
$\lambda_{j,ii}$. Taking measurement errors into account is done by 
replacing the variance-covariance matric $\Sigma$ by
$\Sigma+\Lambda_j$.
The regression equation then becomes:
\begin{equation}
\label{modeltwo}
Y_{ji}=\mu_{ji}(\bfbeta_i)+\varepsilon_{ji}+\varepsilon^\ast_{ji},
\end{equation}
with $\varepsilon_{ji}\sim N(0,\sigma_{ii})$, $\varepsilon^\ast_j\sim
N(0,\lambda_{ji})$, 
and $\lambda_{ji}$ known. 
Note that, for generality, we allow the measurement error to depend on the
component within the vector $\BY$ as well as the actual 
grid point $j$.
Alternatively, $j$ could refer to a star when the
dataset is made up of a collection of stars (cf.\ Problem\,4).
The likelihood then becomes: 
\begin{equation}
\label{likelihoodtwee}
L(\bfbeta_i,\sigma_{ii})=
\prod_{j=1}^q\frac{1}{\sqrt{\sigma_{ii}+\lambda_{ji}}\sqrt{2\pi}}\exp\left\{
-\frac{1}{2}\frac{(y_{ji}-\mu_{ji})^2}{\sigma_{ii}+\lambda_{ji}}
\right\}.
\end{equation}
The kernel of the log-likelihood becomes:
$$
\widetilde{\ell}(\bfbeta_i,\sigma_{ii})=-\frac{1}{2}\sum_{j=1}^q
\ln(\sigma_{ii}+\lambda_{ji})-
\frac{1}{2}\sum_{j=1}^q\frac{(y_{ji}-\mu_{ji})^2}{\sigma_{ii}+\lambda_{ji}}.
$$
The corresponding score equations:
\begin{eqnarray}
\label{scoretweea}
\BS(\bfbeta_i)&=& 
\sum_{j=1}^q\frac{(y_{ji}-\mu_{ji})}{\sigma_{ii}+\lambda_{ji}}\cdot
\frac{\partial\mu_{ji}}{\partial\bfbeta_i},\\
S(\sigma_{ii})&=&-\frac{1}{2}\sum_{j=1}^q\frac{1}{\sigma_{ii}
+\lambda_{ji}}+\frac{1}{2}\sum_{j=1}^q\frac{(y_{ji}-\mu_{ji})^2}{(\sigma_{ii}+\lambda_{ji})^2}.
\label{scoretweeb}
\end{eqnarray}
Expressions for Hessians, expected Hessians, and variance-covariance matrices are:
\begin{eqnarray*}
H(\bfbeta_i)&=-&\sum_{j=1}^q\frac{1}{\st}
\left(\frac{\partial\mu_{ji}}{\partial\bfbeta_i}\right)\left(\frac{\partial\mu_{ji}}{\partial\bfbeta_i}\right)^\top+\sum_{j=1}^q\frac{(y_{ji}-\mu_{ji})}{\st}
\frac{\partial^2\mu_{ji}}{\partial\bfbeta_i\partial\bfbeta_i^\top},\\
{\cal H}(\bfbeta_i)&=-&\sum_{j=1}^q\frac{1}{\st}
\left(\frac{\partial\mu_{ji}}{\partial\bfbeta_i}\right)\left(\frac{\partial\mu_{ji}}{\partial\bfbeta_i}\right)^\top,\\
V(\bfbeta_i)&=&\left[\sum_{j=1}^q\frac{1}{\st}
\left(\frac{\partial\mu_{ji}}{\partial\bfbeta_i}\right)\left(\frac{\partial\mu_{ji}}{\partial\bfbeta_i}\right)^\top\right]^{-1},\\
H(\sigma_{ii})&=&
\frac{1}{2}\sum_{j=1}^q\frac{1}{(\st)^2}-\sum_{j=1}^q\frac{(y_{ji}-\mu_{ji})^2}{(\st)^3},\\
{\cal H}(\sigma_{ii})&=&\frac{1}{2}\sum_{j=1}^q\frac{1}{(\st)^2}-\sum_{j=1}^q\frac{1}{(\st)^2}
=-\frac{1}{2}\sum_{j=1}^q\frac{1}{(\st)^2},\\
V(\sigma_{ii})&=&
\left[\frac{1}{2}\sum_{j=1}^q\frac{1}{(\st)^2}\right]^{-1}.
\end{eqnarray*}
Also here, the asymptotic 
covariance between $\bfbeta_i$ and $\sigma_{ii}$ is equal to zero.

In this case, the score equations corresponding to
Eqs\,(\ref{scoretweea})--(\ref{scoretweeb}) cannot be solved separately. One can
therefore iterate between the two, i.e., first solve
$S(\bfbeta_i)=\mbox{\bfseries 0}$ with $\sigma_{ii}$ fixed and then
$\BS(\sigma_{ii})=0$ with $\bfbeta_i$ fixed. These two steps are then iterated
until convergence. To achieve this, one can use the iterated profile likelihood
method, in which case $\sigma_{ii}$ can be considered just another component of
the parameter vector: $(\beta_{i1},\dots,\beta_{iu_i},\sigma_{ii})^\top$.\\[0.2cm]

\noindent\underline{Special Case: Linear Model}\\
Also here, simplification is possible for linear models.  
We assume the same linear mean structure
as in Section~\ref{linmod1}, with the same design matrices $X_i$ and $Y_i$, and
we additionally create a diagonal weight matrix $W_i$, with elements
$w_{ji}=(\st)^{-1}$ along the diagonal.
The score equation for $\bfbeta_i$ takes the form:
\begin{eqnarray*}
S(\bfbeta_i)=0&
\Longleftrightarrow&\sum_{j=1}^q w_{ji}(y_{ji}-\bx_{ji}^\top\bfbeta_i)\bx_{ji}=0\\
&\Longleftrightarrow&\sum_{j=1}^q w_{ji}\bx_{ji} y_{ji}=
\left(\sum_{j=1}^q \bx_{ji} w_{ji}\bx_{ji}^\top\right)\bfbeta_i\\
&\Longleftrightarrow& X_i^\top W_i Y_i= (X_i^\top W_i X_i)\bfbeta_i\\
&\Longleftrightarrow&\bfbeta_i=\left(X_i^\top W_i X_i\right)^{-1}X_i^\top W_i Y_i.
\end{eqnarray*}
Here, $\bfbeta_i$ depends on $\sigma_{ii}$ through the weights $w_{ji}$. 
This means that the above expression produces the maximum likelihood 
estimator only after the MLE for $\sigma_{ii}$ is found. 
Both optimizations can be coupled in the following iterative scheme:
\begin{itemize}
\item Update the variance: $\sigma_{ii}^{(new)}=\sigma_{ii}^{(old)}-{\cal
    H}^{-1}
\left(\sigma_{ii}^{(old)}\right)S\left(\sigma_{ii}^{(old)}\right)$.
\item With this new $\sigma_{ii}^{(new)}$, the weights can be recomputed.
\item These weights are then used to update $\bfbeta_i$  
using $\bfbeta_i^{(new)}=\left(X_i^\top W_i X_i\right)^{-1}X_i^\top W_i Y_i$.
\end{itemize}
The above triplet of steps is repeated until convergence, 
at which time the variance can be estimated from:
\begin{eqnarray*}
V(\bfbeta_i)&=&\left(X_i^\top W_i X_i\right)^{-1},\\
V(\sigma_{ii})&=&2\left(\sum_{j=1}^q w_{ji}^2\right)^{-1}.
\end{eqnarray*}

In the case where linear statistical models do not offer a proper fit to mimic
the theoretical models $\calm(\bftheta,\bfpsi)$, Appendix\,\ref{FracPol} 
offers a practical
guide to try non-linear models by means of fractional polynomials.

\subsubsection{Statistical model selection}
For the overall set of $n$ frequencies per star, we may want to choose between various
candidate statistical models, where each $\bfbeta_i$ is of dimension $u_i$ for
$i=1,\ldots,n$.  Let
us assume we have two nested statistical models and we want to compare them: a
``null'' model, $H_0$, where each of the estimated parameters $\bfbeta_{0,i}$
has dimension $u_{0,i}$ and a more complex alternative model described by
$\bfbeta_{i}$, each with dimension $u_{i}>u_{0,i}$.  Hence the parameters
$\bfbeta_{0,i}$ constitue a sub-space of that of the more complex model
described by $\bfbeta_{i}$.  In the case of such nested models, the comparison
of their appropriateness can be done using the classical likelihood-based test
statistics: likelihood ratio, score, and Wald test statistics with expressions
given by, respectively:
\begin{eqnarray}
T_{LR} & = & 2  \left[\sum_{i=1}^n \ell(\widehat{\bfbeta_i}) \ 
- \  \sum_{i=1}^n\ell(\bfbeta_{0,i})\right],  \label{tlr}\\[2mm]  
T_S & = & \sum_{i=1}^n \left[\left.\frac{\partial \ell(\bfbeta_i)}{\partial
          \bfbeta_i}\right|_{\bfbeta_i=
\bfbeta_{0,i}}\right]^\top \left[\left. - 
\frac{\partial^2 \ell(\bfbeta_i)}{\partial \bfbeta_i \partial
          \bfbeta_i}\right|_{\bfbeta_i
=\bfbeta_{0,i}}\right]^{-1}
\left[\left.\frac{\partial \ell(\bfbeta_i)}{\partial
          \bfbeta_i}\right|_{\bfbeta_i=
\bfbeta_{0,i}}\right] \label{ts}\\[2mm]
T_W & = & \sum_{i=1}^n \left[\widehat{\bfbeta_i}
-\bfbeta_{0,i} \right]^\top \left[\left. - \frac{\partial^2
          \ell(\bfbeta_i)}{\partial \bfbeta_i \partial
          \bfbeta_i}\right|_{\bfbeta_i=\widehat{\bfbeta_i}}\right]\left[\widehat{\bfbeta_i}
-\bfbeta_{0,i} \right].\label{tw}
\end{eqnarray}
Classical likelihood theory implies that, under $H_0$, $T_{LR}$, $T_S$, and
$T_W$ are asymptotically equivalent and $\chi^2_z$ distributed, with
$z=\sum_{i=1}^n (u_{i}-u_{0,i})$ the overall difference in the total number of
degrees of freedom of the model $H_0$ and the alternative more complex model
\citep{CH90}.  To reduce complexity, the second derivate matrix occurring in
(\ref{ts}) and (\ref{tw}) can be replaced by $-{\cal H}(\bfbeta_i)$.  

A model selection procedure such as that based on the likelihood ratio test
statistic suffers from two limitations. First, it can be used only with nested
models (i.e., when one model is a sub-model of the other). Second, there is the
known tendency to favor the more complex models more than should be the
case. Various information criteria, such as Akaike's and Bayesian
information criteria (AIC and BIC, respectively, but there are several others)
have been formulated to this effect. Several such criteria are in themselves
functions of the likelihood ratio test statistic, often being equal to it, but
augmented with a term that depends on the difference in the number of model
parameters between both models being compared. The AIC is defined as:
\begin{equation}
\label{AIC}
{\rm AIC}\  \equiv\ 2 \sum_{i=1}^n u_i - 2 \sum_{i=1}^n \ell(\widehat{\bfbeta_i})
\end{equation}
\citep[][]{Akaike1987} and the BIC:
\begin{equation}
\label{BIC}
{\rm BIC}\  \equiv\ 
\left(\sum_{i=1}^n u_i\right) \ln\,(n) - 2 \sum_{i=1}^n \ell(\widehat{\bfbeta_i})
\end{equation}
\citep[][]{Schwarz1978}. The latter is a particular type of approximation to
Bayesian posterior probabilities for statistical model selection, which
penalizes stronger for complexity than the AIC \citep{Claeskens2008}.  
Both the 
AIC and BIC are only relevant for relative comparisons, where a lower value
implies a better statistical model. The AIC and 
BIC offer an alternative to the likelihood
ratio, score, and Wald test statistics discussed above, whether the
statistical models are nested or not.

A drawback for the use of AIC and BIC is that 
there is no formal (distribution) theory that can be used to select a
model. Therefore, one often resorts to rules of thumb, such as that based on the
difference between the AIC or BIC values of two models A and B. 
A guide for values of 
$\Delta$AIC$\equiv$AIC(A)-AIC(B) or $\Delta$BIC$\equiv$BIC(A)-BIC(B) 
is as follows:
$[0,2]$ for weak, $]2,6]$ for positive, $]6,10]$ for strong, and
$]10,+\infty[$ for very strong evidence in favor of model B
\citep[][Sect.\,4]{Molenberghs2005}. A related method, Bayes
factors, was described by \citet{KassRaftery1995}.

\subsubsection{\label{chol}Error estimation based on statistical models}
Once the optimal statistical model has been found to replace the ``true'' model
$\calm(\bftheta,\bfpsi)$, we need to assess the error of the corresponding
$\bftheta$ prediction based on the Mahalanobis distance as in Problem\,1. In
this case, the uncertainty of $\bfbeta$ needs to be taken into account.  Assume
that $\bfbeta$ and its variance-covariance matrix have been estimated from MLE,
$V\equiv V(\bfbeta)$, say. It is easiest to proceed in a Monte Carlo way. First,
sample $R$ values $\bfbeta_r\sim N(\widehat{\bfbeta},V)$, $r=1,\dots,R$. Then,
apply (\ref{predictionellipsoid}) for the statistical model with parameter
$\bfbeta_r$ separately, leading to $\calc_r$. Finally, consider
$$\calc=\cup_{r=1}^R \calc_r.$$  
Generating a copy  
$\bfbeta_r\sim N(\widehat{\bfbeta},V)$ can easily be done in the following way:
\begin{itemize}
\item Generate $\gamma_r\sim N(\mbox{\bfseries 0}_p, I_p)$, where $\mbox{\bfseries 0}_p$ is a column vector of zeros of length $p$ and $I_p$ is the $p\times p$ identity matrix.
\item Decompose $V=LL^\top$, where $L$ is the (lower triangular) Cholesky
  decomposition of $V$ \citep{Nash1990}.
\item Define $\bfbeta_r=\widehat{\bfbeta}+L\gamma_r$.  \\[1cm]
\end{itemize}

\subsubsection{Application}

\begin{figure}
\centering
  \includegraphics[width=0.95\linewidth]{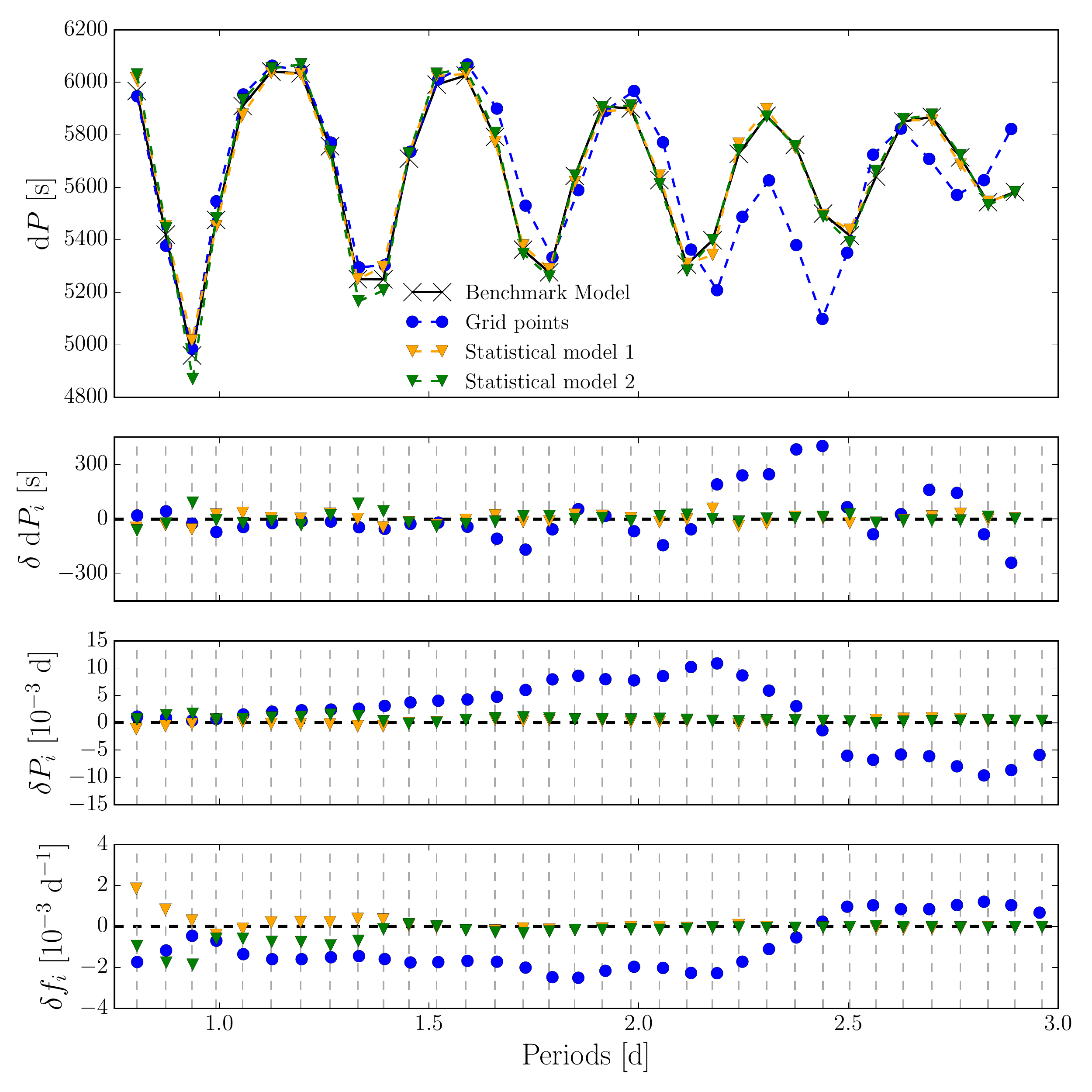}
  \caption{Comparison between ``observed'' period spacing patterns from a
    benchmark model with $(M,X,X_c,f_{\rm ov})$=$(3.25, 0.71, 0.5, 0.015)$ and
    their estimated values (upper panel) from a grid based on step overshoot
    models of similar age and from two linear statistical models deduced from
    this step overshoot grid. The differences of the period spacing (2nd panel),
    of the corresponding mode periods (3rd panel) and of the mode frequencies (lowest
    panel) are also shown.}
\label{OLS-periodspacing}
\end{figure}

\begin{figure}
\centering
\includegraphics[width=0.32\linewidth]{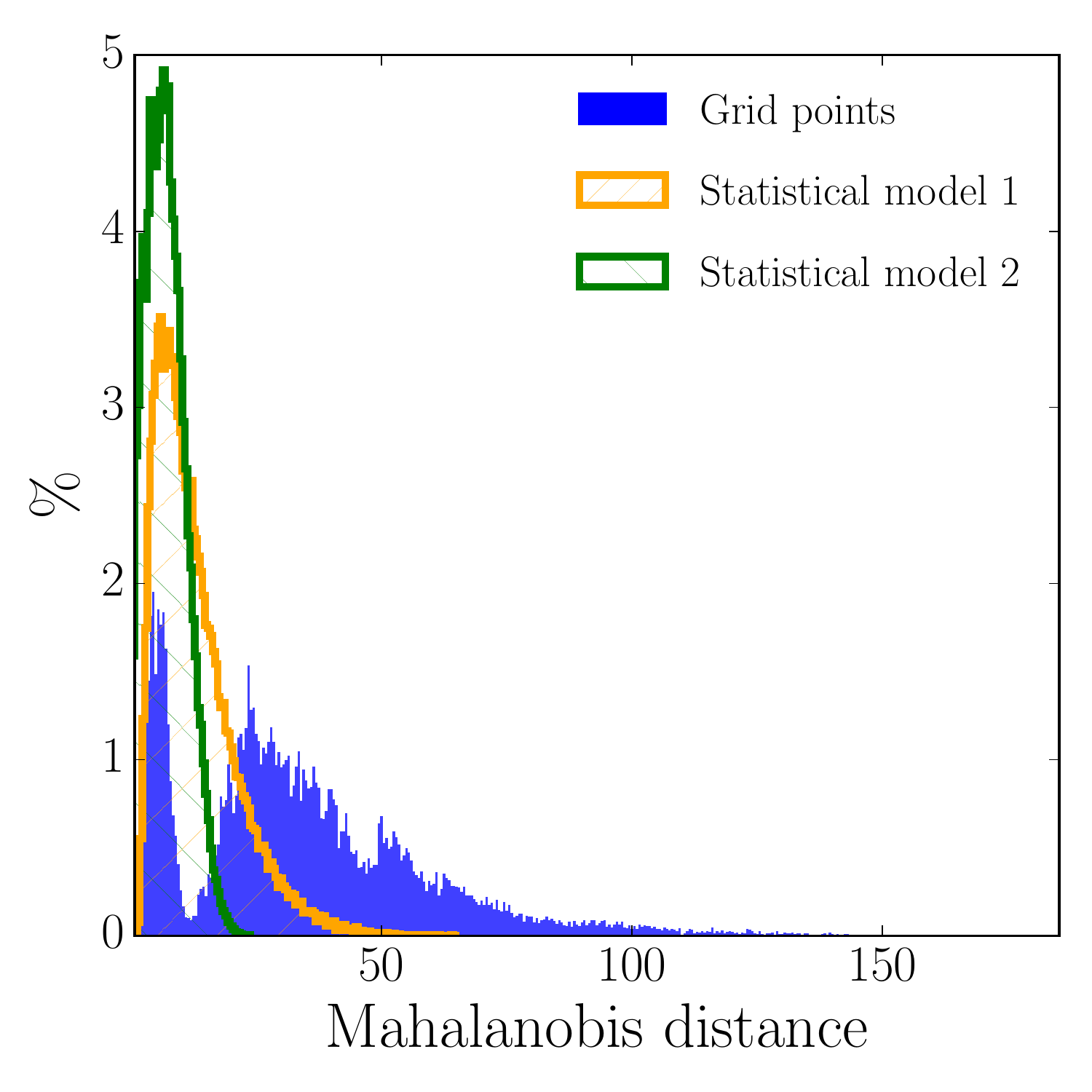}
\includegraphics[width=0.32\linewidth]{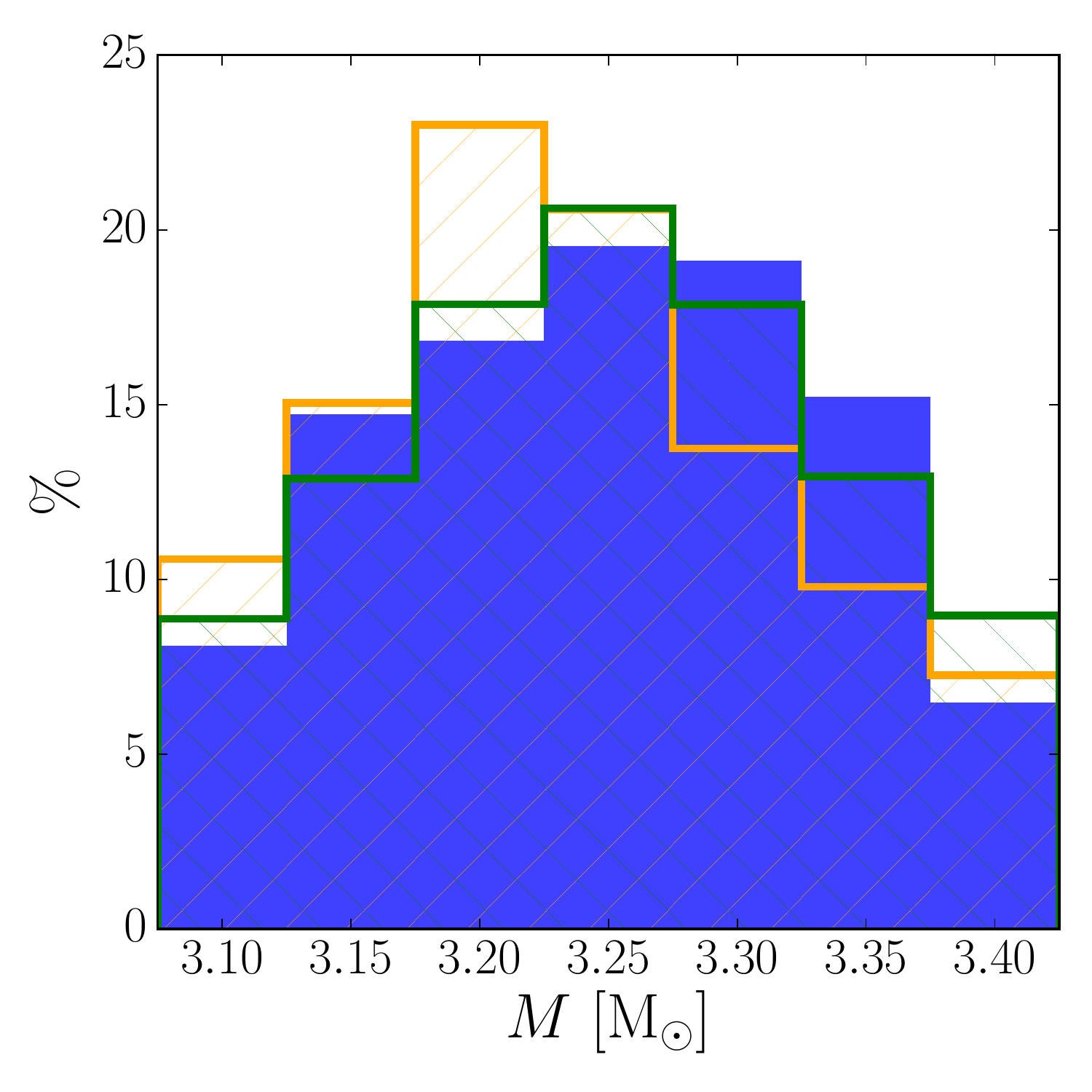}
\includegraphics[width=0.32\linewidth]{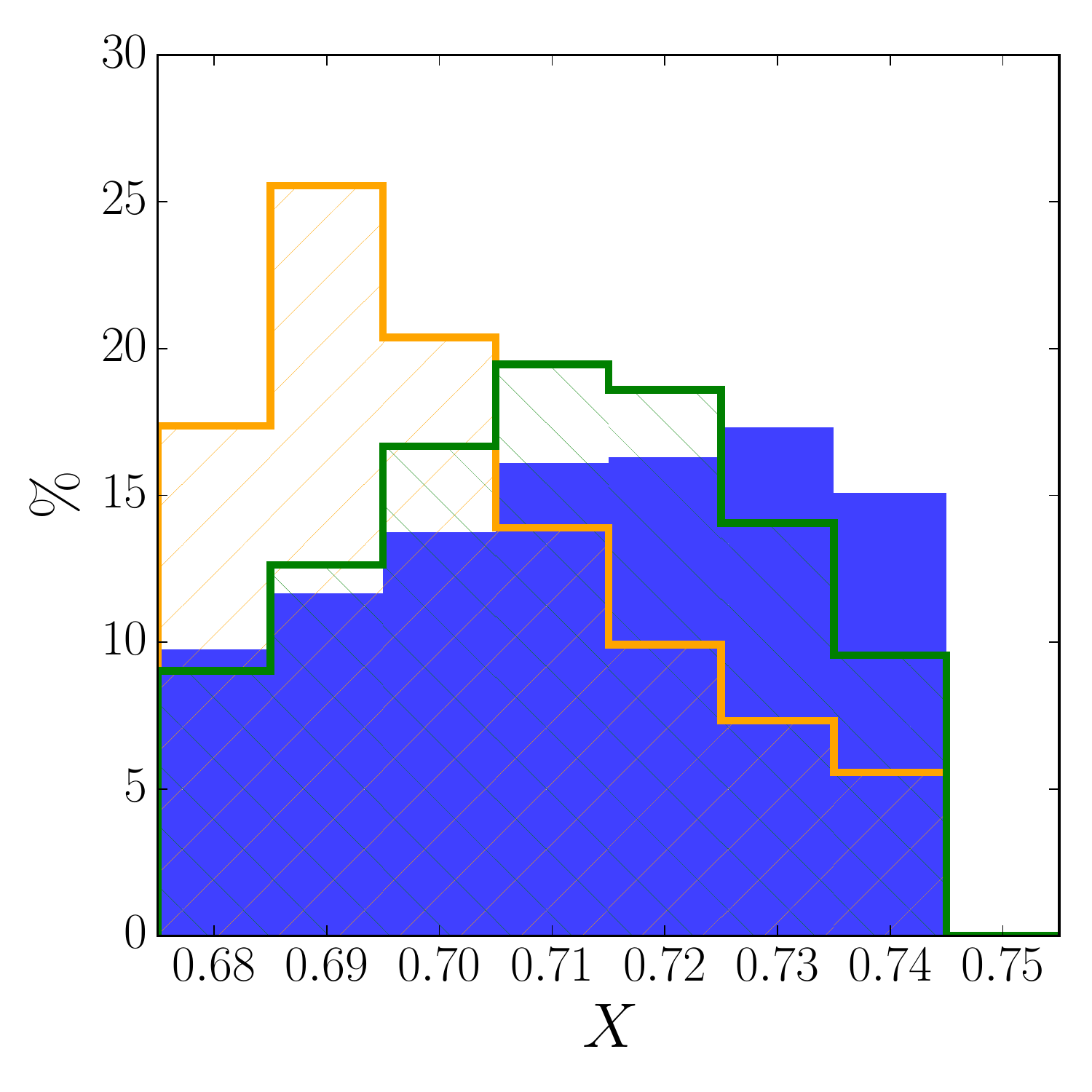}\\
\includegraphics[width=0.32\linewidth]{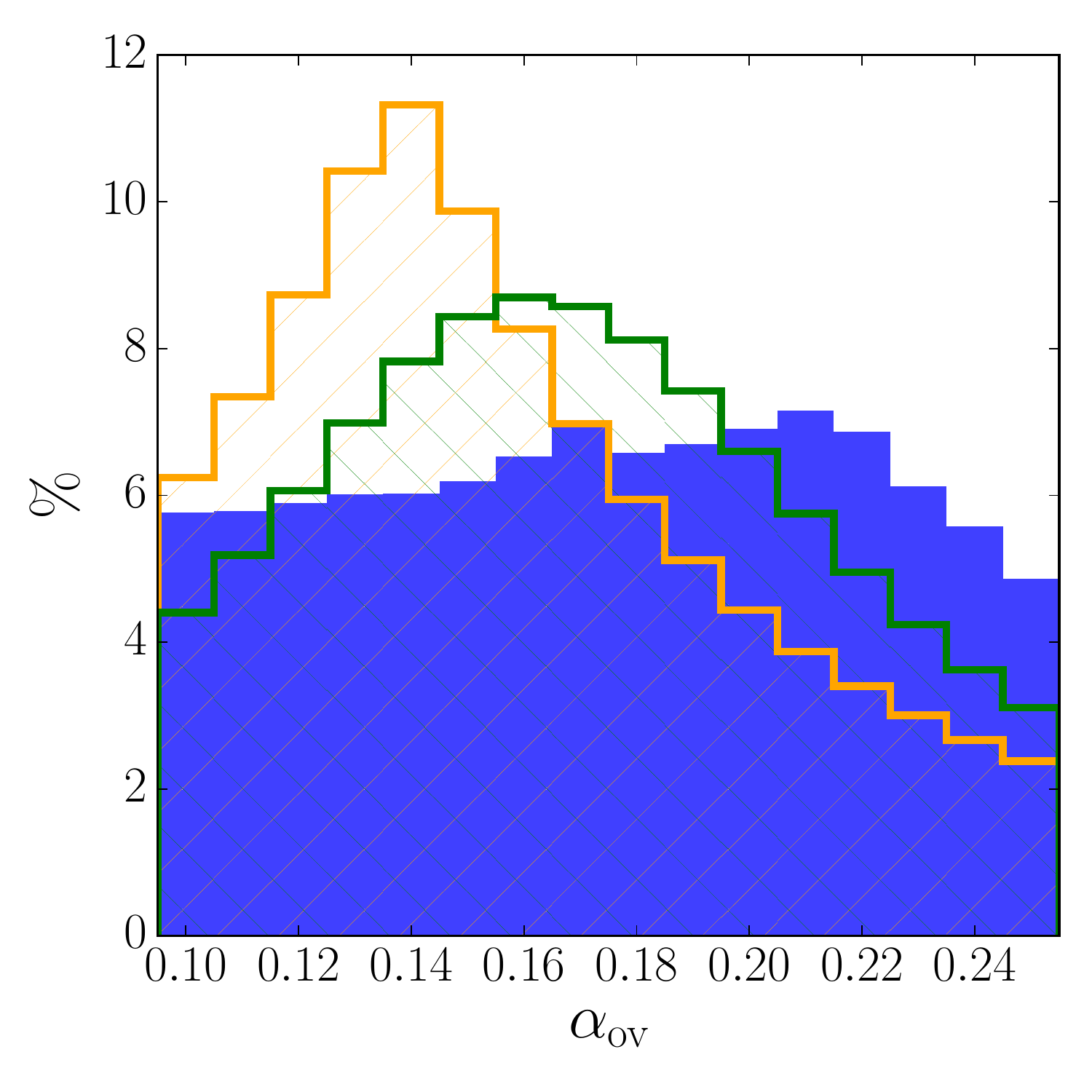}
\includegraphics[width=0.32\linewidth]{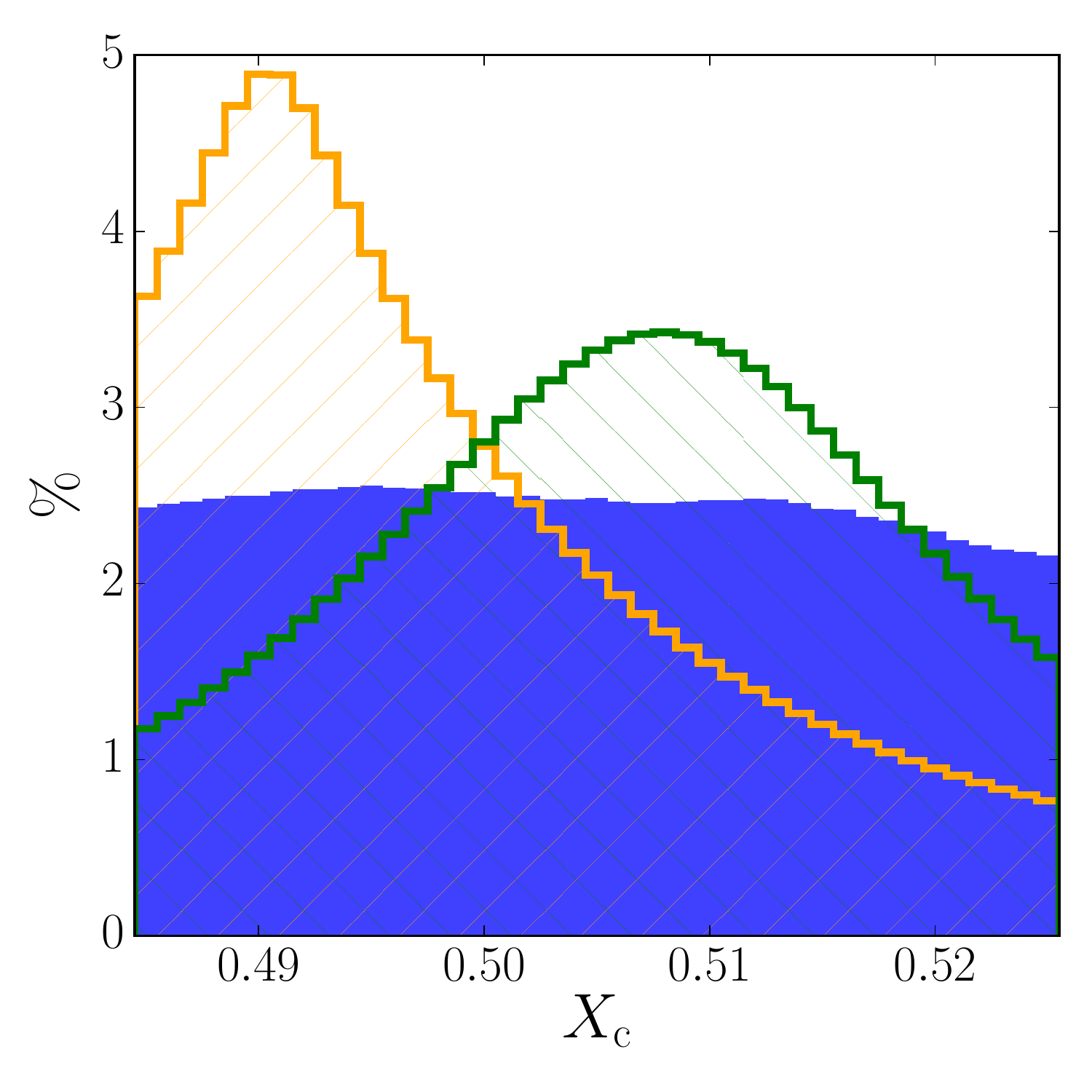}
\caption{Distributions of the Mahalanobis distance and of the four parameter
  estimations for $\bftheta$ resulting from the comparison of 34 prograde dipole
  mode frequencies of a benchmark model with $(M,X,X_c,f_{\rm
    ov})$=$(3.25, 0.71, 0.5, 0.015)$ from a grid based on step overshoot models
  of similar age and for two linear statistical models deduced from this step
  overshoot grid.}
\label{OLS-estimation}
\end{figure}

As an illustration of Problem\,3, we once more rely on the 4D grids of stellar
models by \citet[][Table\,2]{Pedersen2018}, already used in the previous section
(see Fig.\,\ref{BenchStep-Exp}).  We consider the $n=34$ frequencies of the
dipole prograde gravity-mode frequencies of the benchmark model with parameters
$(M,X,X_c,f_{\rm ov})$=$(3.25, 0.71, 0.5, 0.015)$ as an ``observed'' star with $\BY^\ast$
and
we search for the best model in the grid computed for step overshoot described
by $\alpha_{\rm ov}$ that minimizes the difference between $\BY$ and
$\BY^\ast$, where we now search for 
statistical models to replace the values of 
the 34 frequencies (omitting this time the grid models with $X_c\sim\,0.1$). 
For this illustration, we constructed two linear
nested models according to Eq.\,(\ref{modelonebisj}) for $i=1, \ldots, 34$ with:
\begin{itemize}
\item
Statistical model\,1: \\
$\bftheta=(\theta_1, \theta_2, \theta_3, \theta_4) =
(M, X, X_c, \alpha_{\rm ov}$); 
\item
Statistical model\,2: \\
$\bftheta=(\theta_1, \theta_2, \theta_3, \theta_4, \theta_5, \theta_6, 
\theta_7, \theta_8, \theta_9, \theta_{10}) =
(M, X, X_c, \alpha_{\rm ov}, M\cdot X, M\cdot X_c, M\cdot \alpha_{\rm ov},
X\cdot X_c, X\cdot\alpha_{\rm ov}, X_c\cdot\alpha_{\rm ov})$,
\end{itemize}
where we perform OLS to estimate $\bfbeta$ via Eqs\,(\ref{beta-OLS}) and
(\ref{variance-OLS}). For each of the 34 frequencies, this leads to a
$\bfbeta$-vector of dimension 5 in the case of Statistical model\,1 (an
intercept estimate and four regression coefficients for each of the four
components of $\bftheta$) and of dimension 11 for Statistical model\,2. The
choice for Statistical model\,2 was taken on the basis of correlations between
the parameters of the vector $\bftheta$ in previous seismic modeling applications
\citep[e.g.,][Fig.\,5]{Moravveji2015}.  The results from the step overshoot
grid itself and from the two linear statistical models are shown in
Figs\,\ref{OLS-periodspacing} and \ref{OLS-estimation}. The accompanying 
parameter estimates obtained from the two statistical models 
are as follows:
\begin{itemize}
\item
Statistical model\,1: 
$\widehat{M} = 3.20^{3.25}_{3.15} {M}_\odot, 
\widehat{X} = 0.69^{0.70}_{0.68}, 
\widehat{X_c} = 0.490^{0.515}_{0.485}, 
\widehat{\alpha_{\rm ov}} = 0.14^{0.21}_{0.12}$;\\
\item
Statistical model\,2:
$\widehat{M} = 3.25^{3.30}_{3.15} {M}_\odot, 
\widehat{X} = 0.72^{0.73}_{0.68},
\widehat{X_{c}} = 0.512^{0.525}_{0.490},
\widehat{\alpha_{\rm ov}} = 0.14^{0.22}_{0.10}$,
\end{itemize}
\noindent where the super- and subscripts denote the ranges of the parameters by
performing the error determination using the Monte Carlo method discussed in
Sect.\,\ref{chol}.  It can be seen that both statistical models represent the 34
``observed'' frequencies from the benchmark model very well and in both cases
better than the corresponding best set of 34 frequencies retrieved from the grid
points of the step overshoot grid.  The standard deviations of the 34
  frequencies estimated from Eq.\,(\ref{maxlikjj}) are in the range 
$\sqrt{\sigma_{ii}}\in
[0.0022,0.0281]$\,d$^{-1}$ for model\,1 and $[0.0022,0.0264]$\,d$^{-1}$ for
  model\,2.  This application shows the benefit of replacing a model grid by
  statistical models based on such a grid and application of MLE, because it
  allows a refined parameter estimation without having to compute an
  infinitely fine grid. 

  Even though our formalism is suitable for treating non-linear statistical models,
  applications to gravity-mode oscillations of main-sequence stars
will most of the time work well for
  linear statistical models whenever the dimensions of the physical quantities
  in the vector $\bftheta$ are appropriately 
chosen. In practice, one works best with
  $M$ expressed in solar masses, $X$, $X_c$ and $Z$ in percentages, $f_{\rm ov}$
  or $\alpha_{\rm ov}$ in local pressure scale height, $\log\,D_{\rm ext}$ with
  $D_{\rm ext}$ expressed in cm$^2$\,s$^{-1}$ 
and the frequencies of rotation and of
  the oscillations in d$^{-1}$. In that case, it is well known that deviations
  from linearity of the dependence of the frequencies on the stellar parameters
  $\bftheta$ remain modest, except for the more rapid phases near the
  TAMS. Early illustrations of these dependencies can be found in Fig.\,2 of
  \citet{Ausseloos2004} for pressure modes and in Fig.\,5 of
  \citet{Moravveji2015} for gravity modes of B-type pulsators.  

As for the statistical model selection, we compare the two nested linear models
by means of the various test described above. For the likelihood ratio test, we
get $T_{LR}=0.24$. Comparing this with the 5\% cutoff of a $\chi^2_{204}$ (given
that $z=34 \cdot (11-5)$) points out that Statistical model\,2 is better than
Statistical model\,1, in line with Fig.\,\ref{OLS-periodspacing}. This is not
suprising as the 2nd model has many more degrees of freedom.  Given that fact,
it is better to make the model comparison by means of the AIC and BIC, according
to Eqs\,(\ref{AIC}) and (\ref{BIC}), because these penalize for the complexity of
the model. We obtain the following values: AIC(model\,1)=-6.29
and BIC(model\,1)=253.20, while AIC(model\,2)=401.47 and BIC(model\,2)=972.33.  
Hence $\Delta$AIC=407.73 and $\Delta$BIC=719.13.
We
thus conclude 
to have found very strong evidence
that model\,1 outperforms model\,2 according to both the AIC and
BIC, i.e., the improvement for the estimation of the frequencies in the grid
points achieved by the 2nd model is insufficient to justify the more complex
model with the extra degrees of freedom.

The astrophysical conclusion of this application (see Fig.\,\ref{OLS-estimation})
is that the mass and age of a
star can be well estimated from MLE based on model grids at the level of
  variance considered in this example, even if the
shape of the core overshooting is unknown. The resulting uncertainty of
the overshoot parameter is relatively large due to its correlated nature with
the global stellar 
mass and age and with other aspects of the input physics. Note that the
latter correlation structure
is not properly taken into account when estimating overshooting
from binary or cluster modeling in the literature so far.

\subsection{Problem 4: Stellar model selection from 
an ensemble of pulsators}
Suppose that we have not one star as assumed so far, but rather an ensemble of $N$ 
stars, as well as a collection of candidate theories:
\begin{itemize}
\item Observations $\BY^\ast_t$, $t=1,\dots, N$;
\item Theories $\calm(\bftheta^{(m)},\bfpsi^{(m)})$, $m=1,\dots, M$.
\end{itemize}
The goal is, using all stars simultaneously, to select the most appropriate
theory among the $M$ candidate theories.  Using the solution to Problem 1, i.e.,
applying (\ref{maldistgrid}) to every star $t$ for every grid $m$, we find
the pair $\{\BY_{mt},\bftheta_{mt}\}$, with $\bftheta_{mt}$ the fitted stellar
properties for star $t$ from theory $m$, and $\BY_{mt}$ the predicted
observation for star $t$ from theory $m$.

The discrepancy between observed values and values predicted from theory $m$ is:
$$
D_m=\sum_{t=1}^N D_{mt}=\sum_{t=1}^N\left\{
(\BY^\ast_t -\BY_{mt})^\top V_m^{-1}(\BY^\ast_t -\BY_{mt})
\right\},
$$
with $V_m$ the dispersion matrix, calculated as in (\ref{vmat}), for the $m$th theory.
Here, the stellar models can take the form of an astrophysical model, as well as
that of a statistical model built according to the principles laid out in
Problem 3.

As argued above, it is again sensible to take the working view that
$\BY_t^\ast\sim N(Y_{mt},V_m)$. Let $t_m$ be the length of vector
$\bftheta^{(m)}$. Then $D_{mt}$ follows a $\chi^2_{n-t_m}$ distribution, and
hence $D_m$ as a whole follows a $\chi^2_{N(n-t_m)}$ distribution.  Using these
distributions, the tail probability
$p({\calm})=P\left(\chi^2_{N(n-t_m)}\ge D_m\right)$ can be calculated. The theory
with the largest $p({\calm})$ fits the observed data of the ensemble of stars
best.
 
We are not yet able to include an example of such an application, since we first
must apply the above set of Problems\,1, 2, 3 to a sufficient number of observed
gravity-mode pulsators.  This will be the subject of subsequent papers with
applications of the current methodological framework.

\subsection{Bayesian approach for parameter  estimation}

Statistical inference in asteroseismology is nowadays often done with a Bayesian
approach, where the likelihood function based on data is combined with prior
knowledge to update the information on the parameters $\bftheta$ and to revise
the probability associated with the parameters (posterior probability).  In that
sense, an 
MLE is related to Bayesian analysis with a so-called 
non-informative (or  flat) prior.
We refer to \citet{Gruberbauer2012,Appourchaux2014} for thorough
discussions in the case of solar-like oscillations in low-mass stars, where the
input physics of the model $\calm (\bftheta,\bfpsi)$ is usually taken to be
very similar to the one of solar models calibrated from helioseismology.
Indeed, for the ``easy'' case of low-mass stars with solar-like oscillations,
one can make the reasonable assumption that such stars have input physics very
similar to the Sun, because they are slow rotators and have a
  convective envelope.

In the context of gravity modes and the results obtained in
Sections\,\ref{ingredients}, \ref{pulsation-error}, and \ref{models-error}, it
is essential to start any Bayesian analysis with a non-informative
prior about the input physics and parameters to be estimated. Given the results
on theoretical uncertainties of gravity-mode frequencies of rotating stars with
a convective core, as summarized in Table\,\ref{percentages}, knowledge of the
choice of $\bftheta$ to be estimated and of the input physics in the model
$\calm (\bftheta,\bfpsi)$ to be fixed for the stellar models has yet to be
assembled for large samples of stars with a variety in mass, rotation,
metalicity, and binary properties.  As long as we do not know the level and
shape of the interior profiles for core overshooting, rotation and chemical
mixing, one should not impose insecure assumptions about these quantities,
because they have a large effect on the gravity modes. For this reason, a
Bayesian approach is not obvious yet for gravity-mode pulsators with a
convective core, because we essentially do not have any objective prior
knowledge on the choice and properties of $\calm(\bftheta,\bfpsi)$.  In this
case, one should not introduce biases about the ingredients of the stellar
models (cf.\ Table\,\ref{percentages}) but rather complete Problem\,4 for
samples of tens to hundreds of stars first.

However, using prior knowledge in the forward seismic modeling of gravity-mode
pulsators may be applied at the level of identification of the degree $l$ and/or
the azimuthal order $m$ of the gravity modes.  In all of the above, we have
assumed that each observed frequency $Y_i^\ast=f_i^\ast$ has known $(l,m)$. The
occurrence of rotational splitting indeed implies unambiguous labeling of the
mode wavenumbers $(l,m)$ \citep[e.g.,][]{Kurtz2014,Saio2015,Moravveji2015}, but
this may not necessarily be the case when only period spacing patterns are
available for mode identification \citep[e.g.,][]{VanReeth2016,
  Papics2017,Ouazzani2017,Saio2018a}.  In case of ambiguity
  \citep[e.g.][]{Degroote2010,Papics2012,Buysschaert2018}, several options for
  $(l,m)$ should be kept open, just as for the case of solar-like oscillations
  with doubtful identifications \citep[see][for an enlightening
  discussion]{Appourchaux2014}.   In this sense, model selection is
    ubiquitous in asteroseismology, even for the ``well-known'' case of
    solar-like oscillations, not only due to lack of secure mode identification
    \citep[cf.\ the F-type pulsator HD\,49333 as
    in][]{Benomar2009,Gruberbauer2009} but also given 
problems in mode detection
    \citep[e.g.,][]{Deheuvels2010,Corsaro2014,Davies2016}.  For heat-driven
    coherent modes, identification of $l$ and/or $m$ can sometimes be achieved
    from time-resolved long-term multicolour photometry and/or line-profile
    variations \citep[see][Chapter 6, for an extensive discussion and
    methodology to achieve this]{Aerts2010}.  In case of successful empirical
    identification of $(l,m)$, this information can be used as an informative
    prior.  Lack of identification may also occur at the level of a unique value
    for the radial orders $n_{pg,i}$ of each $f_i^\ast$. Indeed, this labeling
    may be ambiguous for the regime of high-order gravity modes, given the
    density of the mode spectrum at low frequencies.  One may then want to
    consider a Bayesian approach instead of MLE, so as to incorporate prior
    probabilities for the $\bftheta$ considering different priors of $(l,m,n)$
    for each of the modes.  The ensuing estimation methods is known as {\em
      maximum a posteriori estimation}.

Given that the posterior involves the likelihood function as well, its
evaluation is commonly approached in a simulation-based fashion, using Monte
Carlo Markov Chain (MCMC) methods \citep{LouisCarlin2009}. 
Care has
to be taken in order not to miss the best
solution of the score equations
(\ref{scoretweea}) and (\ref{scoretweeb}). An optimal approach is to compare MLE
and Bayesian treatments. Such comparisons in asteroseismology have been done
quite extensively before \citep[e.g.][]{Appourchaux2012a, Appourchaux2012b}, but
only in the case of solar-like oscillations, where $\calm(\bftheta,\bfpsi)$ 
is far better known
and the effect of rotation can be ignored or treated in the Ledoux approximation
\citep[e.g.,][]{Appourchaux2008,Benomar2009,Deheuvels2012,Gruberbauer2013,Appourchaux-etal2014,Lagarde2016,Rodrigues2017,Bossini2017,Handberg2017}.

Investigating MLE versus Bayesian MCMC has not been done in the context of
forward seismic modeling of gravity modes.  For applications to
intermediate-mass or high-mass stars as treated here, interior rotation and
convective core overshooting are crucial and cannot be ignored in the parameter
estimation because the gravity-mode frequency values depend heavily on it.
The same is true, although to a lesser extent, for chemical mixing
in the $\mu-$gradient zone near the receding core, as this determines the mode
trapping in that region. This fortunate circumstance in principle allows 
the properties of the core overshooting and of the near-core physics
from gravity modes to be estimated, 
in a similar way that mode trapping properties allow 
the chemical profiles of white dwarf stars to be deduced
\citep{Giammichele2018}. However, it does imply that we must first model an
ensemble of stars without biasing prior knowledge on these uncalibrated physical
phenomena.

An extensive grid-based MLE and stellar model selection procedure based on a
$\chi^2-$type analysis for gravity modes and with estimation of
the core overshooting parameter, was applied to the B-type gravity-mode
oscillator KIC\,10526294 by \citet{Moravveji2015}.  This MLE and model selection
approach was then followed by rotation profile estimation from inversion by
\citet{Triana2015}, who used the AIC to deduce the most likely rotation profile.
This combination of MLE for forward seismic modeling and stellar model selection
from a grid-based approach, followed by statistical model selection to deduce
the optimal rotation profile is the only application of its kind so far, given
that \citet{Kurtz2014,Saio2015,Kallinger2017} and \citet{Szewczuk2018}
performed the modeling with simplistic parameter
estimation, if at all, 
without considering convective overshooting  as a free parameter.

\newpage

\section{\label{scheme}Overall forward modeling scheme for 
gravity modes in rotating stars}

Forward seismic modeling requires observed and identified pulsation mode
frequencies and fits them with theoretically predicted frequencies deduced from
stellar models.  As shown in Fig.\,\ref{TA-nonTA}, it is essential to
take into account the Coriolis force in the computations of the theoretical
gravity-mode frequencies when performing forward seismic modeling of
intermediate- and high-mass stars, even for moderate rotators.  We summarize
here our overall modeling scheme, including the Coriolis force that was so far
ignored in the modeling, with the exception of the studies in
\citet{Moravveji2016} and \citet{SchmidAerts2016}.

While {\it mode frequencies\/} are derived from space photometry, the patterns that lead
to the mode identification of gravity modes as start of the modeling in absence
of rotational splitting rather concern the {\it period spacings of the modes},
$P_i^\ast=2\pi/f_i^\ast$ (cf., Fig.\,\ref{KIC-figure}). Once mode
  identification has been achieved, we perform parameter estimation
  from the observed frequencies, because these are the measured quantities from
  the data. Moreover, this has the major advantage that we can easily
  deal with hybrid pulsators, which reveal both pressure and gravity modes,
and can involve the task of searching for regularities in frequency when
identifying pressure modes in addition to gravity-mode period spacing patterns 
\citep[e.g.,][]{BowmanKurtz2018}.

As shown by \cite{VanReeth2016}, the shape
of the observed period spacings of gravito-inertial modes allows us to break down
the +7D estimation problem in a multi-step approach, where the 
rotation frequency in the
near-core region, $f_{\rm rot}^{\rm core}$, is estimated first by marginalizing
over the other six parameters.  Indeed, the near-core rotation of the star sets
the ``slope'' of the period spacing pattern and the value of this slope turns
out to be sufficiently independent of $(M,X,Z,X_c,D_{\rm ov})$ to apply a
two-stage approach. Any ``dips'' in the tilted period spacing pattern are caused
by mode trapping in the near-core region 
due to the value and shape of $D_{\rm ov}(r)$ and
$D_{\rm ext}(r)$ and these can be estimated after $f_{\rm rot}^{\rm core}$ has
been derived.

\cite{VanReeth2016} and \cite{Ouazzani2017} developed methods to estimate
$f_{\rm rot}^{\rm core}$ from the measured slope of the period spacings of
gravity modes for intermediate-mass stars; the method by \cite{VanReeth2016} was
meanwhile also successfully applied to B-type pulsators \citep{Papics2017}.
With $f_{\rm rot}^{\rm core}$ estimated, the modeling process can be continued
with exploitation of the {\it value\/} and {\it morphology\/} of the observed
period spacing patterns, including the structure of the dips due to mode
trapping in the near-core region.  In summary, we propose the following forward
modeling scheme:
\begin{enumerate}
\item Deduce the frequencies $f^\ast_i$ of pressure and  
gravity modes, and their measurement error, $\varepsilon_i^\ast$ 
in both the super-inertial and sub-inertial regime, from space
photometry. 
Select the modes revealing rotational splitting and
the $P_i^\ast=1/f^\ast_i$ of $i=1, \ldots, n$  that constitute
 period spacing patterns.
\item Identify the degree $l$ and azimutal order $m$ of rotationally-split
  modes, whenever detected.  Use the slope of the observed period spacing
  patterns $(\Delta P_i,P_i^\ast)$ to identify the degree $l$ and azimutal order $m$
  of the gravity modes $P_i^\ast$, by estimating the near-core rotation frequency
  $f_{\rm rot}^{\rm core}$, e.g., with the methods in \cite{VanReeth2016} or
  \cite{Ouazzani2017}. If pressure modes are identified from measured rotational
  splitting, deduce the rotation frequency in the stellar envelope,
  $f_{\rm rot}^{\rm env}$, from the computation of the Ledoux constant
  \citep[Eqs\,(3.354) and (3.361) in][]{Aerts2010}.  The estimation of
  $f_{\rm rot}^{\rm core}$ and/or $f_{\rm rot}^{\rm env}$ requires a sparse grid
  of stellar models $\calm(\bftheta,\bfpsi)$ for an appropriate range in mass, core
  overshoot, and metalicity and the computation of their pressure- and
  gravity-mode frequencies in an inertial frame of reference. This step will
  deliver a rough range for the mass and metalicity, as well as a narrow range
  for $f_{\rm rot}^{\rm core}$ (and of $f_{\rm rot}^{\rm env}$ 
if rotational splitting of pressure modes is
  detected).
\item Compute several dedicated +7D fine grids of stellar models with free parameters
  mass, age, initial chemical composition, for the range of 
the estimated near-core rotation $f_{\rm rot}^{\rm core}$ and its error, 
and for a wide range of 
core overshooting $D_{\rm ov}$ and envelope
  mixing $D_{\rm ext}$. For the latter three quantities, assume particular
  shapes for their profiles: $f_{\rm rot} (r)$, $D_{\rm ov} (r)$,
  $D_{\rm ext}(r)$ and also for the temperature gradient in the overshoot zone --
    radiative or adiabatic or a gradual transition between these two -- 
along with fixed choices of the input physics according to
  Table\,\ref{percentages} and treat
  those as different stellar models $\calm(\bftheta,\bfpsi)$.  
For all the model grids,
  compute their gravity-mode frequencies and identify the radial orders $n_{pg;i}$ of the
  detected modes $f^\ast_i$. For each model grid $\calm(\bftheta,\bfpsi)$,
predict and estimate the parameters $\bftheta$ in
  +7D space, and compute the uncertainties of $\bftheta$ 
from MLE, following the methodology
  developed for Problem\,1 and possibly Problem\,3 in 
Sect.\,\ref{MLE}. The user can get a feel for the effect of different input
physics on the mode frequencies and on the forward modeling 
in the context of correlated parameters $\bftheta$
from attacking Problem\,2 for various model grids  $\calm(\bftheta,\bfpsi)$.
\item Select the most likely equilibrium models $\calm(\bftheta,\bfpsi)$
  following Problem\,4 for each star separately ($N=1$), and iterate over the
  scheme for an ensemble of $N$ gravity-mode pulsators. Select the overall best
  input physics and interpret its shortcomings {\it vis-a-vis\/} the
  ensemble of pulsators from the estimated values of the variance-covariance
  matrices $V(\sigma_{ii})$. The latter offer a guide to improve the input
  physics of stellar models.
\end{enumerate}

\section{Conclusions}

In this work, we provided a proper methodological framework to perform forward
seismic modeling based on detected gravity-mode oscillation
frequencies that constitue a period-spacing pattern.  
We allow for correlations among the parameters of the stellar
models, in order to properly take into account parameter degeneracies. The
method relies on the computation of the Mahalanobis distance in a grid-based
approach.  We focused on stars rotating up to about half their critical
rotation, such that the traditional approximation provides a valid approach to
compute the theoretical oscillation frequencies. The method requires
gravity-mode oscillation frequencies that have been measured with a precision
better than 0.001\,d$^{-1}$ and that can be identified in terms of their
spherical degree $l$ and azimuthal order $m$. The parameter estimation is done
in a +7D parameter space for each of the fixed choices of the input physics.  As
a minimum, it delivers estimation of the seven parameters
$(M,X,Z,X_c,f_{\rm rot}^{\rm core},D_{\rm ov},D_{\rm ext})$ and their uncertainties, as
well as an estimate of the variance of each of the theoretical models
$\calm (\bftheta, \bfpsi)$ and its uncertainty, as a guide to improve the input
physics of future stellar structure and evolution models.  A major
aim of our future work is to achieve an
  accurate mass of the helium core near the TAMS for a representative sample of
  gravity-mode pulsators covering a large mass range.

The methodology presented here also offers a way to replace physical model
predictions of theoretical frequencies by statistical model estimations, such
that cumbersome computations of ultra-fine stellar model grids can be avoided
and that proper error estimation can be achieved from the paradigm of maximum
likelihood estimation.  A hierarchy in the importance of the physical
ingredients of the stellar structure models in terms of gravity-mode frequency
uncertainty was offered as guide of the modeling (Table\,\ref{percentages}).  

We stress that the choice of the model parameters $\bftheta$ to estimate depends
on the type of star, its rotational properties, 
and on the type of mode frequencies detected in the
data. Before the CPU-intensive procedure of forward seismic modeling is
started, the user must assemble a good understanding of the key physical
ingredients of stellar structure and evolution models and their computations for
the star under study. Only after proper understanding of stellar structure
theory can one make a sensible choice of the free parameters to estimate from forward
modeling. Our Figs\,\ref{TA-nonTA} -- \ref{diffusion} along with
Table\,\ref{percentages} offer a practical guideline for this choice.

In this work, we did not cover the case of close binary stars whose
  equilibrium models and oscillation modes are subject to tidal forces. Tides
  introduce deformations and/or instabilities in the interior structure of the
  binary components, which may be accompanied by transport of elements and of
  angular momentum.  Moreover, tidal forces affect the solutions to the
  pulsation equations. This is mainly the case for gravity modes, where
  (near-)resonances between the frequencies of free oscillations and (multiples
  of) the orbit may occur.  Just as for rotation and magnetism, the equilibrium
  models of close binaries due to tides are subject to major uncertainties.
  Also in this case, it is best to start forward modeling under the assumption
  that the tides can be treated as a secondary effect.  An iterative scheme,
in which a light curve and time-resolved
  high-precision spectroscopy covering the orbit is simultaneously solved for
  binarity and oscillations, will allow the oscillation frequencies to be derived. Once
  this has been achieved, one can perform the forward modeling as outlined here,
  where strong(er) (model-independent) constraints on the star can be included
  compared to the case of single stars. In this way, deviations between the
  observed oscillation frequencies and those of the optimal model by means of
  $\sigma_{ii}$ can be used as
  a guideline on how to improve the evolutionary models in terms of the missing
  tidal input physics.  

Our formalism presented here will be applied in the near future to numerous
gravity-mode pulsators found in the {\it Kepler\/} data, such as those in
\citet{VanReeth2016}, \citet{BowmanKurtz2018}, and \citet{Papics2017}. Such future
applications to tens of individual stars will undoubtedly improve our knowledge
of the input physics of stellar evolution models, as well as pave the way to
more automated applications to gravito-inertial asteroseismology. The 
TESS mission will offer suitable data for hundreds of gravity-mode pulsators in
the mass range 1.4 -- 40\,M$_\odot$ from its Continuous Viewing Zone
\citep{Ricker2016}, coupled to spectroscopy \citep[e.g., to be assembled with
SDSS-V,][]{Kollmeier2017}.  On the longer term, the PLATO mission will provide
thousands of suitable targets for gravity-mode asteroseismology
\citep{Rauer2014}.

\acknowledgments We express our sincere appreciation to the developer teams of
the MESA and GYRE codes, for making their software available to the astronomical
community. In particular, we thank Aaron Dotter for his advice on how to
  include 
atomic diffusion in MESA in an optimal way for the mass range considered here. 
Further, we appreciate the numerous
  suggestions from the referee to improve the presentation of our results and to
  include more details of our analyses.  CA, GM, and TVR are grateful
for the kind hospitality at the Kavli Institute of Theoretical Physics,
University of California at Santa Barbara, USA, where this work was initiated in
April 2017.  The authors are also grateful to Christel Faes of the University of
Hasselt for her inspiring tutorial on Bayesian Inference in the framework of the
Scientific Research Network ``Turning images into value through statistical
parameter estimation'' funded by the Research Foundation Flanders under FWO
grant WO.010.16N.  The research leading to these results has received funding
from the European Research Council (ERC) under the European Union's Horizon 2020
research and innovation programme (grant agreement N$^\circ$670519: MAMSIE),
from Interuniversitary Attraction Pole research Network P7/06 of the Belgian
Government (Belgian Science Policy, Belspo), and from the National Science
Foundation of the United States under Grant NSF PHY11--25915.

\software{MESA \citep{Paxton2011,Paxton2013,Paxton2015,Paxton2018}, 
GYRE \citep{Townsend2013,Townsend2018}, FASTWIND \citep{Santolaya1997,Puls2005}}

\bibliographystyle{aasjournal}
\bibliography{ms}


\appendix

\section{\label{Inlists}MESA and GYRE inlists}

Example MESA and GYRE inlists used for this work are available from the MESA
Inlists section of the MESA Marketplace: {\tt 
cococubed.asu.edu/mesa$_{-}$market/inlists.html}

In addition, a MESA/GYRE 
tutorial on gravity-mode asteroseismology can be retrieved from the Education
section, MESA summer school year 2016:
{\tt 
cococubed.asu.edu/mesa$_{-}$market/education.html}\\[0.2cm]

\section{\label{FracPol}Solving the score equations in the case of 
non-linear statistical models with fractional polynomials}
 
The aim in Problem\,3 is to solve the score equations and derive optimal values
for $\bfbeta$. The general case of non-linear statistical models to do so may be
quite challenging, although necessary in the case that linear models do not
offer a
proper fit. Although classical polynomial predictors are still very customary
and relatively easy to use, they are often inadequate 
because quickly very
high-degree polynomials are needed, resulting in poor predictive
properties.
This can in principle be overcome by using predictor functions that are
non-linear in the parameters. However, such models, in turn, are often ridden
with computational challenges. Therefore, we would like to point to a model
family that shares with polynomial methods the ease of fitting because 
it still is part of the generalized linear models family, on the one hand, and that
shares with non-linear models the flexibility of describing a wide array of
functions. This family is called ``fractional polynomials'' and was advocated by
\cite{RA94}.

A broad overview of the use and applications of this
method in a variety of statistical areas can be found in \citet{Royston2008}. 
Several uses in hierarchical data structures are described in
\citet{Verbeke2000} and \citet{Molenberghs2005}.

For a given degree $m$ and a univariate argument $x>0$, fractional polynomials
are defined as $$\beta_0+\sum_{j=1}^m\beta_jx^{[p_j]},$$ where the $\beta_j$ are
regression parameters, $x^{[p]}=x^p$ if $p\ne 0$ and $x^{[0]}=\ln(x)$.
The powers $p_1 <
\dots < p_m$ are either positive or negative integers, or fractions.  
\citet{RA94} argue that polynomials with degree higher than 2 are rarely
required in practice and further restrict the powers of $x$ to a small
pre-defined set of noninteger values:
$$\Pi=\{-2,-1,-1/2,0,1/2,1,2,\dots,\max(3,m)\}.$$ 
The full definition includes possible ``repeated powers"
which involve multiplication with $\ln(x)$.  For example, a fractional polynomial of
degree $m=3$ with powers ($-1$,$-1$,2) is of the form
$\beta_0+\beta_1x^{-1}+\beta_2x^{-1}\ln(x)+\beta_3x^2$ 
\citep{RA94,SR99}.
Setting $m=2$, for example, will
generate:
\begin{itemize}
\item[(1)] $4$ ``quadratics" in powers of $x$, represented by 
\begin{itemize}
\item $\beta_0+\beta_11/x+\beta_21/x^2$,
\item $\beta_0+\beta_11/\sqrt{x}+\beta_21/x$, 
\item $\beta_0+\beta_1 \sqrt{x}+\beta_2 x$, and 
\item $\beta_0+\beta_1 x+\beta_2 x^2$; 
\end{itemize}
\item[(2)] a quadratic in $\ln(x)$: $\beta_0+\beta_1 \ln(x)+\beta_2 \ln^2(x)$; and 
\item[(3)] several other curves with shapes different from those of low
degree polynomials. 
\end{itemize}

For given $m$, we consider as the best set of transformations, the one
producing the highest log-likelihood.  For example, the best first
degree fractional polynomial is the one with the highest 
log-likelihood among the eight models with one regressor ($x^{-2},
x^{-1},\dots,x^3$).  As with conventional polynomials, the degree $m$ is
selected either informally on {\it a priori\/} grounds or by increasing $m$
until no worthwhile improvement in the fit of the best fitting fractional
polynomial occurs.  In the above discussion, it is assumed
that $x$ is strictly positive.  If $x$ can take zero values, a preliminary
transformation of $x$ is needed to ensure positivity (e.g., $x+1$).

Evidently, the $x$ in our case would be a component of $\bftheta$. Given that
there are typically several components in $\bftheta$, the ideas above can be
applied to all components simultaneously. This may give rise to a large set of
possible predictors.


\end{document}